\documentclass[aps,onecolumn,prd,showpacs,nofootinbib]{revtex4}

\usepackage{amsmath}
\usepackage{graphicx}
\usepackage{subfigure}
\usepackage{dcolumn}
\usepackage{bm}
\usepackage{amssymb}
\usepackage{latexsym}

\newcommand{\ie}{\textsl{i.e.~}}

\newcommand{\eg}{\textsl{e.g.~}}

\def\spose#1{\hbox to 0pt{#1\hss}}
\def\lta{\mathrel{\spose{\lower 3pt\hbox{$\mathchar"218$}}
     \raise 2.0pt\hbox{$\mathchar"13C$}}}
\def\gta{\mathrel{\spose{\lower 3pt\hbox{$\mathchar"218$}}
     \raise 2.0pt\hbox{$\mathchar"13E$}}}

\newcommand{\dphi}{\delta\phi}
\newcommand{\mean}[1]{\left\langle #1 \right\rangle}

\newcommand{\mpl}{m_\mathrm{Pl}}
\newcommand{\de}[2]{\kern - #1 em \mathrm{d} #2}
\newcommand{\ini}{\mathrm{in}}
\newcommand{\cl}{\mathrm{cl}}
\newcommand{\tin}{{t_\mathrm{in}}}

\newcommand{\bea}{\begin{eqnarray}}  \newcommand{\eea}{\end{eqnarray}}
\newcommand{\beq}{\begin{equation}}  \newcommand{\eeq}{\end{equation}}
\newcommand{\bef}{\begin{figure}}  \newcommand{\eef}{\end{figure}}
\newcommand{\bec}{\begin{center}}  \newcommand{\eec}{\end{center}}
  
\newcommand{\lmk}{\left(}  \newcommand{\rmk}{\right)}
\newcommand{\lkk}{\left[}  \newcommand{\rkk}{\right]}

\newcommand\be{\begin{equation}}
\newcommand\ba{\begin{eqnarray}}
\newcommand\ea{\end{eqnarray}}
\newcommand\eq{\begin{equation}}           
\newcommand\en{\end{equation}}

\newcommand\sn{ \widetilde{N}_1}
\newcommand\snc{ \widetilde{N}_{\rm cri}}

\newcommand\phim{\phi_-}

\newcommand\GeV{\mbox{GeV}}

\begin{document}

\hskip 15.2cm \mbox{\tt RESCEU-11/09}

\title{Constraints on moduli cosmology from the production of dark
  matter and baryon isocurvature fluctuations}

\author{Martin Lemoine} \email{lemoine@iap.fr} \affiliation{Institut
d'Astrophysique de Paris, UMR 7095-CNRS, Universit\'e Pierre et Marie
Curie, 98bis boulevard Arago, 75014 Paris, France}

\author{J\'er\^ome Martin} \email{jmartin@iap.fr}\affiliation{Institut
d'Astrophysique de Paris, UMR 7095-CNRS, Universit\'e Pierre et Marie
Curie, 98bis boulevard Arago, 75014 Paris, France}

\author{Jun'ichi Yokoyama} \email{yokoyama@resceu.s.u-tokyo.ac.jp}
\affiliation{Research Center for the Early Universe (RESCEU), Graduate
  School of Science, The University of Tokyo, Tokyo, 113--0033, Japan}
\affiliation{Institute for the Physics and Mathematics of the Universe (IPMU)
The University of Tokyo, Kashiwa, Chiba 277-8568, Japan}

\date{\today}

\begin{abstract}
  We set constraints on moduli cosmology from the production of dark
  matter -- radiation and baryon -- radiation isocurvature
  fluctuations through modulus decay, assuming the modulus remains
  light during inflation. We find that the moduli problem becomes
  worse at the perturbative level as a significant part of the
  parameter space $m_\sigma$ (modulus mass) -- $\sigma_{\rm inf}$
  (modulus vev at the end of inflation) is constrained by the
  non-observation of significant isocurvature fluctuations. We discuss
  in detail the evolution of the modulus vev and perturbations, in
  particular the consequences of Hubble scale corrections to the
  modulus potential and the stochastic motion of the modulus during
  inflation.  We show, in particular, that a high modulus mass scale
  $m_\sigma\,\gtrsim\,100\,$TeV, which allows the modulus to evade
  big-bang nucleosynthesis constraints is strongly constrained at the
  perturbative level. We find that generically, solving the moduli
  problem requires the inflationary scale to be much smaller than
  $10^{13}\,$GeV.
\end{abstract}

\maketitle

\section{Introduction}
\label{sec:intro}

The existence of scalar fields with gravitational coupling to the
visible sector appears to be a generic prediction of particle physics
beyond the standard model. This, however, may cause serious
cosmological difficulties, as exemplified by the ``cosmological moduli
problem''~\cite{Coughlan:1983ci,Goncharov:1984qm}. Assuming that the
mass $m_\sigma$ of a modulus is of order of the weak scale, as one
would expect for soft masses induced by supersymmetry breaking, this
field should decay after big-bang nucleosynthesis, on a timescale
$t_\sigma\,\sim\,10^8\,{\rm sec}\,(m_\sigma/100\,{\rm
  GeV})^{-3/2}$. The ensuing high energy electromagnetic and hadronic
cascades would then ruin the success of big-bang nucleosynthesis
predictions (see~\cite{Kawasaki:2004qu} for a recent compilation and
references therein) unless the modulus energy density were extremely
small at that time. By ``extremely small'', it is meant about twenty
orders of magnitude smaller than what is generically expected for a
scalar field oscillating in a quadratic potential with an initial
expectation value of order of the Planck scale. Turned around, this
cosmological moduli problem reveals the power of big-bang
nucleosynthesis when used as a probe of high energy physics and early
Universe cosmology.

\par

Several classes of solutions have been proposed. The first one argues
that the vacuum expectation value (vev) $\sigma_{\rm inf}$ of the
modulus at the end of inflation is much smaller than the Planck
scale~\cite{Nanopoulos:1983cf,Dine:1983ys,Coughlan:1984yk,
  Dvali:1995mj,Dine:1995uk}. This is not a trivial requirement as it
demands that the effective minima of the modulus potential at low
energy (\ie well after post-inflationary reheating) and at high energy
(\ie during inflation) coincide with each
other~\cite{Goncharov:1984qm,Dine:1995uk}. Furthermore, quantum
fluctuations of the scalar field will generally push the field away
from this minimum~\cite{Linde:2005yw}.

\par

An alternative solution to the moduli problem proposes that the
modulus mass is so large that the modulus decays before big-bang
nucleosynthesis, leaving enough time for the high energy cascade to
thermalize before the process of nucleosynthesis actually
starts~\cite{Ellis:1986zt,Kawasaki:1995cy,Moroi:1995jn,Moroi:1999zb,Acharya:2008bk}.
This requires $m_\sigma\,\gtrsim\,100\,{\rm TeV}$. Although this lies
some two orders of magnitude beyond the expected soft scale, such
masses can be accommodated in successful models of supersymmetry
breaking such as anomaly mediation or no-scale supergravity as argued
in the above references.

\par

Finally, it has also been proposed to dilute the energy density
contained in the moduli through an epoch of low scale
inflation~\cite{Randall:1994fr} or thermal
inflation~\cite{Yamamoto:1985rd, Lyth:1995ka}.

\par

As formulated above, this standard moduli problem is directly
expressed as a constraint on the energy density of the modulus field
at the time of its decay. Meanwhile, progress in observational
cosmology has been such that it is now possible to constrain the
nature of density perturbations to a high degree of accuracy. Most
notably, the analysis of microwave background temperature fluctuations
allows to constrain the fraction of isocurvature modes to a quite low
level~\cite{Stompor:1995py,Enqvist:2000hp,Amendola:2001ni,Crotty:2003rz,
  Gordon:2003hw,Beltran:2004uv,Moodley:2004nz,KurkiSuonio:2004mn,
  Beltran:2005gr,Bucher:2004an,Seljak:2006bg,Bean:2006qz,
  Trotta:2006ww,Komatsu:2008hk}. As we argue in this paper (see
also~\cite{Lemoine:2009yu}), such constraints on the spectrum and the
nature of density perturbations can be translated into constraints on
moduli cosmology. We will find that the moduli problem becomes worse
at the perturbative level. The main reason is that a modulus, being
uncoupled to fields of the visible sector, inherits its own
fluctuations through inflation. At the time of reheating, there exists
an isocurvature fluctuation between the modulus and radiation, which
is transformed into a dark matter - radiation or baryon -radiation
isocurvature mode when the modulus decays into radiation, dark matter
and baryons.  In this way, one may thus picture the modulus as a
curvaton field, whose phenomenology has been intensively scrutinized
in the past few years (see notably
Refs.~\cite{Mollerach:1990ue,Lyth:2001nq,
  Enqvist:2001zp,Lyth:2002my,Moroi:2001ct,Moroi:2002rd,Linde:1996gt,
  Lyth:2003ip,Gupta:2003jc,Dimopoulos:2003ss,Ferrer:2004nv,Langlois:2004nn,Lemoine:2006sc,Multamaki:2007hv,Lemoine:2008qj}).

The generation of isocurvature fluctuations by modulus decay has been
noted before, see~\cite{Moroi:2001ct}, but to our knowledge, a
detailed analysis of the ensuing constraints on moduli cosmology has
not been given up to now. The present paper furthermore attempts at
being general and exhaustive with respect to modulus cosmology. In the
course of our discussion, we have thus obtained new results in several
places, such as those related to the evolution of the modulus and its
perturbations in the presence of supergravity corrections to the
modulus potential, or generalized existing discussions, for instance
concerning the stochastic behavior of the modulus during inflation.
The layout of the paper is as follows.  In Section~\ref{sec:general},
we describe in general terms how constraints on isocurvature
fluctuations can be turned into constraints on moduli cosmology. In
Section~\ref{sec:calc}, we then calculate for various modulus
effective potentials (time independent, or accounting for supergravity
corrections) the cosmological consequences and present the constraints
in the modulus parameter space $(m_\sigma ,\sigma_{\rm inf})$. We
summarize our findings and discuss how to evade the modulus problem at
the perturbative level in Section~\ref{sec:disc}. Finally, the paper
ends with three technical appendices, which contain results of
importance to the present study but that also possess interest of
their own. The first one, Appendix~\ref{app:stoch}, is devoted to the
calculation of the quantum behavior of the modulus field during
inflation, for large field and small field models, accounting for
possible supergravity corrections to the modulus potential.  The
second one, Appendix~\ref{app:sol}, discusses in greater details the
evolution of the modulus vev when its potential receives supergravity
corrections. The third one,
Appendix~\ref{sec:appendix:concretemodels}, presents some supergravity
based, concrete particle physics models for the modulus field. All
throughout this paper, $M_{\rm Pl}=2.42\times 10^{18}\,$GeV denotes
the reduced Planck mass.

\section{Generation of isocurvature perturbations}
\label{sec:general}

This section sets the stage for the next Section, which performs a
systematic study of the constraints obtained in moduli parameter
space. Here, we introduce the relevant physical parameters and we
describe how the amount of isocurvature fluctuations produced through
modulus decay can be calculated analytically. The formulae obtained
will be useful to interpret the results of numerical calculations
presented in the following Section.

\subsection{Background evolution}
\label{subsec:generalcons}

Let us first sketch the cosmological scenario and the outline of the
calculation. We assume that inflation proceeds at the energy scale
$H_{\rm inf}$. At the end of inflation, the slow-roll conditions are
violated (or, in the case of hybrid inflation, an instability occurs),
the inflaton field $\phi$ starts oscillating rapidly at the bottom of
its potential and the reheating period begins. If the inflaton
potential is quadratic, then the universe becomes matter-dominated. At
a later stage, the inflaton field decays into radiation, dark matter
and baryons. At this high energy scale, well above the dark matter
mass, baryons, radiation and dark matter are all part of the same
``radiation'' fluid and the universe is effectively
radiation-dominated. The temperature $T_{\rm rh}$ of the radiation
fluid at the beginning of this era, the post-inflationary reheating
temperature, is a direct function of the decay rate of the inflaton
field.

\par

Let us now consider the modulus field $\sigma$. In the following, we
denote its vev at the end of inflation by $\sigma _{\rm inf}$ and
treat this quantity as a free parameter (however, see below the
considerations on the quantum behavior of $\sigma$ during
inflation). In the post-inflationary era, in order to follow the
evolution of $\sigma$, the shape of the potential is needed. In what
follows, one considers two cases: one where the potential is purely
quadratic and one where the potential is affected by Hubble scale
contributions, meaning that a term of the form $c^2H^2(\sigma -\sigma
_0)^2$ is added to the quadratic part. If the potential is purely
quadratic and if the modulus field is a test field, then its vev
remains constant (therefore equal to $\sigma _{\rm inf}$) until
$H=m_{\rm \sigma}$. Let us notice that, when $H=m_{\sigma}$, the
energy density of the background is $3M_{\rm Pl}^2m^2_{\sigma }$ while
$\rho _{\sigma }\sim m_{\sigma }^2\sigma _{\rm inf}^2$. Then, the
condition that $\sigma $ is a test field, that is to say that its
backreaction on the expansion rate is negligible, \ie
$\rho_{\sigma}<\rho _{\phi}$, implies that $\sigma _{\rm inf}\lesssim
M_{\rm Pl}$. We thus do not discuss the possibility raised by Linde
and Mukhanov~\cite{Linde:2005yw} that the modulus becomes itself an
inflaton after having acquired a large vev through quantum jumps. On
the other hand, if the Hubble scale corrections are important, $\sigma
$ can never be considered as light (unless $c\ll 1$) since its
effective mass is always of the order of the Hubble scale. As a
result, the evolution of the modulus between the end of inflation and
the time $H=m_{\sigma }$ can become rather involved since $\sigma $
has no reason to stay constant anymore as it was the case in the
purely quadratic situation. For this reason, we postpone the detailed
discussion of the modulus evolution to the following section and we
always express our results in terms of the modulus energy density when
$H=m_{\sigma }$.

\par

At late times $H\le m_{\rm \sigma}$, Hubble corrections to the mass
term indeed become, by definition, negligible. As a consequence, the
modulus potential is then given by $m_{\sigma }^2\sigma ^2/2$. At
$H=m_{\rm \sigma}$, the modulus starts oscillating at the bottom of
this potential and $\rho_{\sigma }\propto a^{-3}$. This occurs at an
equivalent temperature scale given by
\begin{equation}
T_{\rm osci}\,=\,\left(\frac{\pi^2g_{\star,\rm
    osci}}{90}\right)^{-1/4}\left(m_\sigma M_{\rm
  Pl}\right)^{1/2}\,\simeq\,2.25\times 10^{11}\,{\rm GeV}\,
\left(\frac{g_{\star,\rm osci}}{200}\right)^{-1/4}
\left(\frac{m_\sigma}{100\,{\rm TeV}}\right)^{1/2}\ .
\end{equation}
Of course, $T_{\rm osci}$ can correspond to the temperature of the
radiation bath only if $T_{\rm osci}\,\lesssim\,T_{\rm rh}$.
Otherwise, it should be thought of as the temperature that the
radiation bath would have were the energy density contained in
radiation. Typically, one has $T_{\rm osci}>T_{\rm rh}$ unless the
reheating temperature is very high. At the onset of oscillations, the
$\sigma$ field carries a fraction $\Omega_{\sigma,\rm osci}$ of the
total energy density which can be expressed in terms of $\sigma _{\rm
  inf}$, the vacuum expectation value of $\sigma $ at the end of
inflation.

\par

Modulus eventually decays into radiation, dark matter and baryons when
$H=\Gamma _{\sigma}$, $\Gamma _{\sigma }$ being the gravitational decay
width of $\sigma$:
\begin{equation}
\Gamma_\sigma\,=\,\frac{1}{16\pi}\frac{m_\sigma^{3}}{M_{\rm
    Pl}^2}\,\simeq\, 3.51\times 10^{-24}\,{\rm
  GeV}\,\left(\frac{m_\sigma}{100\,{\rm TeV}}\right)^{3}\ .
\end{equation}
As a consequence, the decay of the $\sigma$ field occurs at a
temperature:
\begin{equation}
\label{eq:Td}
T_{\rm d}\,=\,\left(\frac{\pi^2 g_{\star,\rm
    dec}}{90}\right)^{-1/4}\left(\Gamma_\sigma M_{\rm
  Pl}\right)^{1/2}\,\simeq\,2.77\times 10^{-3}\,{\rm
  GeV}\,\left(\frac{g_{\star,\rm
    dec}}{10.75}\right)^{-1/4}\left(\frac{m_\sigma}{100\,{\rm
    TeV}}\right)^{3/2}\ .
\end{equation}
Clearly, the decay occurs much after the onset of oscillations and the
reheating.

\par

Finally, let us end this short description of the background evolution
by mentioning that we assume all throughout this paper that dark matter
originates from freeze-out of annihilations. The dark matter freeze-out
occurs at a temperature of $\sim m_{\chi }/x_{\rm f}$, where $m_{\chi }$
is the dark matter particle mass and $x_{\rm f}\sim 20-30$. Since,
typically, $m_{\chi }\sim {\cal O}\left(100\right)\, \mbox{GeV}$, one
obtains a temperature of $\sim 1-10 \,\mbox{GeV}$. Therefore, provided
$m_{\sigma}\,\lesssim 10^7\,$GeV, the modulus always decays after dark
matter freeze out. On the other hand, the freeze-out of the baryons
takes place at a temperature of $\sim 20\mbox{MeV}$. Therefore, whether
the modulus decay occurs before or after the baryons freeze-out depends
on the value of $m_{\sigma}$.

\par

The previous considerations imply that two crucial variables in this
study are $m_{\sigma }$ and $\sigma _{\rm inf}$ and, therefore, in the
following, we will express our constraints in the plan $(m_{\sigma
},\sigma _{\rm inf})$. The mass scale $m_\sigma$ should in principle
be fixed by high energy physics with a preferred range around
$10^2-10^6\,$GeV. On the contrary, the vev $\sigma_{\rm inf}$ is
determined by the early cosmological evolution. The mass scale
determines, among others, the decay time of the modulus, and together
with the vev $\sigma_{\rm inf}$, it also determines the magnitude of
the modulus energy density at the time of decay, hence the amount of
isocurvature perturbations transfered to the dark matter and baryon
fluids. The modulus vev $\sigma_{\rm inf}$ can be expressed as the sum
of two parts, one corresponding to the classical trajectory
$\sigma_{\rm cl}$ of the modulus field in its potential during
inflation, and the typical spread $\langle\delta\sigma^2\rangle^{1/2}$
around this trajectory due to quantum
effects~\cite{Starobinsky:1986fx,Starobinsky:1994bd,Finelli:2008zg}. The
standard deviation on scales larger than the Hubble radius
$\langle\delta\sigma^2\rangle^{1/2}$ has been discussed recently by
Linde and Mukhanov~\cite{Linde:2005yw} and Lyth~\cite{Lyth:2006gd}. It
is discussed in greater detail in Appendix~\ref{app:stoch}, along with
the classical trajectory $\sigma_{\rm cl}$ during inflation.

\par

For both large field and small field models, one may summarize the
situation as follows (the case of hybrid inflation is also treated in
Appendix~\ref{app:stoch}). Consider first the case in which the
modulus is effectively massless during inflation, meaning:
\begin{equation}
m_{\sigma,\rm eff}\,\ll\, \frac{H_{\rm inf,in}}{\sqrt{N_{_{\rm T}}}}\ ,
\end{equation}
where $H_{\rm inf,in}$ is the Hubble parameter at the beginning of
inflation and $N_{_{\rm T}}$ the total number of e-folds of
inflation. Let us notice that, strictly speaking, this condition is
not equivalent to $m_{\rm \sigma ,eff}\ll H_{\rm inf}$ if the Hubble
parameter evolves during inflation and/or if the total number of
e-folds is large. The modulus mass is written $m_{\sigma,\rm eff}$ to
encompass two different cases: a fixed mass $m_\sigma$ or a
(supergravity induced) Hubble scale mass $c_{\rm i}H$, with $0<c_{\rm
  i}<1$. Here, we have written $c_{\rm i}$ in order to emphasize (see
also Appendix~\ref{app:stoch} and see
Appendix~\ref{sec:appendix:concretemodels} for concrete examples) that
the effective mass during inflation is not necessarily the same as the
effective mass in the post-inflationary era (in other words, a priori,
$c_{\rm i}\neq c$). For large field $m_\phi^2\phi^2$ inflation,
$H_{\rm inf,in}N_{_{\rm
    T}}^{-1/2}=\sqrt{2/3}m_\phi\,\simeq\,1.4\times10^{13}\,$GeV is
fixed by normalization to the cosmic microwave background
anisotropies. Given the values $m_{\rm \sigma}<10^7\, \mbox{GeV}$
considered in this paper, this means that, in the pure quadratic case,
the modulus field is always massless. Of course, when Hubble scale
corrections are present and $c_{\rm i}\not \ll 1$, this is no longer
the case. For small field inflation, however, the situation is
different. Indeed, the quantity $H_{\rm inf,in}N_{_{\rm T}}^{-1/2}$
can be very small or very large depending on the inflationary scale
(which can be as low as $\sim$TeV) and the number of e-folds
($N_{_{\rm T}}\,\gtrsim\,60$ but is otherwise essentially unbounded,
see Appendix~\ref{app:stoch}). Therefore, for small field inflation,
the pure quadratic case can or cannot correspond to a massless
situation. At the classical level, one finds that, in this
``massless'' field case, $\sigma_{\rm cl}\,\sim\,\sigma_{\rm in}$,
meaning that the classical value has not changed during inflation.

Concerning the contribution of quantum effects, one finds:
\begin{equation}
\left\langle\delta\sigma^2\right\rangle^{1/2}\,\simeq\, 
\frac{H_{\rm inf,in}}{2\pi}\,
N_{_{\rm T}}^{1/2}\ .\label{eq:ds-ul}
\end{equation}
This result holds for small field inflation; for large field
$m^2\phi^2$, it is a factor $\sqrt{2}$ smaller, see
Appendix~\ref{app:stoch}. This value does not depend on $m_{\sigma,\rm
eff}$ and it diverges in the limit of de Sitter spacetime ($N_{_{\rm
T}}\rightarrow +\infty$), as expected for a massless field. This value
can actually be understood simply as follows: every e-fold $H_{\rm
inf}t$, the field performs a random step of length $\pm H_{\rm
inf}/(2\pi)$, which add up randomly, yielding the above random walk
behavior.  Setting $N_{_{\rm T}}\,\gtrsim\,60$ yields the following
lower bound:
\begin{equation}
\label{eq:varlowerlight}
\left\langle\delta\sigma^2\right\rangle^{1/2}\,
\gtrsim\,5\times10^{-6}\,M_{\rm Pl}\,
\left(\frac{H_{\rm inf, in}}{10^{13}\,{\rm GeV}}\right)\ .
\end{equation}
For small field inflation, $H_{\rm inf, in}\sim H_{\rm inf}$, \ie the
Hubble constant does not change much during inflation. For large field
$m_\phi^2\phi^2$ inflation however, as already noticed before, the
numerical prefactor is $1/\sqrt{2}$ times smaller, but $H_{\rm inf,
  in}\,\simeq \sqrt{2N_{_{\rm T}}}H_{\rm inf}$, and $H_{\rm
  inf}\,\sim\,10^{13}\,$GeV, so that overall the above bound is a
factor $\simeq 8$ larger, see Eq.~(\ref{eq:boundchm2}).

\par

Let us now consider the other limit of a massive field, $m_{\sigma,\rm
  eff}\gg H_{\rm inf,in}N_{_{\rm T}}^{-1/2}$, yet not too massive in
the sense that $m_{\sigma,\rm eff}\,\ll\,H_{\rm inf}$. This is
relevant for the pure quadratic case during small field inflation
(depending on the parameters, see above) and when the Hubble scale
corrections are present in both cases. Then, at the classical level,
one finds that the field evolves and rolls down its potential during
inflation.  Regarding the magnitude of quantum effects, one obtains:
\begin{equation}
\left\langle\delta\sigma^2\right\rangle^{1/2}\,\simeq\, 
\sqrt{\frac{3}{8\pi^2}}
\frac{H_{\rm inf}^2}{m_{\sigma,\rm eff}}\ .\label{eq:ds-l}
\end{equation}
For large field $m_\phi^2\phi^2$ inflation and typical moduli masses
$\ll 10^{13}\,$GeV, this case only applies if the modulus receives
Hubble scale mass corrections $m_{\sigma,\rm eff}=c_{\rm i}H$, since
$H_{\rm inf,in}N_{_{\rm T}}^{-1/2}\,\sim\,10^{13}\,$GeV. Furthermore,
the right hand side in the above equation should be multiplied by an
extra factor of $\sqrt{3+c^2_{\rm i}}/c_{\rm i}$ in this case, see
Eq.~(\ref{eq:boundchh2}). Of course, as $c_{\rm i}\rightarrow0$ the
field becomes light and one recovers the previous result, see
Appendix~\ref{app:stoch}.  In the case of small field inflation, the
above value can be quite small and all the more so as the scale of
inflation is lowered. Note that the above value of
$\left\langle\delta\sigma^2\right\rangle^{1/2}$ reproduces the
well-known Bunch-Davies expression for a massive field in de Sitter
spacetime~\cite{Bunch:1978yq,Vilenkin:1982wt,Linde:1982uu,
  Starobinsky:1982ee}. As discussed in Ref.~\cite{Linde:2005yw}, one
can understand this result by considering the same random walk as
before, but noting that modes on large wavelengths redshift away in
proportion to $\exp\left[-m_{\sigma,\rm eff}^2t/(3H_{\rm
    inf})\right]$, which implies that the maximum contribution to the
fluctuations has been generated during the last $\Delta N\,\sim
3H_{\rm inf}^2/m_{\sigma,\rm eff}^2$ e-folds. The product $\Delta N
\times H_{\rm inf}^2/(4\pi^2)$ then reproduces the Bunch-Davies result
(squared) to within a factor 2. In this respect, one should note that
the previous limit $m_{\sigma,\rm eff}\,\ll\,H_{\rm
  inf}/\sqrt{N_{_{\rm T}}}$ that we considered corresponds to a field
so light that modes do not have time to redshift away in $N_{_{\rm
    T}}$ e-folds. In this limit, $\Delta N$ is bounded by $N_{_{\rm
    T}}$, hence Eq.~(\ref{eq:ds-ul}) is recovered.

\par

Finally, the last case of interest is $m_{\sigma,\rm eff}\,\gg\,H_{\rm
  inf}$. In this situation, the field is too massive to be excited,
and consequently $\left\langle\delta\sigma^2\right\rangle^{1/2}$ is
exponentially suppressed.

\subsection{Evolution of perturbations}
\label{subsec:generalcons-pert}

Let us now introduce the scenario at the level of perturbations. As we
have just done for the background quantities, one can also follow the
perturbations of each species throughout the cosmic evolution. To be
more precise, we are interested in the curvature perturbation for the
species ``$\alpha $'' defined
by~\cite{Wands:2000dp,Lyth:2003im,Lyth:2004gb,Martin:1997zd}
\begin{equation}
\zeta _{\alpha}\equiv  -\Phi 
-H\frac{\Delta \rho _{\alpha}}{\dot{\rho}_{\alpha}}
\simeq -\Phi+\frac{\Delta _{\alpha}}{3\left(1+\omega
_{\alpha}\right)}\, ,
\end{equation} 
where $\Phi $ is the Bardeen potential, $\Delta _{\alpha}$ the
gauge-invariant density contrast and $\omega _{\alpha }\equiv p_{\alpha
}/\rho _{\alpha }$ the equation of state parameter.

\par

After the decay of the inflaton field, the fluctuations in $\phi $
have been transmitted to radiation, characterized by $\zeta _{\gamma
}^{\rm (i)}$, dark matter, $\zeta _{\chi }^{\rm (i)}$ and baryons (and
anti-baryons), $\zeta _{\rm b }^{\rm (i)}$, $\zeta _{\rm \bar{b}
}^{\rm (i)}$. Since these fluids share thermal equilibrium, one
has~\cite{Weinberg:2004kf} $\zeta _{\gamma }^{\rm (i)}=\zeta _{\chi
}^{\rm (i)}=\zeta _{\rm b }^{\rm (i)}= \zeta _{\rm \bar{b} }^{\rm
  (i)}$. Let us notice that, at this stage, dark matter and baryons
are still relativistic fluids. Indeed, dark matter becomes
non-relativistic at a temperature of $\sim m_{\chi}={\cal O}(100)\,
\mbox{GeV}$. Regarding the baryons, the situation is more complicated
since, in principle, they become non-relativistic at a temperature of
$\sim 1\, \mbox{GeV}$ (about the same as the dark matter freeze-out
temperature), that is to say well below the reheating temperature, but
in fact at that temperature one still have a quark-gluon plasma. 

We define the above initial conditions, indexed with $^{\rm (i)}$,
well into the modulus oscillations era, at $H\,\ll\,m_\sigma$ and
$T\,<\,T_{\rm rh}$, and before the modulus comes to dominate the
energy density. In this era, the modulus can be considered as a
pressureless fluid, supergravity contributions to its potential have
become negligible, hence previous results on curvaton phenomenology
can be applied, as discussed further below. One needs however to
relate the modulus curvature perturbations at this time,
$\zeta_{\sigma}^{\rm (i)}$ to the modulus perturbations acquired
through inflation. This obviously depends on the modulus potential at
$m_\sigma\,<\,H\,<\,H_{\rm inf}$.

For a simple time independent quadratic modulus potential, one can use
the results of Ref.~\cite{Langlois:2004nn}, which give:
\begin{equation}
\zeta_{\sigma}^{\rm (i)}\,=\,-\frac{3}{2}\Phi_{\rm inf} +
\frac{2}{3}\frac{\delta\sigma_{\rm inf}}{\sigma_{\rm
    inf}}\ .\label{eq:ziniquad}
\end{equation}
The quantity $\Phi_{\rm inf}$ denotes the Bardeen potential at the end
of inflation, $\delta\sigma_{\rm inf}$ denotes the modulus
perturbations on large scales and the calculation assumes that the
modulus behaves as a test field, so that $\Phi$ is approximately
constant: $\Phi_{\rm inf}=\Phi^{\rm (i)}$. It also assumes that
radiation dominates at the time at which $\zeta_{\sigma}^{\rm (i)}$ is
defined. It is important to realize that the radiation curvature
perturbation is related to the Bardeen potential through
$\zeta_{\gamma}^{\rm (i)}\,=\,-3\Phi^{\rm (i)}/2$, so that the initial
modulus -- radiation isocurvature perturbation can be rewritten as
\begin{equation}
S_{\sigma\gamma}^{\rm (i)}\,=\,2\frac{\delta\sigma_{\rm
    inf}}{\sigma_{\rm inf}}\ .\label{eq:Siniquad}
\end{equation}
In Section~\ref{subsec:sugracorrections}, we show that this result
holds even when the modulus potential receives a supergravity inspired
$+c^2H^2$ quadratic mass term. This result is of importance for the
present discussion, since it shows that the modulus -- radiation
isocurvature fluctuation disappears in the limit $\delta\sigma_{\rm
  inf}/\sigma_{\rm inf}\rightarrow 0$. One way to achieve this is to
assume that the modulus is heavy during inflation, either because
$H_{\rm inf}\,\lesssim\,m_{\sigma}$ or because the modulus receives an
effective mass term $+c_{\rm i}^2H^2$ during inflation. Furthermore,
Ref.~\cite{Yamaguchi:2005qm} has shown that the isocurvature mode
between the inflation and any heavy field actually disappears during
inflation because the heavy field is drawn to the minimum of its
potential at every point in space, so that in this case, there would
not even be an isocurvature fluctuation to start with, at the end of
inflation. For this reason, we discard for now this case and assume
everywhere that the modulus has remained light during inflation, in
which case $\delta\sigma_{\rm inf}/\sigma_{\rm inf}\,\simeq\, H_{\rm
  inf}/(2\pi\sigma_{\rm inf})$. In
Appendix~\ref{sec:appendix:concretemodels}, we present several
concrete models of inflationary model building in a supergravity
framework; for both models of D-term inflation, it is found that the
modulus remains light during inflation, but acquires a Hubble
effective mass after inflation.

\par

In the following, we set $\zeta_\gamma^{\rm
  (i)}\,\simeq\,10^{-5}$. From the time at which the initial
conditions are defined, all the $\zeta _{\alpha}$ remain constant
until dark matter freeze-out (in between, the dark matter has become
non-relativistic, see before). During this phase, the radiation,
baryons and modulus fluid perturbations are not affected,
\begin{equation}
\zeta
_{\gamma }^{>_{\rm f}}=\zeta _{\gamma }^{\rm (i)}\, ,
\quad
\zeta
_{\rm b}^{>_{\rm f}}=\zeta _{\rm b}^{\rm (i)}\, ,
\quad
\zeta
_{\rm \bar{b}}^{>_{\rm f}}=\zeta _{\rm \bar{b}}^{\rm (i)}\, ,
\quad
\zeta
_{\rm \sigma}^{>_{\rm f}}=\zeta _{\rm \sigma}^{\rm (i)}\, ,
\end{equation}
but the dark matter perturbations are modified according
to~\cite{Lyth:2003ip,Lemoine:2006sc}
\begin{equation}
\zeta
_{\chi }^{>_{\rm f}}=\zeta _{\chi }^{\rm (i)}
+\frac{(\alpha _{\rm f}-3)\Omega _{\rm \sigma }^{>_{\rm f}}}
{2(\alpha _{\rm f}-3)+\Omega _{\rm \sigma }^{>_{\rm f}}}\left[
\zeta _{\rm \sigma}^{\rm (i)}-\zeta _{\gamma }^{\rm (i)}
\right]\, ,
\end{equation}
with $\alpha _{\rm f}\equiv x_{\rm f}+3/2$. From the above equation,
one sees that the quantity $\zeta _{\chi }$ is not modified if $\Omega
_{\rm \sigma }^{>_{\rm f}}\rightarrow 0$ (\ie the modulus is
negligible at dark matter freeze-out) and/or $\zeta _{\rm \sigma}^{\rm
  (i)}=\zeta _{\gamma }^{\rm (i)}$ in which case the freeze-out
surface exactly coincides with the uniform radiation surface.

\par

Then, the $\zeta_{\alpha}$'s remain constant until the baryons freeze
out (assuming it occurs before modulus decay). Through this stage, it
is clear that $\zeta_{\gamma }$, $\zeta _{\chi}$ and $\zeta _{\sigma}$
remain unaffected. On the contrary, one expects $\zeta _{\rm b}$ and
$\zeta _{\rm \bar{b}}$ to evolve. One can consider a ``net baryon
number'' fluid, the energy density of which is given by $\Omega_{\rm
  b}-\Omega _{\rm \bar{b}}$. Before baryons freeze out, this fluid is
made of baryons and anti-baryons in thermal equilibrium (with a small
excess of baryons) but after the freeze out of annihilations, it is
essentially made of baryons. In the absence of any baryon number
violating process, this fluid of ``net baryon number'' is isolated,
hence its curvature perturbation remains constant. Therefore, after
baryons freeze-out, one has $\zeta_{\rm b}=\zeta_{\rm b}^{>_{\rm
    f}}=\zeta_{\rm b}^{\rm (i)}$ and $\zeta _{\rm \bar{b}}=0$. Notice
that the same reasoning would also be valid in the case where the
freeze-out occurred after modulus decay. In fact, the above discussion
would be modified only if baryon number violation occurred after
modulus decay.

\par

Finally, the modulus decays in dark matter, radiation, baryons and
anti-baryons and it is clear that all the corresponding curvature
perturbations are then modified. We obtain (see
Refs.~\cite{Lemoine:2006sc,Lemoine:2008qj} for details):
\begin{eqnarray}
\label{eq:zetagammafinal}
\zeta_{\gamma }^{>_{\rm d}} &=& \zeta_{\gamma }^{<_{\rm d}}
+r^{<_{\rm d}}\left(
\zeta _{\rm \sigma}^{<_{\rm d}}-\zeta _{\gamma }^{<_{\rm d}}
\right)
=\zeta_{\gamma }^{\rm (i)}
+r^{<_{\rm d}}\left[
\zeta _{\rm \sigma}^{\rm (i)}-\zeta _{\gamma }^{\rm (i)}
\right]\, ,
\\
\zeta_{\chi }^{>_{\rm d}} &=& \zeta_{\chi }^{<_{\rm d}}
+\frac{B_{\chi}\Omega _{\sigma }^{>_{\rm f}}}
{\Omega _{\chi }^{>_{\rm f}}+B_{\chi}\Omega _{\sigma }^{>_{\rm f}}}
\left(
\zeta _{\rm \sigma}^{<_{\rm d}}-\zeta _{\chi }^{<_{\rm d}}
\right)=\zeta_{\chi }^{>_{\rm f}}
+\frac{B_{\chi}\Omega _{\sigma }^{>_{\rm f}}}
{\Omega _{\chi }^{>_{\rm f}}+B_{\chi}\Omega _{\sigma }^{>_{\rm f}}}
\left[
\zeta _{\rm \sigma}^{\rm (i)}-\zeta _{\chi }^{>_{\rm f}}
\right]\, ,\\
\zeta_{\rm b }^{>_{\rm d}} &=& \zeta_{\rm b }^{<_{\rm d}}
=\zeta_{\rm b }^{\rm (i)}\, .
\end{eqnarray}
As explained in Refs.~\cite{Lemoine:2006sc,Lemoine:2008qj}, in order
to obtain these relations, we have assumed that a fraction
$B_\chi\equiv \Gamma_{\sigma \chi}/\Gamma _{\sigma}\,\ll\,1$ of the
$\sigma$ energy density goes into dark matter particles. One can
relate this parameter $B_\chi$ to the branching ratio of modulus decay
into dark matter particles as follows. Assuming that dark matter
particles thermalize instantaneously, then if one modulus produces
through its decay $N_{\chi}$ particles, one finds that $B_\chi \,=\,
N_\chi m_\chi/m_\sigma$. In the range of parameters that we are
interested in, $m_\chi\,\ll\,m_\sigma$ and $N_\chi\,\lesssim 1$, see
Refs.~\cite{Moroi:1999zb,Kohri:2005ru} for a detailed
discussion. Strictly speaking, the thermalization is
quasi-instantaneous only in the high modulus mass range, $m_{\sigma }
\gta 100\, \mbox{TeV}$, while at lower masses some redshifting due to
the cosmic expansion occurs. This can be seen as follows. Using the
results of Refs.~\cite{Hofmann:2001bi,Bertschinger:2006nq}, one can
write down the scattering cross-section of Compton-like processes
$\chi +\ell \rightarrow \chi +\ell $ (with $\ell$ an
ultra-relativistic lepton of the thermal bath, \eg a neutrino or an
anti-neutrino at the BBN epoch):
\begin{equation}
\sigma _{\chi\ell}\simeq \frac{3C}{128\pi}
\frac{\left(s-m_{\chi}^2\right)^2}{m_{\chi}^2s^2}\, ,
\end{equation}
with $C$ a prefactor of order unity defined in
Ref.~\cite{Bertschinger:2006nq}, and $s$ the standard center of mass
energy squared: $s\simeq 2E_{\chi }E_{\ell}+m_{\chi}^2$ with
$E_{\ell}\simeq 3.15 T$ where $T$ is the temperature of the thermal
bath. In the region of interest (at modulus decay), one can check that
$2E_{\chi }E_{\ell}\ll m_{\chi}^2$. The ratio of the interaction rate,
$\Gamma_{\chi \ell} \equiv n_{\ell}\sigma _{\chi\ell}$ to the Hubble
rate at the time of modulus decay can then be expressed as
\begin{equation}
\frac{\Gamma _{\chi \ell}}{H}\biggl \vert _{\rm d}=1.4\times 10^{8}C 
\left(\frac{E_{\chi}}{m_{\sigma}}\right)^2
\left(\frac{m_{\sigma }}{100 \, \mbox{TeV}}\right)^{13/2}
\left(\frac{m_{\chi}}{100 \, \mbox{GeV}}\right)^{-6}\, .
\end{equation}
As a consequence, if $m_{\sigma }\gta 30 \, \mbox{TeV}\, C^{-2/9}
\left(m_{\chi}/100 \, \mbox{GeV}\right)^{8/9}$, one has $\Gamma
_{\chi \ell}\gta H$ for all energies $E_{\chi}>m_{\chi}$, which
implies that the $\chi $-particle becomes non-relativistic through
multiple interactions in less than a Hubble time. In this range, and
for our purpose, one can treat the dark matter fluid as a pressureless
fluid immediately after modulus decay. Inversely, if $m_{\sigma }\lta
7 \, \mbox{TeV}\, C^{-2/13} \left(m_{\chi}/100 \,
  \mbox{GeV}\right)^{12/13}$, $\Gamma _{\chi \ell}\lta H$ for all
energies $E_{\chi}<m_{\sigma}/2$: the particle never interacts and
simply redshifts to non-relativistic velocities within $\ln
[m_{\sigma}/(2m_{\chi})]$ e-folds. Finally, in the intermediate range,
the $\chi $ particle looses its energy through interactions to some
intermediate value $90\, \mbox{GeV}\, C^{-1/2}\left(m_{\sigma }/100 \,
  \mbox{TeV}\right)^{-9/4} \left(m_{\chi}/100 \,
  \mbox{GeV}\right)^{3}$, then redshifts away down to
$m_{\chi}$ through cosmic expansion. 

\par

All in all, our above assumption of ``instantaneous thermalization''
amounts to neglecting this redshifting factor, which in turn slightly
overestimates the abundance of $\sigma $ produced dark matter by a
factor which never exceeds $\simeq 35 $. This value applies at
$m_{\sigma }\simeq 7\, \mbox{TeV}$ and it rapidly decreases to unity
away from this value. This only affects very marginally the results
derived below. 

\par

We have also assumed that the fraction of modulus energy density
$B_{\rm b+\bar b}$ that goes into baryons and anti-baryons is very
much smaller than unity, as one would expect. With respect to
Ref.~\cite{Lemoine:2008qj}, we have also assumed here that modulus
decay preserves baryon number. Finally, the parameter $r^{<_{\rm d}}$
that appears in the first of the above equations has been introduced
in Ref.~\cite{Lyth:2002my}; if dark matter is entirely produced by
modulus decay (\ie $\Omega_\chi^{>_{\rm f}}\,\approx\,0$), then
$1-r^{<_{\rm d}}$ characterizes the amount of initial modulus -
radiation isocurvature mode that is transfered through modulus
decay. We use the simple formula:
\begin{equation}
\label{eq:rd}
r^{<_{\rm d}}\,\simeq\,\Omega_\sigma^{<_{\rm d}}\ ,
\end{equation}
which has been found numerically to be a good approximation~\cite{Gupta:2003jc}.

\par

After modulus decay, the $\zeta_{\alpha}$'s remain constant throughout
the subsequent cosmic evolution. The corresponding values can be
compared to CMB data.

\subsection{Transfer of isocurvature modes}
\label{subsec:transferiso}

As we will see shortly, dark matter - radiation and baryon - radiation
isocurvature modes are generated in different regions of the parameter
space. The constraints obtained are thus complementary to each
other. For this reason, and for the sake of clarity, we discuss the
generation of each mode in turn. Let us also recall that the
isocurvature perturbations between two fluids ``$\alpha$'' and ``$\beta
$'' is defined by $S_{\alpha \beta } \equiv 3\left(\zeta_{\alpha }-\zeta
_{\beta}\right)$.

\par

The dark matter - radiation and baryon isocurvature modes on large
scales are given by~\cite{Lemoine:2006sc}:
\begin{eqnarray}
S_{\chi\gamma}^{>_{\rm
    d}}\,&\simeq &\,\frac{1}{1+\Upsilon_\chi}\left[\frac{(\alpha_{\rm
      f}-3)\Omega_{\sigma}^{>_{\rm f}}}{ 2(\alpha_{\rm
      f}-2)+\Omega_{\sigma}^{>_{\rm f}}}\frac{\Omega_{\chi}^{>_{\rm
        f}}}{ \Omega_{\chi}^{>_{\rm f}} +
    B_\chi\Omega_{\sigma}^{>_{\rm f}}} +
  \frac{B_\chi\Omega_{\sigma}^{<_{\rm d}}}{ \Omega_{\chi}^{<_{\rm d}}
    + B_\chi\Omega_{\sigma}^{<_{\rm d}}}-r^{<_{\rm d}}\right]
S_{\sigma\gamma}^{\rm (i)}\ ,\label{eq:Scr}\\
S_{\rm b\gamma}^{>_{\rm d}}\,&\simeq&\, -\Omega_{\sigma}^{<_{\rm d}}
S_{\sigma \gamma}^{\rm (i)}\ .\label{eq:Sbr}
\end{eqnarray}
These formulae can be straightforwardly deduced from the results quoted
in the previous section, except the presence of the parameter
$\Upsilon_\chi$ in $S_{\chi\gamma}^{>_{\rm d}}$. This parameter
represents the ratio of the dark matter annihilation rate to the
expansion rate immediately after the decay of the modulus field. If this
latter produces sufficiently many dark matter particles, these may
annihilate with each other. As $\Upsilon_\chi\,\gg\,1$, meaning that
annihilations are effective, the isocurvature perturbation transfer is
partially erased (see Ref.~\cite{Lemoine:2006sc} for details). One
finds:
\begin{equation}
\Upsilon_\chi\,\simeq\,B_\chi \frac{m_\sigma}{m_\chi}n_\sigma^{<_{\rm
    d}}\frac{1} {\Gamma_\sigma}\frac{g_{\star,\rm
    f}^{1/2}}{0.076}{\rm e}^{x_{\rm f}}\frac{\sqrt{x_{\rm f}}} {m_\chi
  M_{\rm Pl}}\,\approx\,1.68\times 10^{-8} B_\chi
\Omega_\sigma^{<_{\rm d}} \left(\frac{m_\sigma}{100\,{\rm
    TeV}}\right)^3\left(\frac{m_\chi}{200\,{\rm GeV}}\right)^{-2}
\left(\frac{g_{\star,\rm f}}{100}\right)^{1/2}\times \sqrt{x_{\rm
    f}}{\rm e}^{x_{\rm f}}\ .\label{eq:upsilon}
\end{equation}
The above formula neglects the amount of dark matter initially
present; this is a good approximation insofar as the annihilation rate
of these dark matter particles is very much smaller than the Hubble
rate after freeze-out in the absence of modulus decay, which would
lead to $\Upsilon_\chi\,\ll\,1$ hence to a negligible correction to
the equation for the transfer of isocurvature mode. Note that the last
factor involving $x_{\rm f}$ in the above equation may be quite large,
being $\sim 2.2\times 10^9$ for $x_{\rm f}=20$ and $5.9\times 10^{13}$
for $x_{\rm f}=30$. Finally, note that the above formula is only
approximate (see Ref.~\cite{Lemoine:2006sc} for details). The
constraints that we derive further below are obtained through the numerical 
integration of the full set of equations of motion and are therefore
more accurate.

\par

A significant dark matter -- radiation isocurvature mode is generated if
both following conditions are satisfied:
\begin{equation}
B_\chi\Omega_\sigma^{<_{\rm d}}\,\gg\,\Omega_\chi^{<_{\rm d}}\ ,
\quad \Omega_\sigma^{<_{\rm d}}\,\ll\,1 \ .\label{eq:Scr-cond}
\end{equation}
The former condition expresses the fact that the amount of moduli
produced dark matter particles exceeds that coming from freeze-out of
annihilations, while the latter requires that the modulus energy
density is not sufficient to affect the radiation content. All in all,
this means that the modulus perturbation is transferred to the dark
matter fluid but not to radiation. One could also imagine that the
modulus perturbations are transfered to the radiation sector but not
to the dark matter sector, thereby generating a net isocurvature
perturbations. However, this would require either blocking the decay
of modulus to dark matter, which is unlikely as the dark matter
particle is always much lighter than the modulus in the parameter
space we are interested in, or, having the modulus decay after matter
-- radiation equality, which is forbidden by constraints on cosmic
microwave background distortions.

\par

At this stage, we need to compute explicitly the parameters appearing
in Eqs.~(\ref{eq:Scr}) and~(\ref{eq:Sbr}). In particular, from the
expression of the dark matter annihilation cross-section, one obtains
\begin{equation}
\label{eq:omegachif}
\Omega_\chi^{>_{\rm f}}\,\simeq\, 1.67\times 10^{-3}x_{\rm f}^{3/2}
{\rm e}^{-x_{\rm f}}\ .
\end{equation}
One also needs to evaluate the quantity $\Omega _{\sigma}^{>_{\rm f}}$
and $\Omega _{\sigma}^{<_{\rm d}}$. From the fact that the energy
density of the modulus scales as $a^{-3}$, one obtains that $\Omega
_{\sigma}$ just after the dark matter freeze out can be expressed as
\begin{equation}
\label{eq:omegasigmaf}
\Omega _{\sigma}^{>_{\rm f}}=\Omega _{\rm \sigma , osci}
\frac{x_{\rm f}}{m_{\chi}}{\rm min}\left(T_{\rm
    osci}\ , \,\, T_{\rm rh}\right)\Omega _{\gamma}^{>_{\rm f}}\, .
\end{equation}
In this expression, the minimum of $T_{\rm osci}$ and $T_{\rm rh}$
appears because, if the oscillations start before the end of
reheating, then the inflaton and modulus energy densities have the
same scaling until the reheating stage is completed. Only below
$T_{\rm osci}$ or $T_{\rm rh}$, whichever is smaller, the energy
density of the $\sigma $ oscillations increases relatively to
radiation energy density. In order to obtain the above formula, we
have also assumed that the modulus can never start a new phase of
inflation. In the same manner, immediately prior to decay, the ratio
of the energy density contained in $\sigma$ oscillations to that
contained in radiation reads:
\begin{equation}
\Omega _{\sigma}^{<_{\rm d}}
=
\Omega_{\sigma,\rm osci}\,\frac{{\rm min}\left(T_{\rm
    osci}\ , \,\, T_{\rm rh}\right)}{T_{\rm
    d}}\Omega _{\rm \gamma}^{<_{\rm d}}\ .\label{eq:rsig}
\end{equation}

\par

We are now in a position where one can calculate the transfer of
isocurvature perturbation from modulus - radiation to dark matter -
radiation and baryon - radiation. This is done in the Section that
follows.

\section{Constraints in the modulus parameter space}
\label{sec:calc}

Our present goal is to compute the amount of isocurvature perturbation
produced through the differential decay of the modulus into dark
matter and radiation, as well as between the baryon and the radiation
fluid, which can then be compared to existing bounds obtained through
the analysis of cosmic microwave background
fluctuations~\cite{Stompor:1995py,Enqvist:2000hp,Amendola:2001ni,
  Crotty:2003rz,Gordon:2003hw,Beltran:2004uv,Moodley:2004nz,
  KurkiSuonio:2004mn,Beltran:2005gr,Bucher:2004an,Seljak:2006bg,
  Bean:2006qz,Trotta:2006ww}. We have chosen to express the amount of
isocurvature fluctuation in this matter sector as follows:
\begin{equation}
\delta_{\rm m\gamma}\,\equiv\, \frac{\zeta_{\rm
    m}-\zeta_\gamma}{\left(\zeta_{\rm
    m}+\zeta_{\gamma}\right)/2},\label{eq:delta}
\end{equation}
where the subscript ``$\mbox{m}$'' comprises all of non-relativistic
matter, \ie dark matter and baryons (so that, for instance, $\Omega
_{\rm m}\equiv \Omega _{\chi}+\Omega _{\rm b}$). The quantity
$\zeta_{\rm m}$ can be written in terms of the baryon and dark matter
curvature perturbations $\zeta_{\rm b}$ and $\zeta_{\chi}$:
\begin{equation}
\zeta_{\rm m}\,\equiv\,\frac{\Omega_{\rm b}}{\Omega_{\rm m}}\zeta_{\rm
  b} + \frac{\Omega_{\chi}}{\Omega_{\rm m}}\zeta_{\chi}\ .
\end{equation}
The definition~(\ref{eq:delta}) can be justified by the fact that the
data are in fact sensitive to the quantity defined by $S_{\rm b\gamma
}^{\rm eff}=S_{\rm b\gamma } +\Omega _{\chi}S_{\rm \chi \gamma
}/\Omega _{\rm b}$, see Ref.~\cite{Gordon:2002gv}. In this reference,
the quantity $B\equiv S_{\rm b\gamma }^{\rm eff}/\zeta_{\gamma }$ is
constrained using various cosmic microwave background data (including
WMAP1). At $95\%$ C.~L. it was found that $-0.53<B<0.43$, see
Ref.~\cite{Gordon:2002gv}. Our quantity $\delta _{\rm m \gamma}$ is
related to $B$ through
\begin{equation}
\delta _{\rm m\gamma}=\frac{2B}{6\Omega_{\rm m}/\Omega _{\rm b}+B}\, ,
\end{equation}
which implies $-0.12\lta \delta _{\rm m\gamma}\lta 0.089$ where we
have taken $\Omega _{\chi}h^2\simeq 0.12$ and $\Omega _{\rm
  b}h^2\simeq 0.0225$. The choice~(\ref{eq:delta}) is also motivated by
the fact that the most recent analysis of Wilkinson Microwave
Anisotropy Probe fifth year (WMAP5) data has constrained the same
quantity for dark matter only (\ie ${\rm m}\rightarrow
\chi$)~\cite{Komatsu:2008hk}. These results give an upper bound as low
as 2.0\% (95\% C.L.) for fully anti-correlated isocurvature modes and
8.6\% (95\% C.L.)  for uncorrelated modes. In the following, we will
present contours on the quantity $\delta_{\rm m\gamma}$ and emphasize
the loci of 1\% and 10\%. 

\par

The quantity $\delta_{\rm m\gamma}$ must be calculated at the time of
recombination, well after baryon and dark matter freeze-out of
annihilations and curvaton decay (and big bang nucleosynthesis). The
individual gauge invariant curvature perturbations $\zeta_{\rm b}$,
$\zeta_{\chi}$ and $\zeta_\gamma$ are then constant since the fluids
can be considered as isolated at that time. Given the results of the
previous section, in order to compute the constraints in modulus
parameter space and to evaluate $\delta_{\rm m\gamma}$, the value of
$\Omega_{\sigma,\rm osci}$, which directly controls the energy density
parameter of the modulus at the time of decay, is the only remaining
quantity which remains to be specified.

\par

Before embarking on a detailed discussion of these above
considerations, one should note that two other cosmological
constraints are to be satisfied. One concerns the present day
abundance of dark matter and the other the overall amplitude of the
total curvature perturbation. Regarding the former, dark matter is
produced both thermally (through the freeze-out of annihilations) with
present day abundance $\Omega _{\chi, 0}^{\rm f}$, and non-thermally
(through modulus decay) with present day abundance $\Omega _{\chi,
  0}^{\sigma}$ (immediately after decay $\Omega
_{\chi}^{\sigma}=B_{\chi}\Omega_{\sigma}^{<_{\rm d}}$). Hence, the
final abundance is controlled by the parameters $x_{\rm f}$,
$m_{\chi}$, $N_{\chi}$ and $\Omega_{\sigma}^{<_{\rm d}}$. In order to
minimize the dimensionality of our parameter space, we have chosen to
proceed as follows. We set $x_{\rm f}=21$ and $m_{\chi}=100\,
\mbox{GeV}$, which implies that $\Omega _{\chi, 0}^{\rm f}=0.2$; we
then tune $N_{\chi }$ for each value of our main parameters $m_{\sigma
}$ and $\sigma _{\rm inf}$ such that the total $\Omega_{\chi, 0}$ lies
within a factor of three of its observed value. More precisely, we
maintain $N_{\chi}=1$ (see above and
Refs.~\cite{Moroi:1999zb,Kohri:2005ru}) whenever $\Omega _{\chi,
  0}^{\sigma}<0.5$ and decrease it in order to saturate this last
bound otherwise. This is somewhat arbitrary, but given the remaining
freedom in $x_{\rm f}$ and $m_{\chi}$, one could always tune slightly
the parameters to achieve a better agreement with the known value. In
any case, this procedure hardly modifies the constraints obtained in
this article, as we have checked. Furthermore, this approach is
conservative in the sense that we forbid the non-thermal channel to
exceed twice the thermal channel which slightly reduces the
isocurvature perturbations. In this way, in all of the parameter space
scanned in the subsequent figures, the dark matter abundance is
correct to within a factor of two to three.

\par

Concerning the overall amplitude of the curvature perturbation, one
needs to require that $\zeta _{\gamma}^{>_{\rm d}}\simeq
10^{-5}$. This quantity is determined by Eq.~(\ref{eq:zetagammafinal})
and it can be rewritten as:
\begin{equation}
\zeta_{\gamma}^{>_{\rm d}}\,\simeq\,\zeta_{\gamma}^{\rm (i)} +
\frac{1}{3}\Omega_{\sigma}^{<_{\rm d}}S_{\sigma\gamma}^{\rm (i)}\ .
\end{equation}
Therefore the magnitude of the total curvature perturbation is
controlled by several parameters, including $\sigma_{\rm inf}$ and
$H_{\rm inf}$ which determine the scalings of $\Omega_{\sigma}^{<_{\rm
    d}}$ and $S_{\sigma\gamma}^{\rm (i)}$. In principle, it would be
possible to rescale $H_{\rm inf}$ in order to reach the correct
magnitude for $\zeta_{\gamma}^{>_{\rm d}}$ at each value of
$\sigma_{\rm inf}$. However, this would make the interpretation of the
figures rather complex. In the following, we have rather chosen to
plot the constraints obtained for two values of $H_{\rm inf}$ in each
case, in order to gauge the effect of $H_{\rm inf}$ on these
constraints. These two values are $H_{\rm inf}=10^{13}\,$GeV, which
provides a natural scale for inflation since it corresponds to the
simplest inflaton potential $m_\phi^2\phi^2$ (see also the discussion
about naturalness in Ref.~\cite{Boyle:2005ug}), and $H_{\rm
  inf}=10^9\,$GeV. The latter is chosen arbitrarily, but it is such
that the total curvature perturbation is of the right order of
magnitude at every point of the modulus parameter space. For the
former value of $H_{\rm inf}$, a significant region of modulus
parameter space is excluded by the normalization of the total
curvature perturbation; however, this region is entirely contained in
the region which is excluded by the isocurvature constraints.

\subsection{Quadratic potential}
\label{subsec:quadratic}

Here, we assume that the potential of the modulus is a simple quadratic
potential $V(\sigma)=m_\sigma^2\sigma^2/2$ from the end of inflation
onwards. At the onset of oscillations, the $\sigma$ field carries a
fraction $\Omega_{\sigma,\rm osci}$ of the total energy density:
\begin{equation}
\label{eq:osciquadratic}
\Omega_{\sigma,\rm osci}\,=\,\frac{1}{6}\,
\left(\frac{\sigma_{\rm inf}}{M_{\rm Pl}}\right)^2\ .
\end{equation} 
Therefore, one can now explicitly evaluate the quantities $\Omega
_{\sigma}^{>_{\rm f}}$, $\Omega _{\sigma}^{<_{\rm d}}\sim r^{<_{\rm
    d}}$, see
Eqs.~(\ref{eq:omegachif}),~(\ref{eq:omegasigmaf}),~(\ref{eq:rsig})
and~(\ref{eq:rd}). In particular, using Eqs.~(\ref{eq:rsig})
and~(\ref{eq:Td}), immediately prior to decay, the ratio of the energy
density contained in $\sigma$ oscillations to that contained in
radiation now reads:
\begin{equation}
\Omega _{\sigma}^{<_{\rm d}}
\,\simeq\,6\times10^{10}\alpha_{\rm
  osc/rh}\,\left(\frac{\sigma_{\rm inf}}{M_{\rm
    Pl}}\right)^2\left(\frac{m_\sigma}{100\,{\rm
    TeV}}\right)^{-3/2}\left(\frac{g_{\star,\rm
    dec}}{10.75}\right)^{1/4}\left(\frac{T_{\rm rh}}{10^9\,{\rm
    GeV}}\right)\Omega _{\gamma }^{<_{\rm d}}\ .
\label{eq:rsig-quad}
\end{equation}
The parameter $\alpha_{\rm osci/rh}$ is defined as follows:
\begin{equation}
\alpha_{\rm osci/rh}\,\equiv\,\,{\rm min}
\left(1,\,\frac{T_{\rm osci}}{T_{\rm rh}}\right)\ .
\end{equation}
This parameter is most likely $1$ if one relies on the constraints on
the reheating temperature that result from the influence of a
moderately massive gravitino on big-bang nucleosynthesis. In models in
which the gravitino is very massive, $m_{3/2}\,\gtrsim\,100\,$TeV,
however, such constraints can be evaded. We thus treat $\alpha_{\rm
  osci/rh}$ as a free parameter. Obviously, looking at
Eq.~(\ref{eq:rsig-quad}), unless $\sigma_{\rm
  inf}\,\lesssim\,10^{-5}M_{\rm Pl}$, the modulus is bound to dominate
the energy density of the Universe at the time of its decay. 

\par

\begin{figure*}
  \centering
  \includegraphics[width=0.49\textwidth,clip=true]{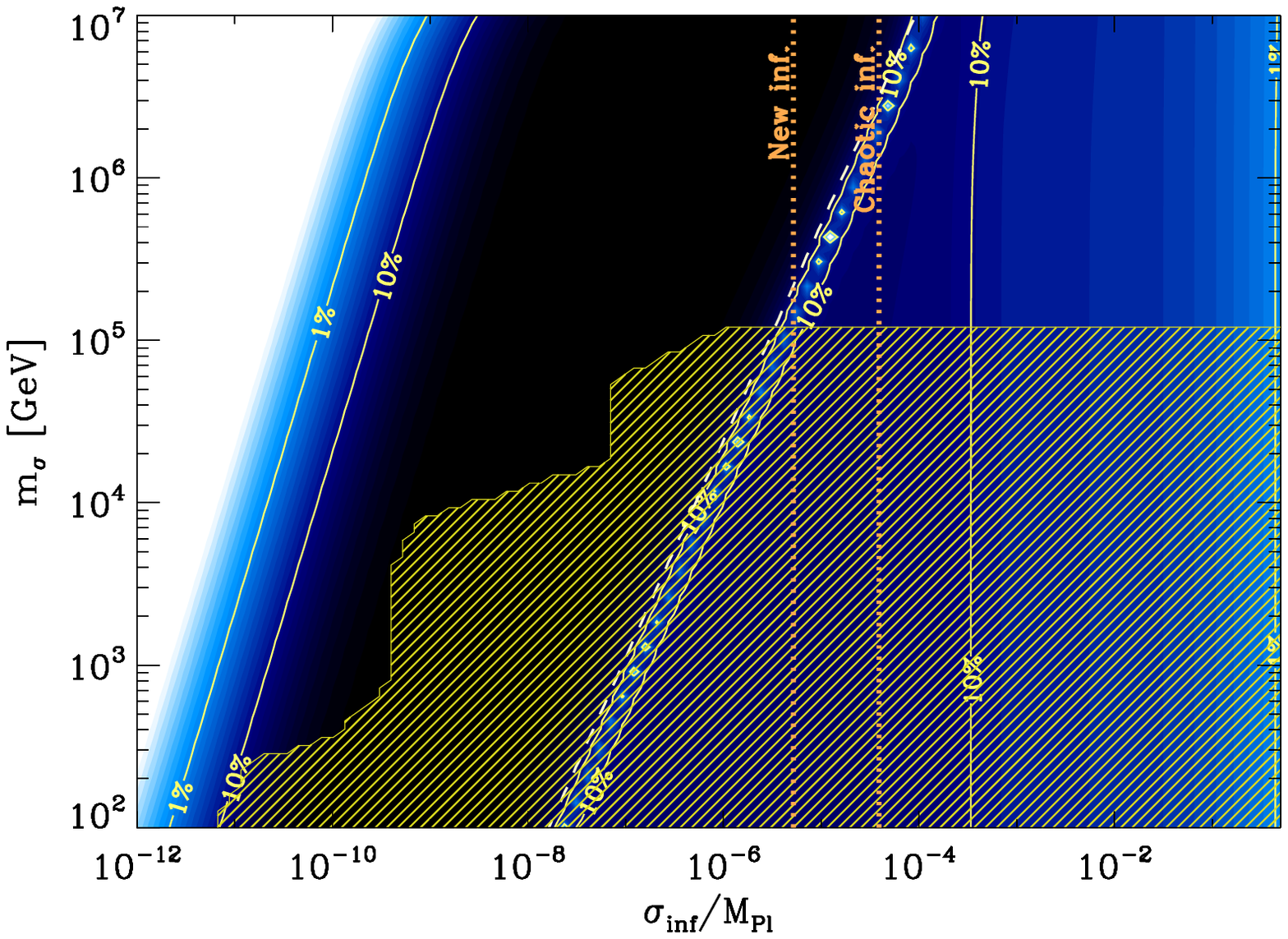}
  \includegraphics[width=0.49\textwidth,clip=true]{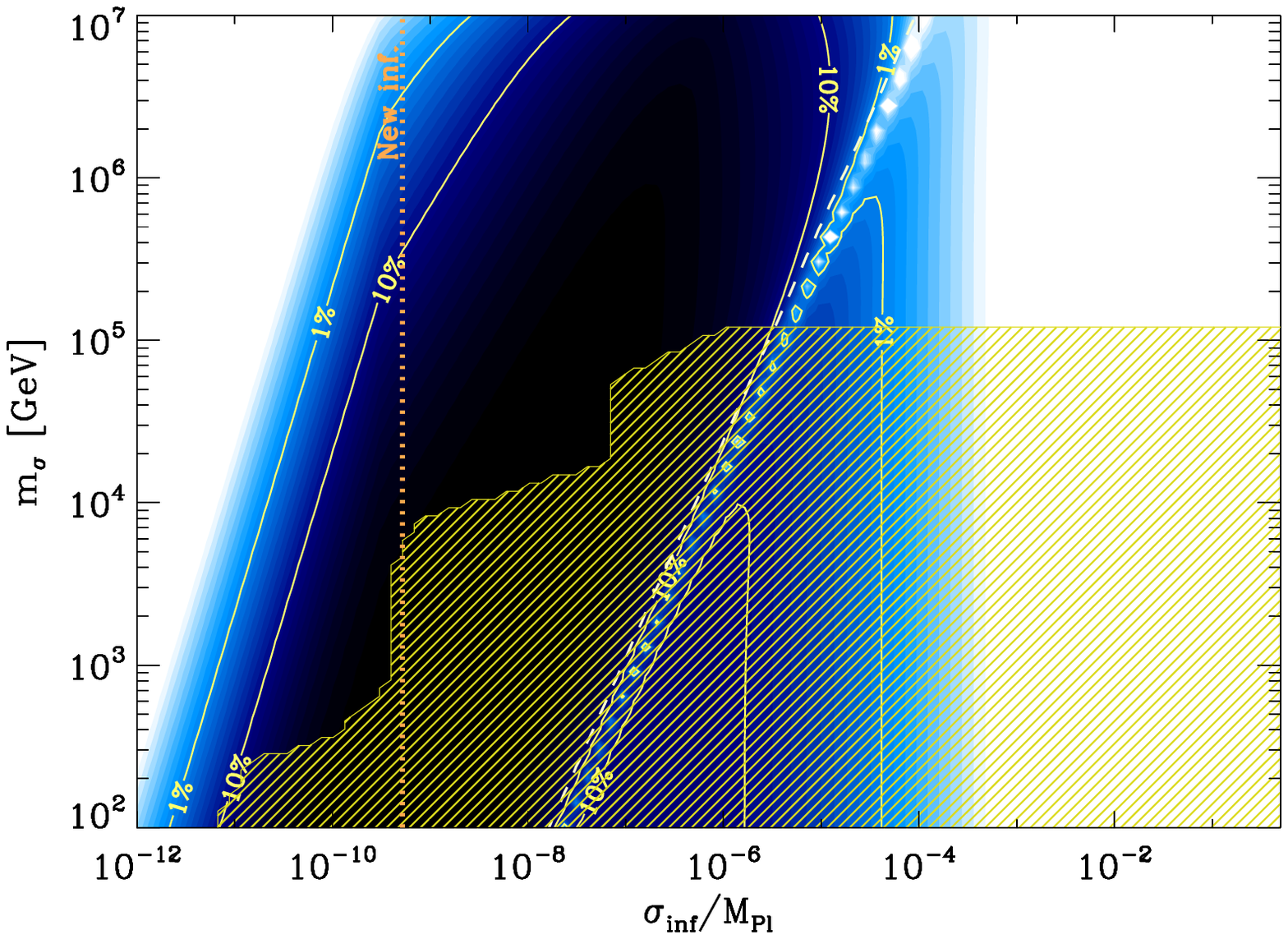}
  \caption[...]{Constraints in the $(m_\sigma,\sigma _{\rm inf})$
    plane. The yellow shaded region shows the region excluded by the
    effect of moduli decay on big-bang nucleosynthesis. The blue
    region give the contours of $\delta_{\rm m \gamma}$ for the matter
    -- radiation isocurvature mode (1\% and 10\% contours
    indicated). The white dashed line indicates the place where
    $\Omega_\sigma^{<_{\rm d}}=0.5$. The orange dotted lines indicate
    the standard deviation of the modulus expected from stochastic
    evolution in its potential during inflation, for new and chaotic
    inflation, as indicated. Left panel: $H_{\rm inf}=10^{13}\,$GeV
    and $T_{\rm rh}=10^9\mbox{GeV}$. This value corresponds to the
    energy scale of inflation during chaotic inflation. The same value
    is achieved for small field inflation if $\mu\simeq M_{\rm
      Pl}$. In the case of chaotic inflation, the standard deviation
    is given by Eqs.~(\ref{eq:varchao2}) and~(\ref{eq:boundchm2})
    since the condition $m_{\rm \sigma}<H_{\rm inf}N_{_{\rm
        T}}^{-1/2}\simeq 1.4\times 10^{13}\mbox{GeV}$ is always
    satisfied for $m_{\sigma}<10^7\mbox{GeV}$. In the case of small
    field inflation, the standard deviation is given by
    Eq.~(\ref{eq:variancenonBD}) assuming the above mentioned
    condition is also satisfied in this case which amounts requiring
    that $N_{_{\rm T}}<10^{12}$. Right panel: $H_{\rm inf}=10^9\,$GeV
    and same reheating temperature. For small field inflation, this
    corresponds to $\mu \simeq 0.22 M_{\rm Pl}$ assuming $p=3$. The
    chaotic inflation standard deviation does not appear in this panel
    because $H_{\rm inf}=10^9\,$GeV cannot be realized in this
    case. For small field inflation the standard deviation is again
    given by Eq.~(\ref{eq:variancenonBD}) which, this time, requires
    $N_{_{\rm T}}<10^{4}$.} \label{fig:iso-quad-m2-tot}
\end{figure*}

Let us now analyze the constraints presented in
Fig.~\ref{fig:iso-quad-m2-tot}. This figure shows the contours of the
$\delta_{\rm m\gamma}$ quantity calculated numerically, assuming
$T_{\rm rh}=10^9\,$GeV, $H_{\rm inf}=10^{13}\,$GeV in the left panel
and $H_{\rm inf}=10^9\,$GeV in the right panel, $N_\chi=1$. This
figure assumes initial (gauge invariant) density contrasts
$\Delta_{\gamma}^{\rm (i)}=2\times10^{-5}$ and $\delta\sigma_{\rm
  inf}/\sigma_{\rm inf}=H_{\rm inf}/(2\pi\sigma_{\rm inf})$ on large
scales. The dashed yellow area is excluded by big-bang
nucleosynthesis; in order to draw this region, we used the results of
Ref.~\cite{Kawasaki:2004qu} for a hadronic branching ratio of
$10^{-3}$ and initial jet energy $1\,$TeV. The white dashed line
indicates the place where $\Omega_\sigma^{<_{\rm d}}=0.5$ and
separates roughly the regions in which either the dark matter or the
baryon isocurvature mode dominates (see below). There is actually an
accidental cancellation of these two modes close to that line. it has
also been assumed that $\zeta_{\sigma}^{\rm (i)}$ saturates at $0.5$
in order for the numerical computations to proceed without errors.  In
any case, the region in which $\zeta_\sigma^{\rm (i)} \gg 1$,
corresponding to $\sigma_{\rm inf}\,\ll\, H_{\rm inf}$, is an
``unlikely'' region, in the sense that $\sigma_{\rm inf}$ is expected
to be typically larger than $H_{\rm inf}/(2\pi)$ due to quantum
effects, see Eqs.~(\ref{eq:ds-ul}) and~(\ref{eq:ds-l}) for
$m_{\sigma,\rm eff}\,<\,H_{\rm inf}$. The dotted orange lines in
Fig.~\ref{fig:iso-quad-m2-tot} show the standard deviations
$\langle\delta\sigma^2\rangle^{1/2}$ expected from the stochastic
motion of the inflaton, following Appendix~\ref{app:stoch} and the
formulae Eqs.~(\ref{eq:ds-ul}) and (\ref{eq:ds-l}), for two
inflationary scenarios and two inflationary scales. It is important to
stress the following: these standard deviations are measured
relatively to the ``instantaneous'' classical value of the modulus
field.

\par 

Figure~\ref{fig:iso-quad-m2-tot} provides a clear example of the power
of constraints obtained at the perturbative level on moduli cosmology,
as the region excluded by the production of isocurvature fluctuations
significantly exceeds that constrained by big-bang
nucleosynthesis. For instance, at a natural scale
$m_\sigma\,\sim\,1\,$TeV, the upper bound on $\sigma_{\rm inf}$
(equivalently, on the modulus energy density) is more stringent that
those from big-bang nucleosynthesis by some two orders of magnitude.
This figure also clearly shows that taking
$m_{\sigma}\,\gtrsim\,100\,$TeV allows to evade the constraints from
big-bang nucleosynthesis, but not those from cosmological density
perturbations if the inflationary scale is large, $H_{\rm
  inf}\sim10^{13}\,$GeV.  One may note that this region is also
constrained by the possible overproduction of gravitinos through
modulus decay~\cite{Nakamura:2006uc,Endo:2006zj,Dine:2006ii}. It is
thus mandatory to require that the modulus potential suffers
corrections, in such a way as to reduce considerably the modulus
energy density at the time of decay, or that the inflationary scale is
much lower. Indeed, if $H_{\rm inf}\,\ll\,10^{13}\,$GeV, a region
devoid of constraints opens up at large values of $\sigma_{\rm inf}$
and large masses $m_\sigma\,\gtrsim\,100\,$TeV. The large modulus mass
then allows to evade the constraints on entropy injection around
big-bang nucleosynthesis, while the large value of $\sigma_{\rm inf}$
ensures that the initial modulus -- radiation isocurvature fluctuation
has become negligible at such a small inflationary scale. More
specifically, using Eq.~(\ref{eq:Siniquad}) one finds that
$S_{\sigma\gamma}^{\rm (i)}\,\ll\,\zeta_{\gamma}^{\rm (i)}$ when
$\sigma\,\gtrsim\, M_{\rm Pl}\,(H_{\rm inf}/10^{13}\,{\rm GeV})$, and
the isocurvature fraction constrained by microwave background
anisotropies is directly proportional to this ratio.  The constraints
obtained in the limit of a high inflationary scale also exclude the
possibility of late time entropy production through modulus
decay. This result is in itself significant as such entropy production
is often invoked to dilute unwanted relics.

\begin{figure*}
  \centering
  \includegraphics[width=0.49\textwidth,clip=true]{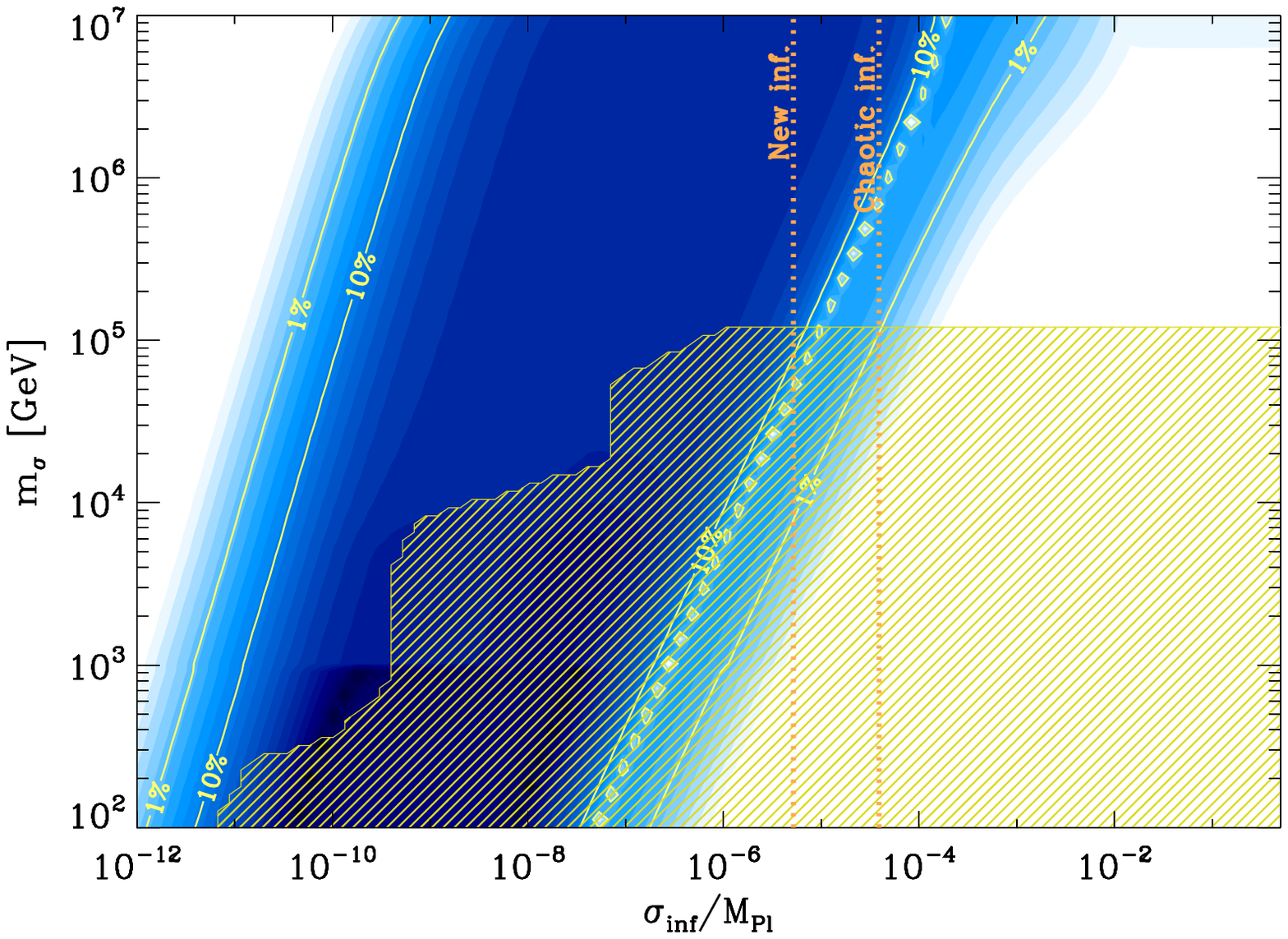}
  \includegraphics[width=0.49\textwidth,clip=true]{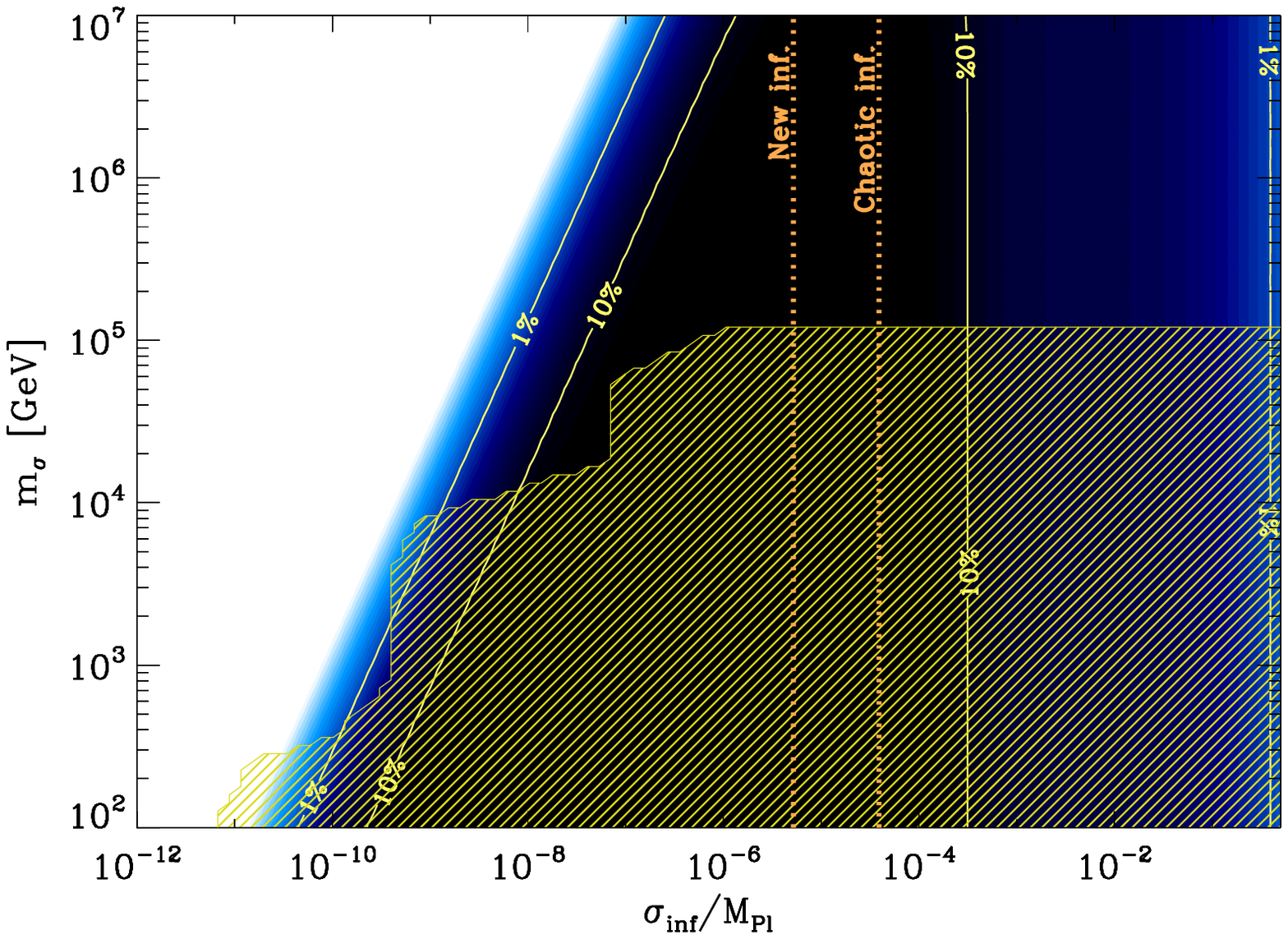}
  \caption[...]{Same as the left panel of
    Fig.~\ref{fig:iso-quad-m2-tot}, but showing only the contribution
    of dark matter isocurvature modes in the left panel (\ie setting
    arbitrarily the baryon isocurvature mode to zero) and the baryon
    isocurvature mode only in the right panel (setting the dark matter
    isocurvature mode to zero).} \label{fig:iso-quad-m2-dmb}
\end{figure*}

In order to understand the relative contributions of dark matter -
radiation and baryon - radiation isocurvature fluctuations, it is
useful to break Fig.~\ref{fig:iso-quad-m2-tot} into two subfigures,
each showing one of the two contributions. In the left panel of
Fig.~\ref{fig:iso-quad-m2-dmb} we plot the same contours as in
Fig.~\ref{fig:iso-quad-m2-tot}, assuming arbitrarily (for the sake of
demonstration) $\zeta_{\rm b}^{>_{\rm d}}\,\sim\,\zeta_{\rm
  \gamma}^{>_{\rm d}}$ (\ie setting the baryon isocurvature mode to
zero). Similarly, in the right panel of
Fig.~\ref{fig:iso-quad-m2-dmb}, we plot the contours of $\delta_{\rm
  m\gamma}$ assuming $\zeta_\chi^{>_{\rm d}}\,\sim\,\zeta_{\rm
  \gamma}^{>_{\rm d}}$ (\ie no dark matter isocurvature mode). These
two plots illustrate the complementarity of the constraints coming
from these two isocurvature modes. It is possible to understand each
of them as follows.

\par

Assuming for the time being, that $\zeta_{\rm b}^{>_{\rm
    d}}\,\sim\,\zeta_{\rm \gamma}^{>_{\rm d}}$ (\ie no baryon
isocurvature mode), the constrained fraction $\delta_{\rm m\gamma}$
can be written in the following way:
\begin{eqnarray}
\delta_{\rm m\gamma}&\,\simeq\,& 
\frac{\Omega_\chi \zeta_\chi^{>_{\rm d}}/\Omega _{\rm m} 
-\Omega_\chi\zeta_{\gamma}^{>_{\rm d}}/\Omega _{\rm m}}
{\left[\Omega_{\chi}\zeta_\chi^{>_{\rm d}}/\Omega _{\rm m} 
+\left(2 -\Omega_{\chi}/\Omega_{\rm m}\right)
\zeta_\gamma^{>_{\rm d}}\right]/2}=
\frac{2x_{\chi\gamma}}{6+x_{\chi\gamma}}\, ,
\end{eqnarray}
where 
\begin{eqnarray}
x_{\chi\gamma}\,\equiv\,\frac{\Omega_\chi}{\Omega_{\rm m}}
\frac{S_{\chi\gamma}^{>_{\rm d}}}{\zeta_\gamma^{\rm >_{\rm d}}}\ .
\end{eqnarray}
Therefore, the actual quantity that constrains the magnitude of the
final dark matter isocurvature mode is $x_{\chi\gamma}$. This is not
surprising since $x_{\chi \gamma }$ is, up to a factor $\Omega
_{\chi}/\Omega _{\rm m}$, exactly equal to the quantity $B$ introduced
above. Using Eq.~(\ref{eq:Scr}), one can rewrite $x_{\chi\gamma}$ as:
\begin{equation}
x_{\chi\gamma}\,\simeq\,3\frac{\Omega_\chi}{\Omega_{\rm
    m}}\frac{1}{1+\Upsilon_\chi} \left(\frac{B_\chi\Omega_\sigma^{<_{\rm
      d}}}{\Omega_\chi^{<_{\rm d}}+B_\chi\Omega_\sigma^{<_{\rm d}}}-r^{<_{\rm d}}\right)
\frac{\zeta_\sigma^{\rm (i)}-\zeta_\gamma^{\rm (i)}}
{\zeta_\gamma^{\rm (i)}}\ .
\label{eq:cond-iso-dm}
\end{equation}
It is straightforward to verify that the conditions
(\ref{eq:Scr-cond}) for the generation of isocurvature perturbations
lead, if satisfied, to a non-zero value of $x_{\chi\gamma}$. It is
however important to note that the magnitude of $x_{\chi\gamma}$ also
increases with the ratio $\zeta_{\sigma}^{\rm (i)}/\zeta_{\gamma}^{\rm
  (i)}$. One may rewrite the conditions of existence of a dark matter
- radiation isocurvature mode as follows, neglecting the effect of
$\zeta_\sigma^{\rm (i)}/\zeta_\gamma^{\rm (i)}$ for clarity:
\begin{equation}
\label{eq:existenceiso}
10^{-3}\,x_{\rm f}^{1/4}{\rm e}^{-x_{\rm
    f}/2}\left(\frac{m_\sigma}{100\,{\rm
    TeV}}\right)^{1/2}\left(\frac{T_{\rm rh}}
{10^9\,{\rm GeV}}\right)^{-1/2}\,\ll\,
\frac{\sigma_{\rm inf}}{M_{\rm Pl}}\,\ll\,
4\times 10^{-6} \left(\frac{m_\sigma}{100\,{\rm
    TeV}}\right)^{3/4}
\left(\frac{T_{\rm rh}}{10^9\,{\rm GeV}}\right)^{-1/2}\ .
\end{equation}
Out of simplicity, the dependence on the numbers of degrees of freedom
has been omitted in these equations. In the above expression, the
upper bound comes from the condition $\Omega_\sigma^{<_{\rm d}}\ll 1$
using Eq.~(\ref{eq:rsig-quad}) while the lower bound originates from
the condition $B_\chi\Omega_\sigma^{<_{\rm
    d}}\,\gg\,\Omega_\chi^{<_{\rm d}}$, see the first formula in
Eqs.~(\ref{eq:Scr-cond}), using the definition of $B_{\chi}$, the fact
that $\Omega _{\chi}^{<_{\rm d}}=\Omega _{\chi}^{>_{\rm f}}T_{\rm
  f}/T_{\rm d}$ (with $T_{\rm f}\sim m_{\chi}/x_{\rm f}$) and
Eqs.~(\ref{eq:Td}), (\ref{eq:omegachif})
and~(\ref{eq:rsig-quad}). These constraints delimit a stripe in the
$(\sigma_{\rm inf}, m_\sigma)$ plane which is consistent with what is
observed in Fig.~\ref{fig:iso-quad-m2-dmb}. This stripe is actually
broader at small values of $\sigma_{\rm inf}$ due to the large ratio
$\zeta_\sigma^{\rm (i)}/\zeta_\gamma^{\rm (i)}$ which enhances the
modulus perturbations relative to those of dark matter. The above
formula also neglects the effect of annihilations, which is a good
approximation as long as $\Upsilon_\chi\,\ll\,1$, or:
\begin{equation}
\left(\frac{\sigma_{\rm inf}}{M_{\rm Pl}}\right)\,\ll\,0.7 
\left(\frac{m_\sigma}{100\,{\rm
    TeV}}\right)^{-1/4}\left(\frac{T_{\rm rh}}{10^9\,{\rm GeV}}\right)^{-1/2}
\left(\frac{m_\chi}{200\,{\rm GeV}}\right)^{1/2}
x_{\rm f}^{-1/4}{\rm e}^{-x_{\rm f}/2}\ .
\end{equation}
In practice, annihilations will play a role in suppressing the amount of
isocurvature mode in the high mass region $m_\sigma \,\gg\, 100\,$TeV
and $\sigma_{\rm inf}\,\sim\, 10^{-6}-10^{-4}M_{\rm Pl}$, as can be
checked from the above formulas using $x_{\rm f}\sim 20$.

\par

Turning to the generation of a baryon isocurvature mode, we now assume
that $\zeta_{\chi}^{\rm (f)}\sim \zeta_\gamma^{\rm (f)}$, in which
case Eq.~(\ref{eq:delta}) can be rewritten as:
\begin{equation}
\delta_{\rm m \gamma}\,\simeq\, 
\frac{2x_{\rm b\gamma}}{6+x_{\rm b\gamma}}\ ,
\end{equation}
where 
\begin{equation}
x_{\rm b\gamma}\,\simeq\,-3\frac{\Omega_{\rm b}}{\Omega_{\rm m}}
\frac{\Omega_\sigma^{<_{\rm d}}
\left[\zeta_\sigma^{\rm (i)}-\zeta_\gamma^{\rm (i)}\right]}
{\Omega_\sigma^{<_{\rm d}}\left[\zeta_\sigma^{\rm (i)}
-\zeta_\gamma^{\rm (i)}\right]
+\zeta_\gamma^{\rm (i)}}\ .
\label{eq:cond-iso-quad}
\end{equation}
Therefore the magnitude of the isocurvature mode is proportional to
the fraction of energy density of the modulus at the time of decay,
times the ratio of initial modulus--radiation isocurvature mode to the
initial radiation curvature perturbation. The baryon constraints thus
lie at high values of $\sigma_{\rm inf}$, since the production of
isocurvature fluctuations become dominant when $\Omega_\sigma^{<_{\rm
    d}}$ is of order unity. However, as $\Omega_\sigma^{<_{\rm d}}$
becomes smaller than unity, its weakness can be compensated to some
level by a large value of $\zeta_\sigma^{\rm (i)}/\zeta_\gamma^{\rm
  (i)}$.

 \subsection{Supergravity corrections to the potential}
\label{subsec:sugracorrections}

In supergravity, one expects the potential of the modulus to be lifted
by an effective term of the form $\pm c^2H^2\sigma^2/2$, where the
factor $c^2$ may change between different eras of the thermal history
of the Universe, see Appendix~\ref{sec:appendix:concretemodels} where
concrete models are discussed. Including such supergravity
corrections, the post-inflationary modulus potential may be written as
follows:
\begin{equation}
\label{eq:potsugra}
V(\sigma)\,\simeq\, \frac{1}{2}m_\sigma^2\sigma^2 \pm
\frac{1}{2}c^2H^2\left(\sigma-\sigma_0\right)^2\ .
\end{equation}
One should recall that we assume the modulus field to be light during
inflation, hence the potential~(\ref{eq:potsugra}) refers to the
post-inflationary epoch only. Moreover, in the following, we denote by
``high energy'' the regime in which the corrections proportional to
$c^2H^2$ dominate the term proportional to $m_{\sigma }^2\sigma ^2$
(as mentioned before, although we use the expression ``high energy'',
these considerations apply to the post-inflationary epoch only). Then,
the quantity $c$ controls the mass of the field while $\sigma _0$
represents its minimum, the so-called ``high energy minimum''. One may
expect that the vev at the end of inflation ($\sigma_{\rm inf}$) be
different from $\sigma_0$ since the effective potentials during and
after inflation a priori differ from one another. We also use the
terminology ``low energy'' to characterize the regime in which the
corrections $c^2H^2$ become sub-dominant and where the potential
reduces to $m_{\sigma }^2\sigma ^2/2$. Let us recall that we set the
``low energy minimum'' of the potential at $\sigma =0$. As a result,
the high energy and low energy minima do not coincide if $\sigma
_0\neq 0$ (as one should expect on general grounds).

\par

In order to derive the amplitude of the isocurvature perturbations
produced in the supergravity case, we need to follow the evolution of
the modulus vev and of its perturbations from the end of inflation
until the time at which we set the initial conditions. In particular,
we need to calculate $S_{\sigma\gamma}^{\rm (i)}$ in terms of the
inflaton and modulus perturbations during inflation. In order to do
so, we model the introduction of the effective Hubble mass through a
potential that accounts for the coupling between the modulus and the
inflation, such as $V(\sigma,\phi)\,=\,m_\sigma^2\sigma^2/2 +
m_\phi^2\phi^2\left[1 + c^2\sigma^2/\left(3M_{\rm
      Pl}^2\right)\right]/2$, which produces the desired $+c^2H^2$
effective mass squared. However, the discussion that follows is not
restricted to this particular potential. In order to study this two
field system, we use the formalism of Gordon, Wands, Bassett and
Maartens~\cite{Gordon:2000hv} (notice that we have changed notations
with respect to Ref.~\cite{Gordon:2000hv} since, now, $\sigma$ now
longer represents the adiabatic field but the modulus one). The
entropy field is given by the following expression
\begin{eqnarray}
\delta s
&\,\equiv\,&\cos\theta\delta\sigma -
\sin\theta\delta\phi\,\simeq\,\delta\sigma -
\frac{\dot\sigma}{\dot\phi}\delta\phi\ ,
\end{eqnarray}
where we have used that
\begin{eqnarray}
\cos\theta&\,\equiv\,&\frac{\dot \phi}{\sqrt{\dot \phi^2 +
    \dot\sigma^2}}\,\simeq\,1\ ,\quad \sin\theta \,\equiv\,
\frac{\dot\sigma}{\sqrt{\dot
    \phi^2 +
    \dot\sigma^2}}\,\simeq\,\frac{\dot\sigma}{\dot\phi}\, ,
\end{eqnarray}
the last equalities following from the test field approximation:
$\dot\sigma\,\ll\,\dot\phi$. Then, using the perturbed Einstein
equation for the Bardeen potential on large scales
\begin{equation}
\dot\Phi + H\Phi\,=\,\frac{1}{2M_{\rm Pl}^2}\left(\dot\phi\delta\phi +
\dot\sigma\delta\sigma\right)\ ,
\end{equation}
one can rewrite the entropy perturbation in terms of $\delta \sigma
$ and $\Phi $ only. One obtains
\begin{equation}
\delta s\,\simeq\,\delta\sigma  - \frac{H\,M_{\rm Pl}^2}{\dot\phi^2/2}
\dot\sigma \Phi=\delta \sigma -\frac{2\dot{\sigma }}{3H}\Phi,
\end{equation}
where we have used that $\dot\phi^2/2\,\simeq\,3H^2\,M_{\rm Pl}^2/2$
on average. In the above equation, one has neglected
$\dot\sigma^2/\dot\phi^2$ in front of unity. On large scales, the
equation of evolution of the entropy perturbation
reads~\cite{Gordon:2000hv}:
\begin{equation}
\delta\ddot s + 3H\delta\dot s + \left(V_{ss} +
3\dot\theta^2\right)\delta s\,=\,0\ ,\label{eq:sevol}
\end{equation}
where the expression of $V_{ss}$ can be found in
Ref.~\cite{Gordon:2000hv}. In the test field approximation
$\dot\sigma\,\ll\,\dot\phi$, and assuming powerlaw behaviors of
$\sigma$ and $\phi$, one can check that
$V_{ss}\,\simeq\,V_{\sigma\sigma}$ and
$\dot\theta^2\,\ll\,V_{\sigma\sigma}$ in Eq.~(\ref{eq:sevol}), so that
$\delta s$ follows the same quadratic equation than $\sigma$, and
therefore $\delta s/\sigma$ is conserved. This implies
\begin{equation}
\left .\left[\frac{\delta \sigma}{\sigma} - \frac{2\dot\sigma}{3
    H\sigma}\Phi\right]\right\vert_{t}\, = \,
\left .\left[\frac{\delta \sigma}{\sigma} - \frac{2\dot\sigma}{3
    H\sigma}\Phi\right]\right\vert_{\rm inf}\ ,
\end{equation}
hence, with $t=2/(3H)$, $\Phi\simeq\Phi_{\rm inf}$ and
assuming $\dot\sigma=0$ initially,
\begin{equation}
\label{eq:solentropy}
\left .\frac{\delta\sigma}{\sigma}\right\vert_{t}\,=\,
\frac{\delta\sigma_{\rm inf}}{\sigma_{\rm inf}} +
\frac{t\dot\sigma(t)}{\sigma(t)}\Phi_{\rm inf}\ .
\end{equation}
This solution happens to match that obtained for a pure time
independent modulus potential, see
Ref.~\cite{Langlois:2004nn}. However, in the present case, it accounts
for the sourcing of the modulus perturbation by the effective Hubble
mass. To our knowledge, this result has not been obtained before.  In
the limiting case $\Phi_{\rm inf}\rightarrow 0$, one recovers the
result obtained in Ref.~\cite{Dimopoulos:2003ss} that
$\delta\sigma/\sigma$ is constant. One can now calculate the modulus
-- radiation isocurvature fluctuation at the initial time, \ie in the
radiation era and after the onset of modulus oscillations, and one
recovers the result given in Eq.~(\ref{eq:Siniquad}).

\par

Two additional remarks are in order here. Firstly, in the derivation
above, we have neglected the preheating effects. This is justified by
the following considerations. It turns out that the model investigated
here is exactly similar to the two field model $g^2\phi^2\chi ^2$
studied in Ref.~\cite{Kofman:1997yn} with a dimensionless coupling
constant given by $g^2=c^2m_{\phi}^2/(6 M_{\rm Pl}^2)$. This means
that the quantity $g^2\Phi^2/(4m_{\phi}^2)=c^2\Phi^2/(24 M_{\rm
  Pl}^2)\ll 1$ and that we are never in the ``broad resonance'' regime
where preheating effects are important~\cite{Kofman:1997yn}. Secondly,
we have also numerically integrated the exact equations of motion for
different cases and have checked that the approximations used above
are verified. Above all, we have compared the numerical solution for
$\delta \sigma $ in the post-inflationary epoch to the
solution~(\ref{eq:solentropy}) and have found an almost perfect
agreement.

\subsubsection{Case $+c^2H^2/2$}
\label{subsubsec:potwithplus}

As mentioned previously, the details of the calculations that follow
can be found in Appendix~\ref{app:sol}. In order to re-compute $\Omega
_{\rm \sigma ,osci}$, one must follow the evolution of $\sigma $
between the end of inflation (where $\sigma =\sigma _{\rm inf}$) and
$H=m_{\rm \sigma}$ in the situation where the potential is dominated
by the Hubble scale corrections. This evolution is characterized by
$c$, $\sigma _0$ and $p$, the latter defining the evolution of the
scale factor: $a\propto t^p$, so that $H=p/t$.  These parameters enter
in the following combinations:
\begin{equation}
\mu\,=\,\frac{3p-3}{2}\, ,\quad\nu^2\,=\,(\mu+1)^2 - p^2c^2\ ,
\end{equation}
Depending on the magnitude of $c$, $\nu$ may be real or imaginary,
which gives rise to different evolutions. We examine each of these
cases in turn. 

\par

Let us first start with the case $c < (\mu+1)/p$ (real $\nu$). In
Appendix~\ref{app:sol}, it is shown that $\Omega_{\rm \sigma,osci}$
can be expressed as follows, see Eqs.~(\ref{eq:omnup1})
and~(\ref{eq:omnup2})
\begin{eqnarray}
\label{eq:omeganureal1}
\Omega_{\rm \sigma,osci} &\simeq & \frac{1}{6}{\cal A}_1^2
\left({\sigma_0 \over M_{\rm Pl}}\right)^2
+{1\over 6}{\cal B}_1^2\left({\sigma_{\rm
        inf}-\sigma_0 \over M_{\rm Pl}}\right)^2
\left(\frac{p_{\rm MD}m_{\rm \sigma}}{H_{\rm
      inf}}\right)^{2(\mu_{\rm MD}+1-\nu_{\rm MD})} ,\quad T_{\rm osci}>T_{\rm rh}\, ,\\
\label{eq:omeganureal2}
\Omega_{\rm \sigma,osci} &\simeq &\frac{1}{6}{\cal A}_2^2
\left({\sigma_0 \over M_{\rm Pl}}\right)^2
+{1\over 6}{\cal B}_2^2\left({\sigma_{\rm
        inf}-\sigma_0 \over M_{\rm Pl}}\right)^2
\left(\frac{H_{\rm rh}}{H_{\rm inf}}\right)^{2(\mu_{\rm MD}+1-\nu_{\rm MD})}
\left(\frac{p_{\rm MD}m_{\rm \sigma}}{H_{\rm rh}}\right)^{2(\mu_{\rm RD}+1-\nu_{\rm RD})}\, ,
\quad T_{\rm osci}<T_{\rm rh}\, . 
\end{eqnarray}
The quantities indexed with ``$_{\rm RD}$'' (resp. ``$_{\rm MD}$'')
refer to the radiation dominated (resp. matter dominated) era. These
equations are the supergravity counterparts of
Eq.~(\ref{eq:osciquadratic}) in the case where $\nu $ is real. In the
above expressions, the first term is the contribution originating from
the particular solution of the equation of motion while the second
term is due to the homogeneous solution. A crucial difference between
the scalings of the homogeneous and the particular solution is the
redshift factor $(m_\sigma/H_{\rm inf})^{2(\mu_{\rm MD}+1-\nu_{\rm
    MD})}\,\ll\,1$ for the latter. The prefactors ${\cal A}_1$, ${\cal
  A}_2$, ${\cal B}_1$ and ${\cal B}_2$ are all of order unity, see
Appendix~\ref{app:sol}. The quantity $\Omega_{\rm \sigma,osci}$ will
be dominated by the homogeneous solution contribution whenever:
\begin{equation}
\left\vert\sigma_{\rm inf}-\sigma_0\right\vert\left({m_\sigma\over
  H_{\rm inf}}\right)^{\mu_{\rm MD}+1-\nu_{\rm MD}}\,\gg\,\left\vert\sigma_0\right\vert
\ ,\label{eq:homdom1}
\end{equation}
assuming for simplicity $T_{\rm rh}\,<\,T_{\rm osci}$ which is the
most generic situation. 

\par

Let us now discuss the cosmological consequences for the two cases in
which the particular and the homogeneous solution dominates at late
times, starting with the former case. Then, the constraints in modulus
parameter space are straightforward to derive. Using the results
obtained for the purely quadratic case, for which $\Omega_{\sigma,\rm
  osci}\,=\,(\sigma_{\rm inf}/M_{\rm Pl})^2/6$, one can put an upper
limit on $\sigma_0$ since, in the case considered here,
$\Omega_{\sigma,\rm osci}$ has the same form, $\sigma _{\rm inf}$
being simply replaced with $\sigma _0$ . In
Eq.~(\ref{eq:existenceiso}), we have established the conditions for
the existence of a dark-matter isocurvature mode. It is clear that if
$\sigma _0$ [$\sigma _{\rm inf}$ in Eq.~(\ref{eq:existenceiso})] is
smaller than the lower bound, then there is no isocurvature mode and
the scenario is compatible with the cosmic microwave background
data. 
For $x_{\rm f}\sim 20$, $m_\sigma\,=\,10^5\,$GeV and $T_{\rm
  rh}\,=\,10^9\,$GeV, this gives
\begin{equation}
\sigma_0\,\lesssim\, 10^{-10}M_{\rm Pl}\ .
\end{equation}
From previous analytical calculations, one expect this bound to scale
with the parameters as $H_{\rm inf}^0T_{\rm rh}^{-1/2}m_\sigma^{1/2}$,
see Eq.~(\ref{eq:existenceiso}) replacing $\sigma_{\rm inf}$ by
$\sigma_0$. However, such a value remains well below the typical
displacement expected from the quantum jumps of the modulus in its
potential during inflation, unless $H_{\rm inf}\,\ll\,10^9\,$GeV (see
Eq.~\ref{eq:ds-ul}).

\par

\begin{figure*}
  \centering
  \includegraphics[width=0.49\textwidth,clip=true]{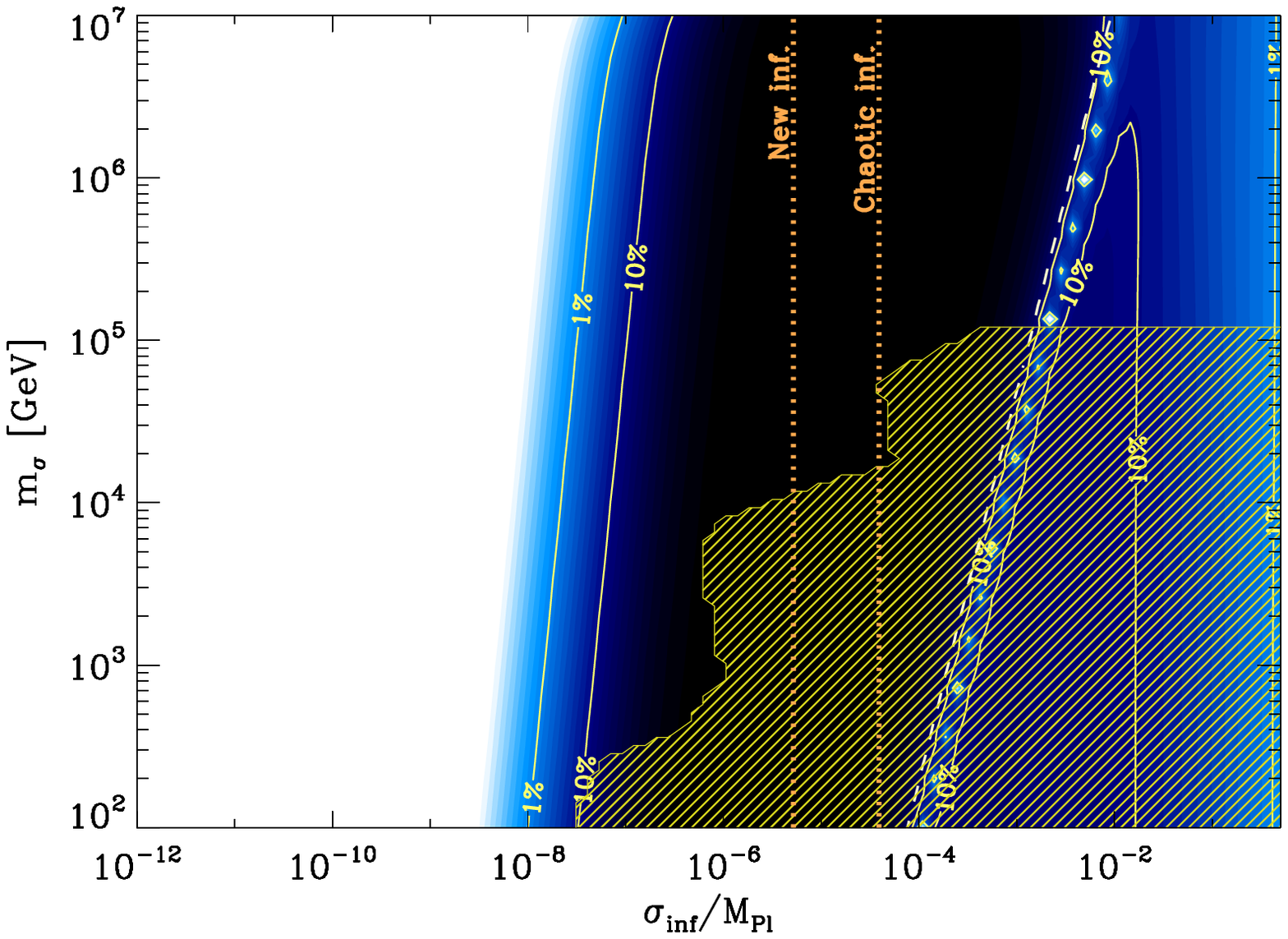}
  \includegraphics[width=0.49\textwidth,clip=true]{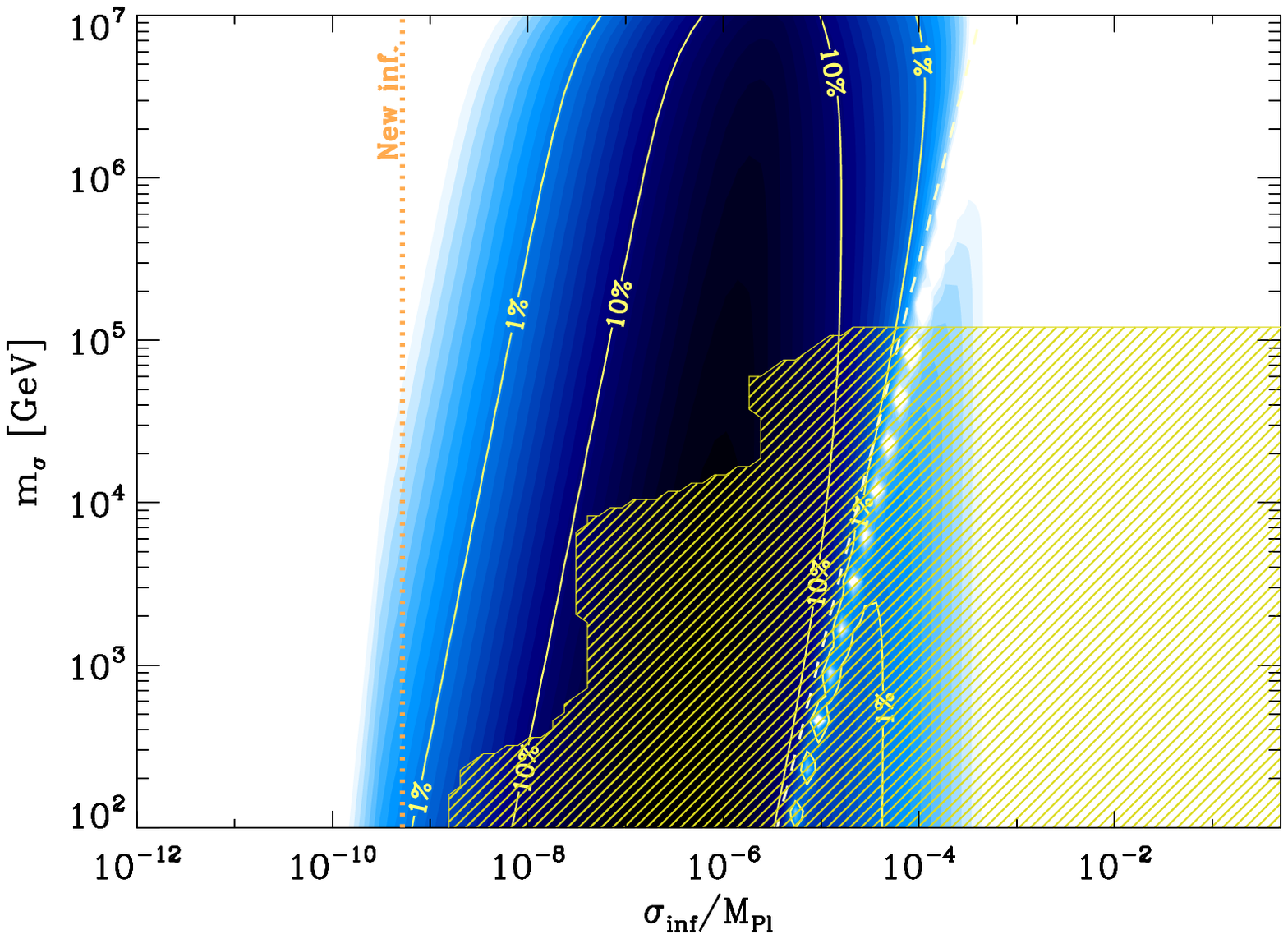}
  \caption[...]{Same as Fig.~\ref{fig:iso-quad-m2-tot}, but
    considering positive squared Hubble scale corrections to the
    modulus potential, with $c^2=0.5$ and $\sigma_0=0$.  The parameter
    $\nu$ is given by $\nu_{\rm MD}^2\simeq 0.027$ and $\nu _{\rm
      RD}^2\simeq -0.063$. For the reheating temperature considered in
    this figure, one always has $T_{\rm osci}>T_{\rm rh}$ and, hence,
    only the value of $\nu _{\rm MD}^2$ matters, see
    Eq.~(\ref{eq:omeganureal1}). Therefore, one is indeed in the case
    $\nu>0$. Left panel: $H_{\rm inf}=10^{13}\, $GeV, $T_{\rm
      rh}=10^9$\,GeV.  Right panel: $H_{\rm inf}=10^9\,$GeV, $T_{\rm
      rh}=10^9$\, GeV. This case is only relevant for small field
    inflation. The standard deviations due to stochastic motion are
    calculated as in Fig.~\ref{fig:iso-quad-m2-tot}.  }
\label{fig:iso-H2-1}
\end{figure*}

This brings us to the other extreme case, in which the homogeneous
solutions dominates the evolution at late times.  As a clear example
of this situation, consider $\sigma_0=0$, but $\sigma_{\rm
  inf}\,\neq\,0$. Physically, this corresponds to the situation in
which the minima of the effective potential after inflation coincides
at high ($H_{\rm inf}> H > m_\sigma$) and low energy ($m_\sigma >
H$). The vev $\sigma_{\rm inf}$ is here non-zero, either because the
effective minimum during inflation does not coincide with that at
latter times, or because $\sigma_{\rm inf}$ is subject to quantum
effects. We thus treat $\sigma_{\rm inf}$ as a free parameter as
before, and one obtains the constraints in the modulus parameter space
presented in Fig.~\ref{fig:iso-H2-1}.

This figure clearly confirms the power of constraints obtained at the
perturbative level. With respect to Fig.~\ref{fig:iso-quad-m2-tot},
which corresponds to the purely quadratic case, one can see that the
constraints from big-bang nucleosynthesis and from the production of
isocurvature perturbations have moved towards higher values of
$\sigma_{\rm inf}$. This is understood easily: in the purely quadratic
case, the modulus energy density remains constant between the end of
inflation and the onset of oscillations, while it decreases in the
present case.  The white dashed line, which indicates the locus of
$\Omega_{\sigma}^{<_{\rm d}}\,=\,0.5$, serves to delimit the
constraints derived from dark matter - radiation and from baryon -
radiation isocurvature modes.

\par

Nevertheless, the constraints obtained still preclude the possibility
of having a heavy modulus with an arbitrary vev at the end of
inflation. In particular, even if the effective minima of the
potential during inflation coincides with that lower energy,
$\left\vert\sigma_{\rm inf}-\sigma_0\right\vert$ should depart from
zero by the value expected from quantum motion during inflation.  In
the left panel, for $H_{\rm inf}=10^{13}\,$GeV, it is found that the
constraints from the production of isocurvature fluctuations extend
significantly below this bound, hence there is no apparent solution to
the moduli problem. As $H_{\rm inf}$ becomes much smaller, some region
of parameter space opens up at large values of $\sigma_{\rm inf}$ and
large values of $m_\sigma$ as in the quadratic case, and the typical
stochastic displacement also decreases. For $H_{\rm inf}=10^9\,$GeV,
corresponding to the right panel of Fig.~\ref{fig:iso-H2-1}, there is
however little room between this lower limit for $\sigma_{\rm inf}$
and the region excluded by isocurvature fluctuations.

\par

It is noteworthy to recall that the situation depicted in the above
figure is realized by two concrete models of inflation discussed in
Appendix~\ref{sec:appendix:concretemodels}.

\par

Let us now turn to the case $c > (\mu+1)/p$ (imaginary $\nu$). For
$p=2/3$, this case corresponds to $c>3/4$. Writing $\nu=i\hat\nu$, one
has $\hat\nu>0$ growing with $c$. In this case the energy density
stored in the modulus at the onset of oscillations can be expressed
as, see Appendix~\ref{app:sol}, especially Eqs.~(\ref{eq:omnum1})
and~(\ref{eq:omnum2})
\begin{eqnarray}
\label{eq:omeganucomplex1}
\Omega_{\rm \sigma,osci} &\simeq & \frac{1}{6}{\cal A}_3^2
\left({\sigma_0 \over M_{\rm Pl}}\right)^2
+{1\over 6}{\cal B}_3^2\left({\sigma_{\rm
    inf}-\sigma_0 \over M_{\rm Pl}}\right)^2
\left({p_{\rm MD} m_\sigma\over H_{\rm
    inf}}\right)^{2(\mu_{\rm MD}+1)} \, , \quad T_{\rm osci}>T_{\rm rh}\, ,
\\
\label{eq:omeganucomplex2}
\Omega_{\rm \sigma,osci} &\simeq & \frac{1}{6}{\cal A}_4^2
\left({\sigma_0 \over M_{\rm Pl}}\right)^2
+\frac{1}{6}{\cal B}_4^2\left(\frac{\sigma _{\rm inf}-\sigma _0}{M_{\rm Pl}}\right)^2
\left(\frac{H_{\rm rh}}{H_{\rm inf}}\right)^{2\mu _{\rm MD}+2}
\left(\frac{p_{\rm MD}m_{\sigma }}{H_{\rm rh}}\right)^{2\mu _{\rm RD}+2}
\, , \quad T_{\rm rh}> T_{\rm osci}\, .
\end{eqnarray}
These equations are the counterparts of Eqs.~(\ref{eq:omeganureal1})
and~(\ref{eq:omeganureal2}) in the case where the quantity $\nu$ is
complex. The coefficients ${\cal A}_3$ and ${\cal A}_4$ are defined in
Eq.~(\ref{eq:prefacA3}). The prefactors ${\cal B}_3$ and ${\cal B}_4$
are defined in Eqs.~(\ref{eq:prefacB3}) and~(\ref{eq:prefacB4}) and
are of order unity. The difference with the case $\nu>0$ (with
prefactors ${\cal A}_1$ and ${\cal A}_2$) comes from the fact that,
now, the numerical prefactors ${\cal A}_3$ and ${\cal A}_4$ may become
quite small if $\nu$ is pure imaginary and its modulus is large. The
prefactor ${\cal A}_3$ indeed scales as:
\begin{equation}
{\cal A}_3\,\simeq\,2\pi\left({\hat\nu_{\rm MD}\over
    2}\right)^{\mu_{\rm MD}+2}{\rm e}^{-\hat\nu_{\rm MD}\pi/2 -
  \mu_{\rm MD}-3} \quad \vert\nu_{\rm MD}\vert\,\gg\,1\ .\label{eq:A3scale}
\end{equation}
Hence, as $c$ grows beyond $1$, the value of $\Omega_{\sigma,\rm part,
  osci}$ decreases exponentially. As argued by
Linde~\cite{Linde:1996cx}, this could alleviate the moduli problem,
although it is notoriously difficult to construct explicit models in
which $c^2\,\gtrsim\,10$.

\par

Let us now study in more details the physical consequences of the
above expressions. Ignoring the factors of order one for simplicity,
and using the explicit expression of the coefficient ${\cal A}_3$, see
Eq.~(\ref{eq:prefacA3}), the particular solution dominates whenever
the following condition is valid
\begin{equation}
  \left\vert\sigma_{\rm inf}-\sigma_0\right\vert\left({m_\sigma\over
      H_{\rm inf}}\right)^{\mu_{\rm MD}+1}\,\ll \,\left\vert\sigma_0\right\vert
  \Gamma\left({\mu_{\rm MD}+3\over2}+i{\hat{\nu}_{\rm MD}\over2}\right)\Gamma
  \left({\mu_{\rm MD}+3\over2}-i{\hat{\nu}_{\rm MD}\over2}\right)\ ,\label{eq:homdom2}
\end{equation}
assuming for simplicity $T_{\rm osci}>T_{\rm rh}$. Then, one finds that
the bound on $\sigma_0$ obtained previously in the case of a real
$\nu$ for $H_{\rm inf}\sim10^{13}\,$GeV is now loosened by:
\begin{equation}
\sigma_0\,\lesssim\, 10^{-10}\left[(2\pi)^2\left({\hat\nu_{\rm MD}\over
  2}\right)^{2(\mu_{\rm MD}+2)}{\rm e}^{-\hat\nu_{\rm MD}\pi - 2(\mu_{\rm MD}+3)}\right]^{-1}M_{\rm
  Pl}\ ,
\end{equation}
for $m_\sigma=10^5\,$GeV, scaling approximately as $m_\sigma^{1/2}$.
To provide concrete estimates, for $c^2=10$ ($\nu\simeq 2.04$ assuming
matter domination), the r.h.s.  becomes $2\times 10^{-7}M_{\rm Pl}$;
for $c^2=20$ ($\nu\simeq2.94$), it is $1.2\times 10^{-6}M_{\rm
  Pl}$. Interestingly, these values always remain smaller or are at
most (for $c^2=20$) comparable to the standard deviation expected from
stochastic motion of the modulus if $H_{\rm
  inf}\sim10^{13}\,$GeV. Therefore, solving the moduli problem in this
way would require not only a large value of $c^2$ in order to lessen
the modulus energy density, hence the transfer of isocurvature
perturbations, but also a small inflationary scale
$\,\ll\,10^{13}\,$GeV in order to diminish the magnitude of stochastic
motion.  Note also that, for smaller values of $H_{\rm inf}$, some
parameter space opens up at large values of $\sigma_{\rm inf}$ and
large $m_\sigma$, as discussed before.

\begin{figure*}
  \centering
  \includegraphics[width=0.49\textwidth,clip=true]{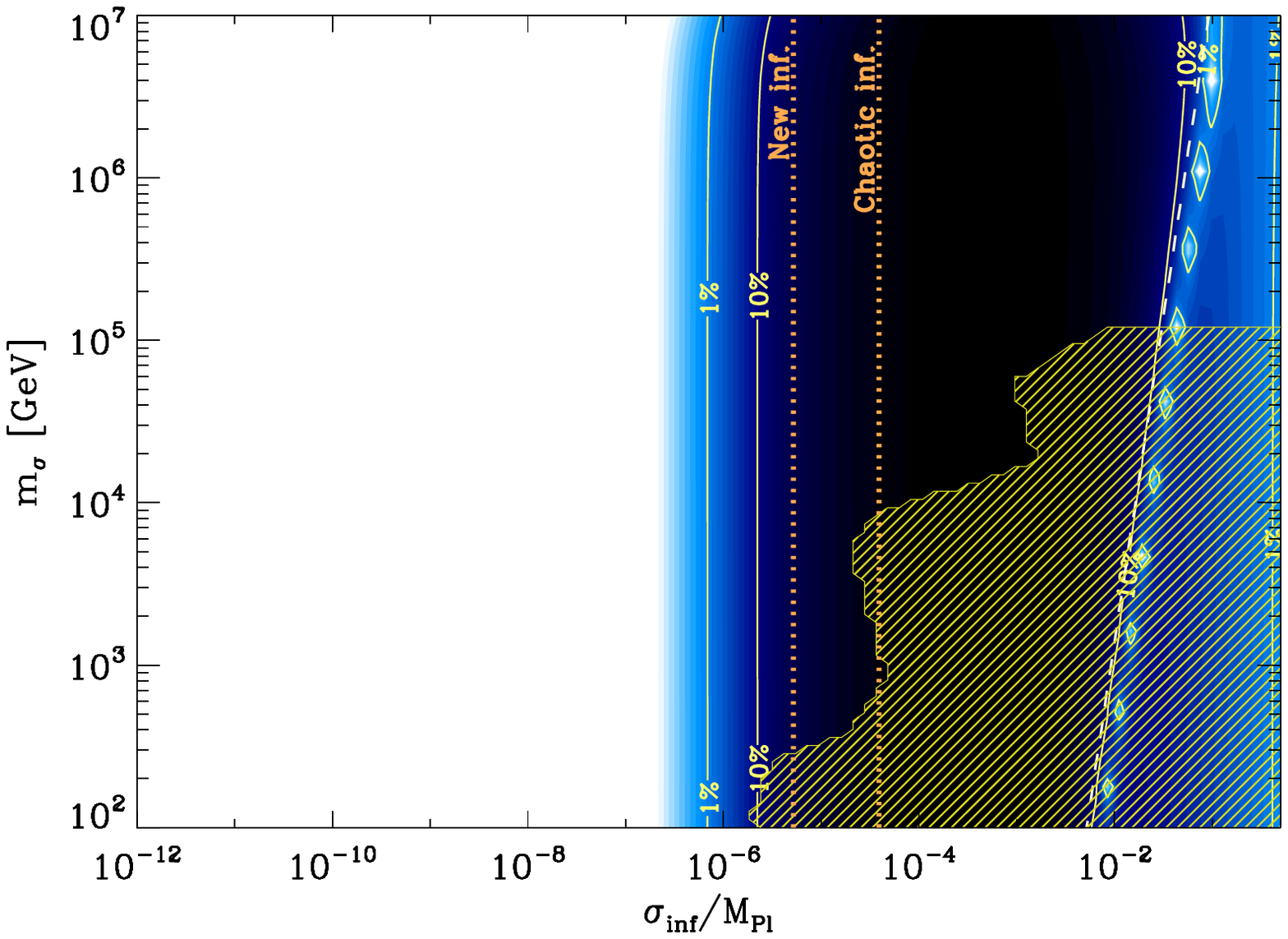} 
  \includegraphics[width=0.49\textwidth,clip=true]{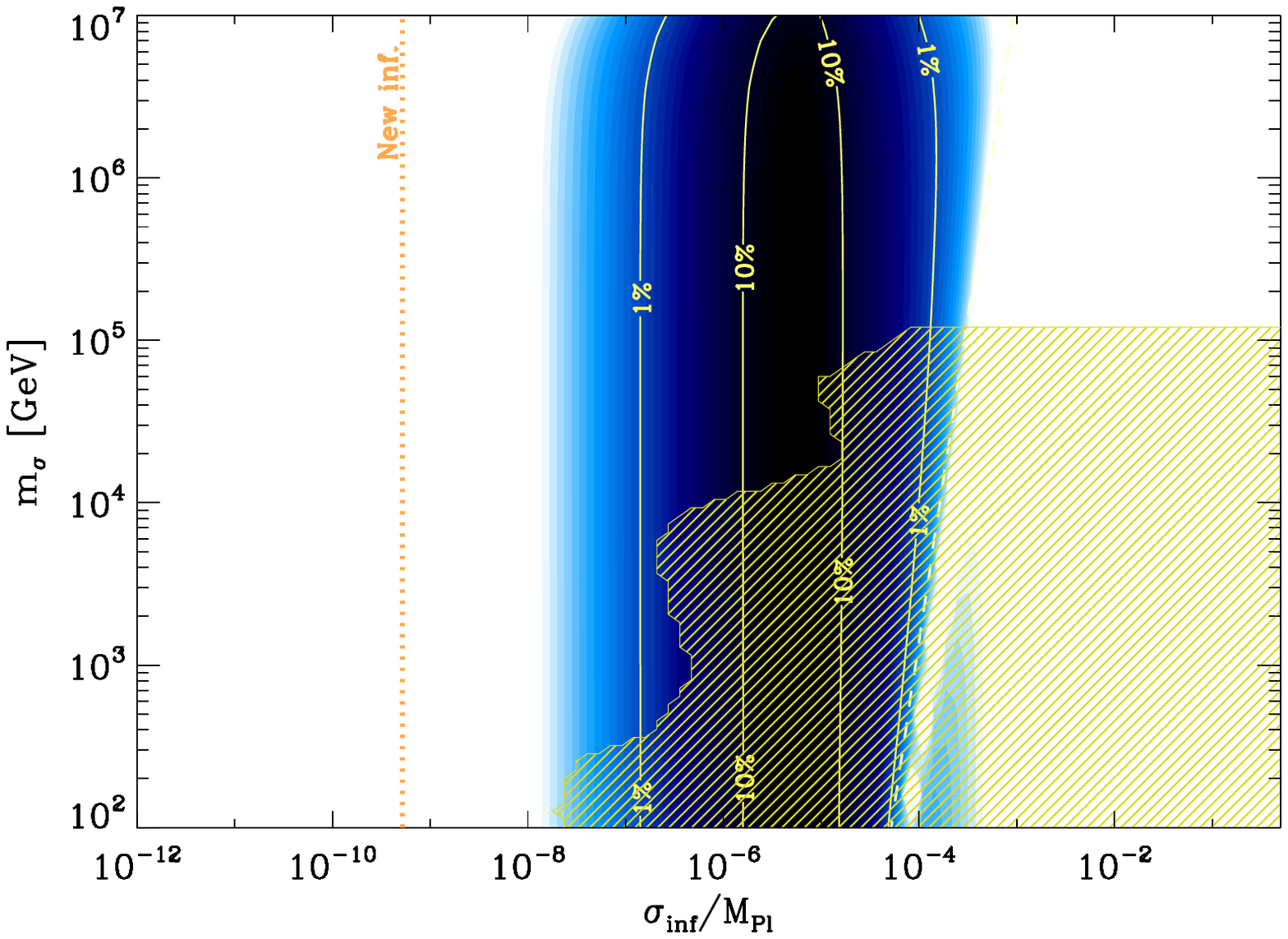} 
  \caption[...]{Same as Fig.~\ref{fig:iso-quad-m2-tot}, but for a
    modulus potential receiving a positive Hubble mass squared
    correction with $c^2=2$, and $\sigma_0=0$. This implies $\nu _{\rm
      MD}^2\simeq -0.64$ and $\nu_{\rm RD}^2\simeq -0.44$.  Left
    panel: $H_{\rm inf}=10^{13}\,$GeV, $T_{\rm rh}=10^9$\, GeV. Right
    panel: $H_{\rm inf}=10^9\,$GeV, $T_{\rm rh}=10^9$\, GeV. For this
    value of the reheating temperature, one always have $T_{\rm
      osci}>T_{\rm rh}$ in the parameter space considered in this
    figure.  } \label{fig:iso-H2-3}
\end{figure*}

If, on the contrary, the homogeneous solution dominates, then the
situation is slightly different and the constraints in the modulus
parameter space are presented in Fig.~\ref{fig:iso-H2-3}. The
comparison with Fig.~\ref{fig:iso-H2-1}, which presented the
constraints for real $\nu$ (and $\sigma_0=0$) is straightforward. One
still obtains an upper bound on $\left\vert\sigma_{\rm
    inf}-\sigma_0\right\vert$ if $H_{\rm inf}\sim 10^{13}\,$GeV,
albeit displaced to larger values due to the fact that
$\Omega_{\sigma}$ now scales as $H^{\mu+1}$ instead of
$H^{\mu+1-2\nu}$, \ie the energy density contained in the modulus
decreases faster. This upper bound can be written as:
\begin{equation}
\left\vert\sigma_{\rm inf}-\sigma_0\right\vert\,\lesssim\, 
3\times10^{-6}M_{\rm Pl}\ ,\label{eq:bbH2}
\end{equation}
with the following set of parameters: $H_{\rm inf}=10^{13}\,$GeV,
$T_{\rm rh}=10^9\,$GeV, $c^2=2$ ($\nu\simeq0.8$ assuming matter
domination after inflation), and the prefactor has been evaluated for
$m_\sigma\,=100\,$TeV. As before, one finds that this bound scales as
$H_{\rm inf}^{1/2}m_\sigma^{0}T_{\rm rh}^{-1/2}$. However, even for
$c^2=2$, this upper bound is smaller than the amplitude of stochastic
motion of the modulus in its potential during inflation, so that this
cannot be considered as a viable solution. One therefore has to
require $H_{\rm inf}\,\ll\,10^{13}\,$GeV, in which case some parameter
space opens up at both small values of $\sigma_{\rm inf}$ (because the
amplitude of stochastic motion is smaller at smaller $H_{\rm inf}$)
and at large values of $\sigma_{\rm inf}$ (where the isocurvature
fluctuation becomes much smaller than the curvature perturbation).

\par

Finally, let us end this section with the following remark. Above we
have discussed the cases of $\nu $ real or pure imaginary in both the
radiation dominated and the matter dominated epochs. Of course, there
are also two mixed cases corresponding to a real $\nu$ during the
matter dominated era and a pure imaginary $\nu $ in the radiation
dominated era and the opposite situation. This would be relevant for
the discussion above only if $T_{\rm rh}>T_{\rm osci}$, since the
matchings would have to be modified. Since this is not the most
generic case, we do not consider this situation in this paper.

\subsubsection{Case $-c^2H^2/2$}
\label{subsubsec:potwithminus}

If the modulus receives a negative Hubble mass squared contribution
after inflation, the minimum of its potential is destabilized and as a
result, the modulus will move until this negative contribution is
balanced by higher order terms in the potential, $\propto \sigma^4$ or
even non-renormalizable contributions. Let us assume for instance that
the next term in the modulus potential takes the form:
\begin{equation}
\frac{\lambda_n}{(n+4)!}\frac{\sigma^{4+n}}{M_{\rm Pl}^{n}}\ .
\end{equation}
Then, as shown in Ref.~\cite{Dimopoulos:2003ss} the effective
potential may be approximated at high energy $H\gg m_\sigma/c$ by:
\begin{equation}
V(\sigma)\,\simeq\,\frac{1}{2}\tilde c^2
H^2\left(\sigma-\sigma_n\right)^2 \, +\,
{\lambda_n\over(n+4)!}{\left(\sigma-\sigma_n\right)^{4+n}\over M_{\rm
    Pl}^n} \ ,\label{eq:effpot}
\end{equation}
where the time-dependent quantity $\sigma _n$ can be expressed as
\begin{equation}
\sigma_n(t)\,=\,\left[\frac{(n+3)!}{\lambda_n}c^2\,
H^2M_{\rm Pl}^n\right]^{1/(n+2)}\ , 
\end{equation}
and $\tilde c^2\,\equiv \,(n+2)c^2$. The value $\sigma_n$ corresponds
to the local minimum of the potential. Constant terms in $V(\sigma)$
have been omitted in Eq.~(\ref{eq:effpot}), as well as subleading
terms when compared to the last term on the right hand side,
see~Ref.~\cite{Dimopoulos:2003ss}. One crucial difference between this
effective potential and that obtained for $+c^2H^2$ is the fact that
the local minimum now depends on $H$ and thus evolves in time.

\par

One should distinguish three phases of evolution depending on which
term in the potential dominates. If $\sigma_{\rm inf} \gg \sigma
_{n,\rm inf}\equiv \sigma _n(H=H_{\rm inf})$, then the field initially
evolves in the high order part of the potential given by $V(\sigma
)\simeq \lambda _n\sigma ^{4+n}/[(n+4)!M_{\rm Pl}^n]$. As shown in
Appendix~\ref{app:sol}, the field is then driven to $\sigma_n$ within
a fraction of e-fold of order $(\sigma_{\rm inf}/\sigma _{n,\rm
  inf})^{-(n+2)/2}$. At this stage, the effective $\tilde
c^2H^2(\sigma-\sigma_n)^2$ of the potential takes over the high order
part. In order to model this case $\sigma_{\rm inf} \gg \sigma _{n,\rm
  inf}$, we simply assume that, starting from the end of inflation,
the field evolves in the $\tilde c^2H^2(\sigma-\sigma_n)^2$
part with an initial value of order $\sigma _{n,\rm inf}$.  In this
situation, we also neglect the initial kinetic energy since kinetic
energy is strongly damped when the field evolves in the high order
part of the potential. Furthermore, this is conservative in the sense
that it underestimates the energy density contained in modulus
oscillations at late times, and therefore tends to loosen slightly the
constraints derived. If $\sigma_{\rm inf} \lta \sigma _{n,\rm inf}$,
then one should directly approximate the potential by $\tilde
c^2H^2(\sigma-\sigma_n)^2$ with $\sigma _{\rm inf}$ as the initial
condition. All in all, it suffices to solve for the evolution of
$\sigma $ in the potential $\tilde
c^2H^2(\sigma-\sigma_n)^2$ with an initial condition $\sigma _{\rm
  eff,min}\equiv \min(\sigma _{\rm inf},\sigma _{n,\rm inf})$.

\par

Thus ignoring the high order part of the potential, at high energy
$H\gg m_\sigma/c$, the solution for $\sigma$ comprises a solution to
the homogeneous equation $\sigma_{\rm hom}$ and a particular solution
$\sigma_{\rm part}$, as before, see Appendix~\ref{app:sol}. The
scaling of the homogeneous solution is similar to that obtained in the
previous section with a $+c^2H^2$ effective squared mass term, but the
particular solution scales differently due to the time dependence of
$\sigma_n$. As shown in Appendix~\ref{app:sol}, $\sigma_{\rm
  part}\,\propto\,\sigma_n$ with a prefactor $\alpha_n$ of order
unity. The explicit expression of $\alpha _n$ can be found in
Appendix~\ref{app:sol}, see Eq.~(\ref{eq:solminushcorr}). Then
$\sigma_{\rm part}\,\propto\, t^{-2/(n+2)}$ so that the energy density
of the modulus associated to this particular solution scales as
$\rho_{\sigma,\rm part}\,\propto\, H^2\sigma _{\rm part}^2\, \propto
\, H^{2(n+4)/(n+2)}$. Consequently, the particular solution
contribution to the energy density at the onset of oscillations can be
written as
\begin{equation}
\Omega_{\sigma,\rm part, osci}\,\approx\, \frac{1}{6}\left({\sigma_{\rm
    part, inf}\over M_{\rm Pl}}\right)^2\left({m_\sigma\over H_{\rm
    inf}}\right)^{4/(n+2)}\ ,\label{eq:sp3}
\end{equation}
where we have ignored all factors of order one. In the above
expression, $\sigma _{\rm part, inf}$ represents the initial value of
the particular solution at the beginning of the era driven by $\tilde
c^2H^2(\sigma-\sigma_n)^2$, that is to say
\begin{equation}
\sigma_{\rm part, 
  inf}\,=\,\alpha _n\sigma _{n,\rm inf}=\alpha_n
\left[\frac{(n+3)!}{\lambda_n}c^2\,H_{\rm
    inf}^2M_{\rm Pl}^n\right]^{1/(n+2)}\,\simeq\, H_{\rm
  inf}^{2/(n+2)}M_{\rm Pl}^{n/(n+2)}\ .\label{eq:sigpartH}
\end{equation}
Concerning the homogeneous solution, its evolution is the same as in
the previous section (see also Appendix~\ref{app:sol}). This means
that one should again distinguish the case where $\nu $ is real or
imaginary and should treat separately the situation where the onset of
oscillations occurs before or after the reheating. Straightforward
calculations, similar to the ones already performed in the previous
sections, lead to our final expression of $\Omega_{\rm \sigma,osci}$
\begin{eqnarray}
\Omega_{\rm \sigma,osci} &\simeq & \frac{1}{6}\left({\sigma_{\rm
    part, inf}\over M_{\rm Pl}}\right)^2\left({m_\sigma\over H_{\rm
    inf}}\right)^{4/(n+2)}
+\frac{1}{6}\left({\sigma_{\rm eff,
    inf}-\sigma_{\rm part, inf} \over M_{\rm Pl}}\right)^2
\left({m_\sigma\over H_{\rm
    inf}}\right)^{2[\mu_{\rm MD}+1-{\rm Re}(\nu_{\rm MD})]}\, ,\quad
T_{\rm osci}>T_{\rm rh}\, ,
\label{eq:oosci_minus1}\\
\Omega_{\rm \sigma,osci} &\simeq & 
\frac{1}{6}\left({\sigma_{\rm
    part, inf}\over M_{\rm Pl}}\right)^2\left({m_\sigma\over H_{\rm
    inf}}\right)^{4/(n+2)}\nonumber \\ & & \,\,\,
+\frac{1}{6}\left({\sigma_{\rm eff,
    inf}-\sigma_{\rm part, inf} \over M_{\rm Pl}}\right)^2
\left({m_\sigma\over H_{\rm
    rh}}\right)^{2[\mu_{\rm RD}+1-{\rm Re}(\nu_{\rm RD})]}
\left({H_{\rm rh}\over H_{\rm
    inf}}\right)^{2[\mu_{\rm MD}+1-{\rm Re}(\nu_{\rm MD})]}\,,
\quad T_{\rm osci}<T_{\rm rh}\, , \label{eq:oosci_minus2}
\end{eqnarray}
where we have neglected the factors of order one and where the
appearance of the real part of $\nu$ accounts for both possibilities
($\nu$ real or imaginary).

\begin{figure*}
  \centering
  \includegraphics[width=0.49\textwidth,clip=true]{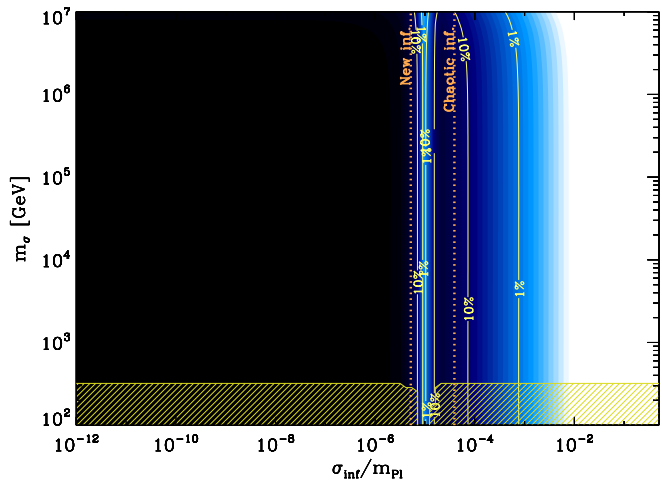}
  \caption[...]{Same as Fig.~\ref{fig:iso-quad-m2-tot}, for a
    potential receiving a negative Hubble mass squared correction with
    $c^2=0.5$, and $n=0$. $H_{\rm inf}=10^{13}\,$GeV (for $H_{\rm
      inf}=10^9\,$GeV, constraints from isocurvature fluctuations and
    from big-bang nucleosynthesis vanish).
} \label{fig:iso-H2m-1}
\end{figure*}

When discussing the case $+c^2H^2$, we considered two cases, one in
which the particular solution dominates, the other in which the
homogeneous solutions dominates. We cannot do so here, because the
particular solution is entirely determined by $H_{\rm inf}$ and no
longer dependent on the magnitude of $\sigma_0$. Both contributions
have to be considered together. Furthermore, we recall that
$\sigma_{\rm eff, inf}=\min(\sigma _{\rm inf},\sigma _{n,\rm
  inf})$. Therefore, whether the particular or the homogeneous
solution dominates in Eqs.~(\ref{eq:oosci_minus1})
and~(\ref{eq:oosci_minus2}) depends on $n$, $\mu$, $\nu$ and
$m_\sigma/H_{\rm inf}$ (assuming $T_{\rm osci}>T_{\rm rh}$).

\par

Regarding the fluctuations of $\sigma$, it is not possible to follow
analytically the evolution of $\delta\sigma/\sigma$ from the end of
inflation until $H=m_\sigma$ due to the non-linearity of the
potential. For the sake of the argument, we thus assume that the
initial conditions are the same as in the previous cases, namely
Eqs.~(\ref{eq:ziniquad}),(\ref{eq:Siniquad}).

\par

Let us first discuss the case $n=0$ and assume for the sake of
discussion that $c^2=0.5$. Then, $\nu_{\rm MD}$ is imaginary since
$\tilde c^2=1$. Furthermore, $2(\mu_{\rm MD}+1)=1$ and $4/(n+2)=2$ in
this case, so that the particular solution is always negligible in
front of the homogeneous solution. The contribution of this latter to
the energy density is nevertheless suppressed by $m_\sigma/H_{\rm
  inf}$ and $\Omega_\sigma^{<_{\rm d}}$ is so small that the
constraints from big-bang nucleosynthesis are significantly weakened,
see Fig.~\ref{fig:iso-H2m-1}; they now allow moduli masses above $300
\mbox{GeV}$ for all $\sigma _{\rm inf}$. This constraint does not
depend on $\sigma _{\rm inf}$ because $\Omega_\sigma^{<_{\rm d}}$
hardly depends on $\sigma_{\rm inf}$ since $\left(\sigma_{\rm
  eff,inf}-\sigma_{\rm part, inf}\right)^2\sim \sigma_{\rm
  part,inf}^2$ in both limits $\sigma_{\rm inf}\,\ll\,\sigma_{\rm
  part,inf}$ and $\sigma_{\rm inf}\,\gg\,\sigma_{\rm
  part,inf}$. However, the constraints from the production of
isocurvature fluctuations are quite significant in this case because
the modulus can perturb significantly the perturbation spectrum of
dark matter eventhough it does not dominate the energy density at the
time of decay. Strictly speaking, this is true as long as $H_{\rm
  inf}\sim10^{13}\,$GeV, as these isocurvature constraints are less
stringent for $H_{\rm inf}=10^9\,$GeV.

\par

\begin{figure*}
  \centering
  \includegraphics[width=0.49\textwidth,clip=true]{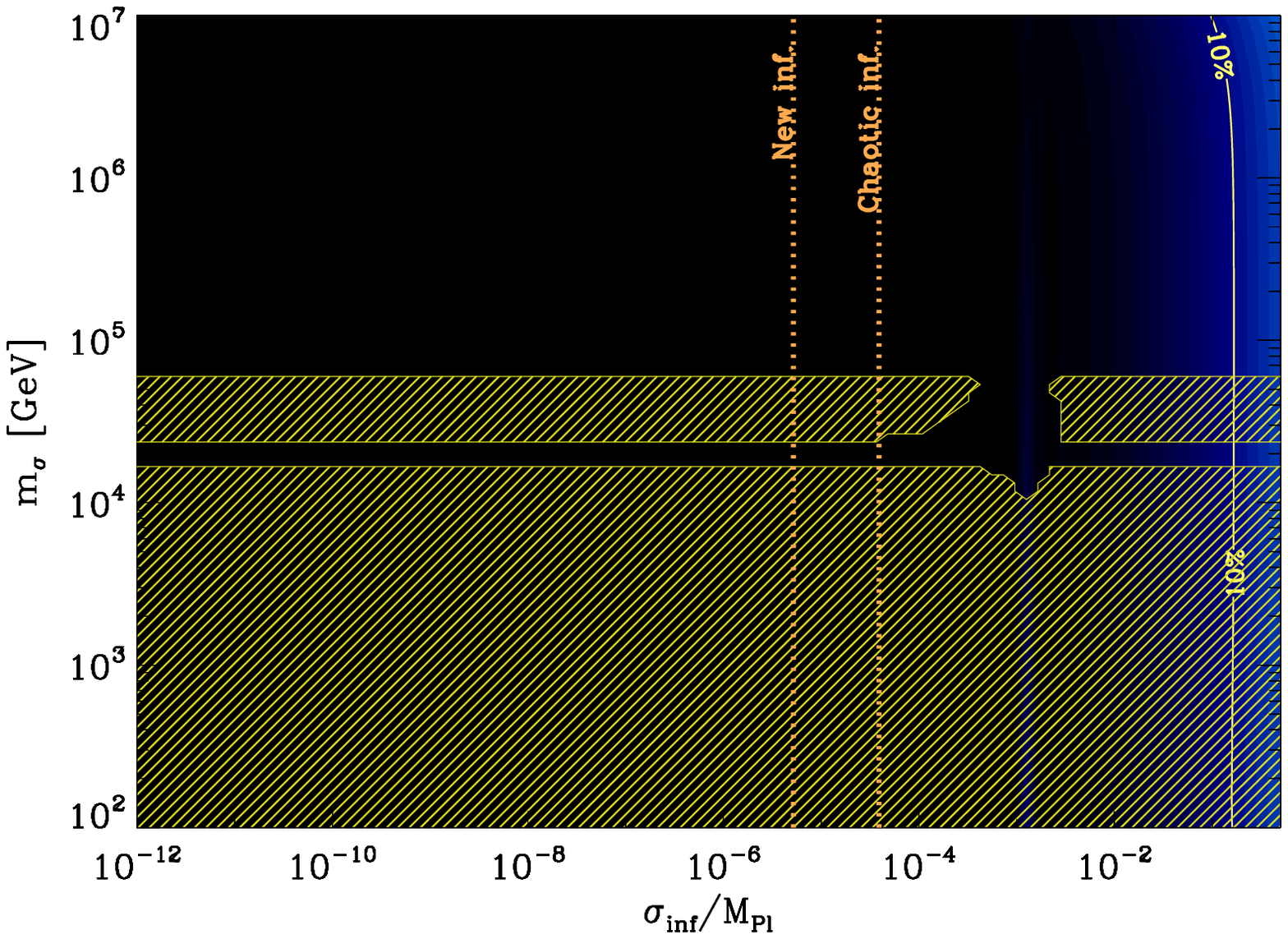}
  \includegraphics[width=0.49\textwidth,clip=true]{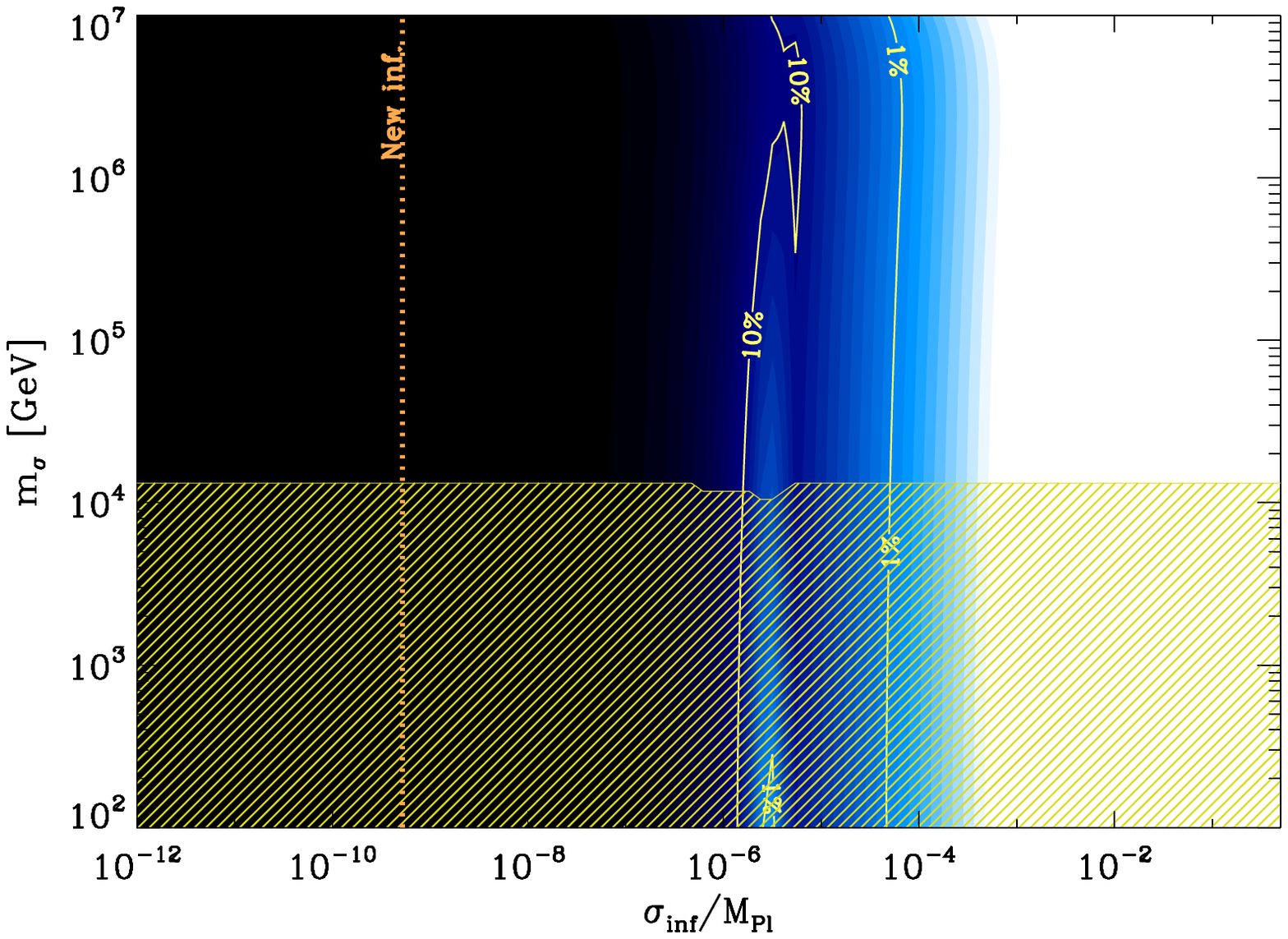}
  \caption[...]{Same as Fig.~\ref{fig:iso-quad-m2-tot}, for a
    potential receiving a negative Hubble mass squared correction with
    $c^2=0.5$, and $n=1$. Left panel: $H_{\rm
      inf}=10^{13}\,$GeV. Right panel: $H_{\rm inf}=10^9\,$GeV.  The
    standard deviations due to stochastic motion are calculated as in
    Fig.~\ref{fig:iso-quad-m2-tot}.  } \label{fig:iso-H2m-2}
\end{figure*}

When $n\geq 1$, $\Omega_{\sigma,\rm osci}$ becomes large enough to
produce substantial isocurvature fluctuations in nearly all of the
parameter space, see Fig.~\ref{fig:iso-H2m-2}. For $n=1$, the
homogeneous solution still dominates over the particular solution. The
constraints can be derived from the discussion of
Sec.~\ref{subsubsec:potwithplus} (\ie $+c^2H^2$ potential, in the case
in which the homogeneous solution dominates) provided one replaces
$(\sigma _{\rm inf}-\sigma _0)^2$ with $(\sigma _{\rm inf}-\sigma
_{\rm part,inf})^2\simeq \sigma _{\rm part, inf}^2$, which is a
function of $n$ and $H_{\rm inf}$, but which does not depend on
$\sigma _{\rm inf}$. Accordingly, $\Omega_\sigma^{<_{\rm d}}$ is
sufficiently large to produce significant isocurvature perturbations
because of the scaling of $\sigma_{\rm part, inf}$ with $n$: for
$n=0$, $\sigma_{\rm part,inf}\,\sim\,H_{\rm inf}$, but for $n\geq 1$,
$\sigma_{\rm part,inf}\,\gg\,H_{\rm inf}$, see
Eq.~(\ref{eq:sigpartH}). As before, we find that the isocurvature mode
becomes small enough to satisfy the constraints from cosmic microwave
background data if $H_{\rm inf}\,\ll\,10^{13}\,$GeV and $\sigma_{\rm
  inf}\,\sim\, M_{\rm Pl}$.

\subsection{Modulus production by inflaton decay}
\label{sec:inflp}

The above discussion has implicitly assumed that no modulus was
produced after inflation. However it seems reasonable to assume that
the inflaton can decay into the modulus sector, through possibly
Planck suppressed interactions. We take a branching ratio such that
each inflaton produces $N_\sigma$ moduli. Unless the modulus and the
inflaton are coupled one to the other, one should expect $N_\sigma
\,\lesssim\, 1/g_{\star,\rm rh}$, which means that at most, moduli are
produced at the same rate than other light particles. If the inflaton
is more strongly coupled to the visible sector than to the modulus
sector, one should expect a much lower value of $N_\sigma$.

\par

Since the modulus mass is generically much smaller than the inflaton
mass, the moduli produced through inflaton decay are
ultra-relativistic, with energy
$E_{\sigma\leftarrow\phi}\,\sim\,m_\phi/2$. These particles do not
thermalize but redshift to non-relativistic velocities. As their
momentum redshifts away according to $p_{\sigma}=m_{\phi}/2(a_{\rm
  rh}/a)$, the particles become non relativistic, when $p\sim
m_\sigma$ or $a_{\rm rh}/a_{\rm n-rel}=2m_{\rm \sigma }/m_{\phi}$. The
corresponding temperature $T_{\rm n-rel}$ is given by
\begin{equation}
g_{\star\rm n-rel}^{1/3}T_{\rm n-rel}\,=\,g_{\star\rm rh}^{1/3}T_{\rm
  rh}\frac{2m_\sigma}{m_\phi}\ .
\end{equation}
For temperatures $T>T_{\rm n-rel}$,
$\rho_{\sigma\leftarrow\phi}\propto 1/a^4$ \ie it scales as radiation
while for $T<T_{\rm n-rel}$, it scales as a pressureless fluid,
$\rho_{\sigma\leftarrow\phi}\propto 1/a^3$. Just after reheating, the
energy density of the moduli that were produced by inflaton decay is
given by
\begin{equation}
\rho_{\sigma\leftarrow\phi}^{\rm rh}\simeq N_{\sigma} 
n_{\phi}\frac{m_{\phi}}{2}=\frac{N_{\sigma}}{2}\rho _{\phi}^{\rm rh}\, .
\end{equation}
If the temperature $T_{\rm n-rel}$ is smaller than $T_{\rm d}$, then
$\rho_{\sigma\leftarrow\phi}$ will scale as the radiation until the
modulus decay. As a consequence
\begin{equation}
\label{eq:rhosigmaphisimple}
\frac{\rho_{\sigma\leftarrow\phi}^{<_{\rm d}}}{\rho_{\gamma}^{<_{\rm
      d}}}\,\simeq\, \frac{N_{\sigma }}{2}\, ,\quad T_{\rm n-rel}<T_{\rm
      d}\, .
\end{equation}
If, on the contrary, $T_{\rm n-rel}>T_{\rm d}$, the energy density
$\rho_{\sigma\leftarrow\phi}$ will then increase with respect to that of
radiation in the era following $T_{\rm n-rel}$ and preceding modulus
decay. Therefore, the ratio of inflaton produced moduli energy density
to radiation energy density immediately before modulus decay can be
written as:
\begin{eqnarray}
\label{eq:rhosigphi}
\frac{\rho_{\sigma\leftarrow\phi}^{<_{\rm d}}}{\rho_{\gamma}^{<_{\rm
      d}}}&\,=\,& \frac{N_\sigma}{2}\frac{T_{\rm n-rel}}{T_{\rm d}}\nonumber\\
&\,\simeq\,&3.61\,
\left(\frac{N_\sigma}{10^{-3}}\right)\left(\frac{g_{\rm \star,rh}}
{g_{\rm \star,n-rel}}\right)^{1/3}
\left(\frac{g_{\rm \star,dec}}{10.75}\right)^{1/4}
\left(\frac{m_\sigma}{100\,{\rm
    TeV}}\right)^{-1/2} \left(\frac{m_\phi}{10^{13}\,{\rm GeV}}\right)^{-1}
\left(\frac{T_{\rm rh}}{10^9\,{\rm GeV}}\right)\ ,\quad T_{\rm n-rel}>T_{\rm
      d}\, ,\nonumber\\
& &
\end{eqnarray}
where one has used the expression~(\ref{eq:Td}) of $T_{\rm d}$. Let us
notice that, in most of parameter space, $T_{\rm n-rel}>T_{\rm d}$.

\begin{figure*}
  \centering
  \includegraphics[width=0.49\textwidth,clip=true]{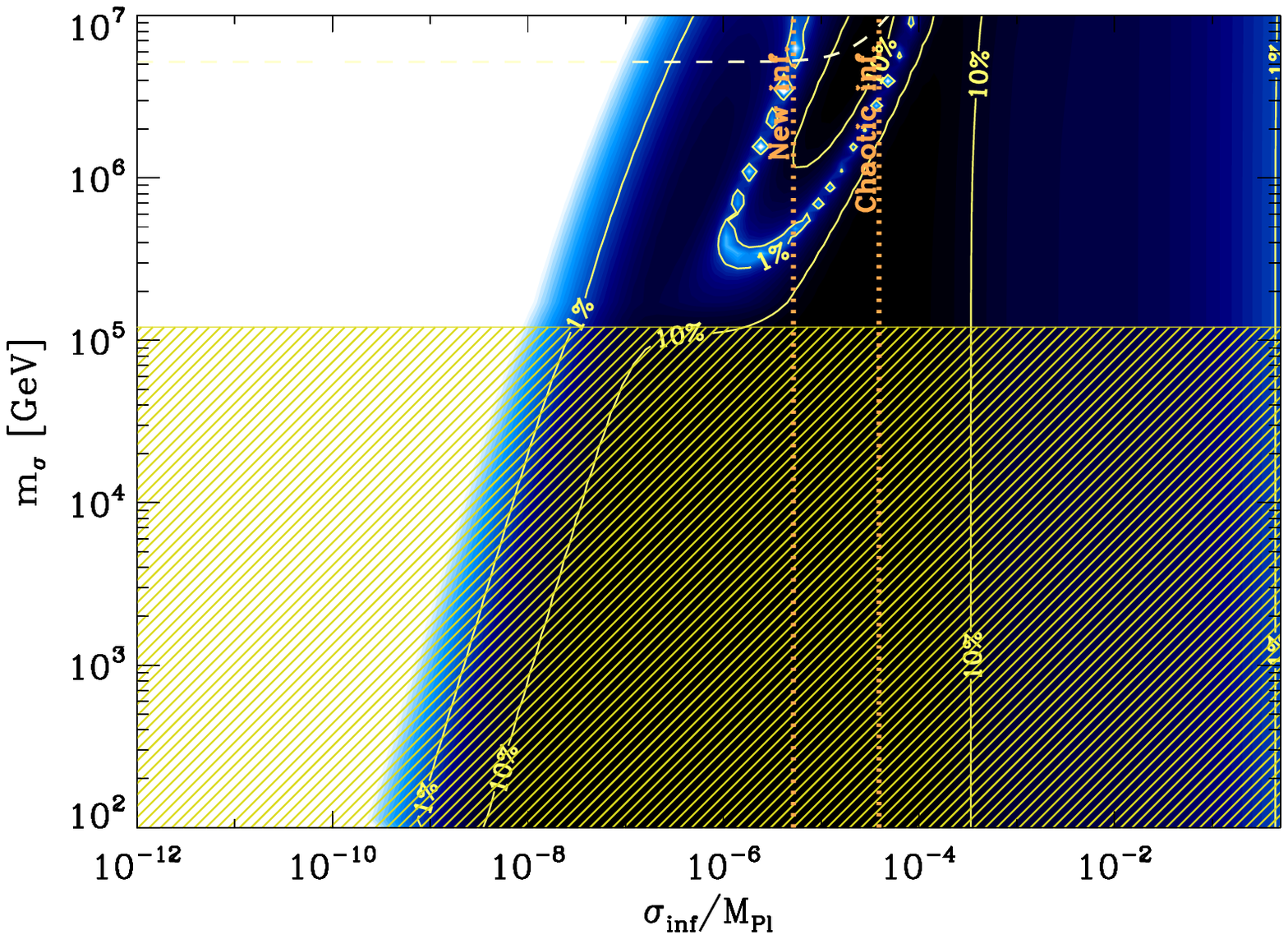}
  \includegraphics[width=0.49\textwidth,clip=true]{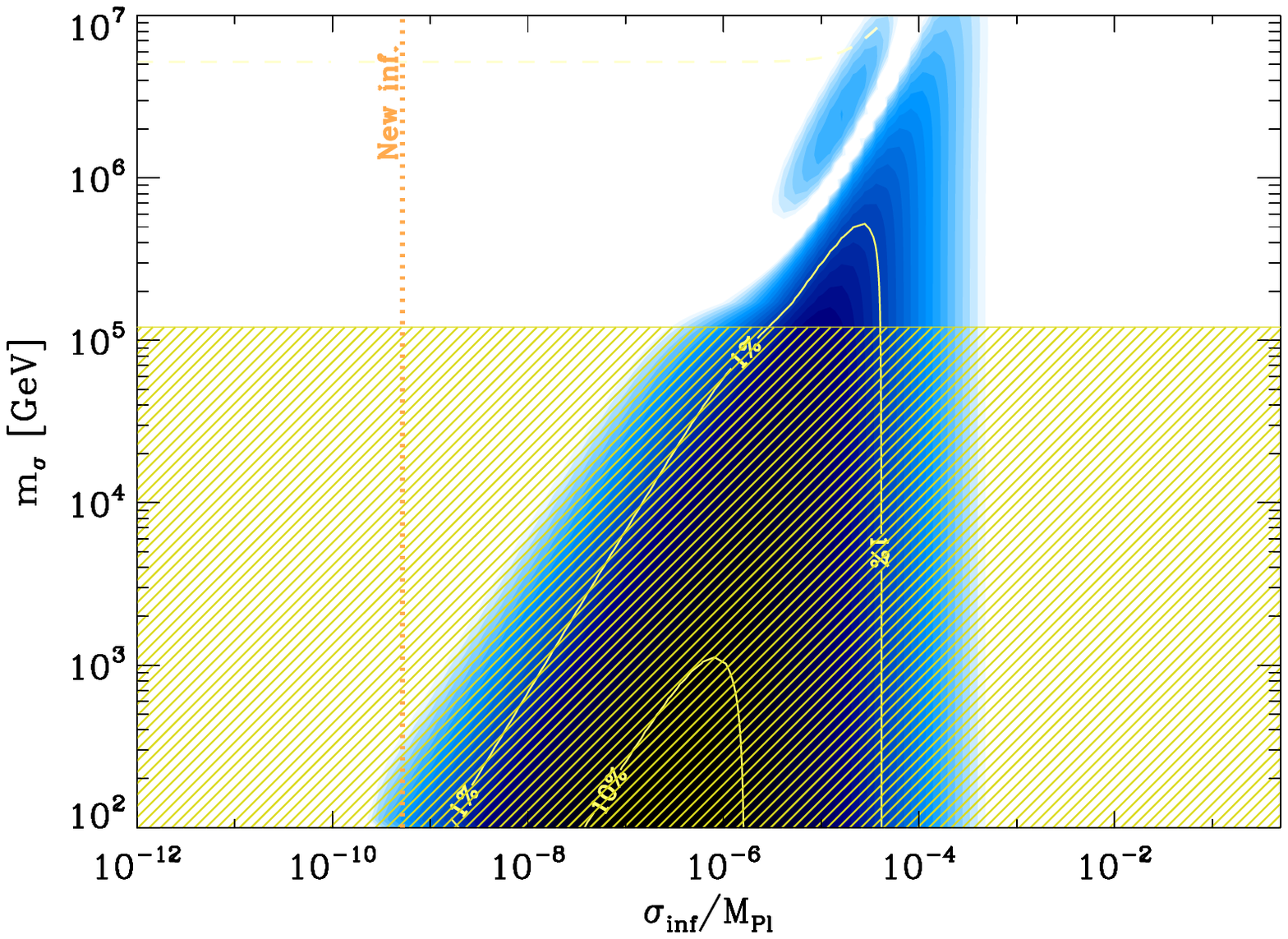}
  \caption[...]{Same as Fig.~\ref{fig:iso-quad-m2-tot}, for a
    quadratic modulus potential, but now accounting for inflaton
    produced moduli with $N_\sigma=10^{-3}$ and an inflaton mass
    $m_\phi=10^{13}\,$GeV. Left panel: $H_{\rm inf}=10^{13}\,$GeV,
    $T_{\rm rh}=10^9$\,GeV. Right panel: $H_{\rm inf}=10^9\,$GeV,
    $T_{\rm rh}=10^9$\,GeV.} \label{fig:iso-H2ip-1}
\end{figure*}

The next step is to study how the previous considerations will impact
the calculations developed in the previous sections. Obviously the
amount of energy density stored in the modulus oscillations remains
unchanged. However the amount of isocurvature fluctuation produced at
a same value of $\Omega_\sigma^{<_{\rm d}}$ is
reduced~\cite{Linde:2005yw}. Indeed, inflaton produced moduli inherit
the same spectrum of perturbations than radiation, and therefore there
is no initial isocurvature perturbation between those moduli and the
radiation fluid. Including these inflaton produced moduli, the initial
value of the modulus curvature perturbation now reads:
\begin{equation}
\zeta_\sigma^{\rm (i)\, \prime}\,=\,\gamma\zeta_\sigma^{\rm (i)} +
(1-\gamma)\zeta_{\phi}^{\rm (i)}\ .
\end{equation}
The quantities in this equation should be understood as follows:
$\zeta_\sigma^{\rm (i)}$ and $\zeta_\phi^{\rm (i)}$ correspond to the
curvature perturbations of the modulus and the inflaton acquired
during inflation, as before. The curvature perturbation
$\zeta_\sigma^{\rm (i)\, \prime}$ is the curvature perturbation for
the modulus that should be used in Eqs.~(\ref{eq:cond-iso-dm})
and~(\ref{eq:cond-iso-quad}) for the calculation of the final
isocurvature perturbations. Finally, $\gamma$ denotes the ratio of the
energy density of moduli initially present at decay to the total
amount of moduli at decay (those initially present together with the
inflaton produced moduli). The final effect is to modify the initial
modulus - radiation isocurvature perturbation by a factor $\gamma$ in
Eqs.~(\ref{eq:cond-iso-dm}) and~(\ref{eq:cond-iso-quad}), namely
$x_{\chi \gamma}\rightarrow \gamma x_{\chi \gamma}$. It is also useful
to define $\gamma'\equiv \rho_{\sigma\leftarrow\phi}^{<_{\rm
    d}}/\rho_{\sigma}^{<_{\rm d}}$ (with $\rho_{\sigma}^{<_{\rm d}}$
the amount of energy density stored in the modulus oscillations
immediately before decay):
\begin{equation}
\gamma\,\equiv\,\frac{1}{\gamma'+1}\ ,
\end{equation}
so that, if $T_{\rm n-rel}>T_{\rm d}$:
\begin{equation}
\label{eq:gammaprime}
\gamma' \simeq  10^{-11}\,
\left(\frac{N_\sigma}{10^{-3}}\right)\frac{1}{\Omega _{\rm \sigma, osci}}
\left(\frac{g_{\rm \star,rh}}
{g_{\rm \star,n-rel}}\right)^{1/3}
\left(\frac{m_\sigma}{100\,{\rm
    TeV}}\right) \left(\frac{m_\phi}{10^{13}\,{\rm GeV}}\right)^{-1}
\ ,\quad T_{\rm n-rel}>T_{\rm
      d}\, .
\end{equation}
In order to obtain this formula, we have used Eq.~(\ref{eq:rhosigphi})
in order to express $\rho_{\sigma\leftarrow\phi}^{<_{\rm d}}$ and
Eq.~(\ref{eq:rsig}) to express $\rho_{\sigma}^{<_{\rm d}}$ [or,
alternatively, Eq.~(\ref{eq:rsig-quad}) with the term $\sigma _{\rm
  inf}^2/M_{\rm Pl}^2$ replaced by $6\Omega _{\rm \sigma ,osci}$ in
order not to be restricted to the quadratic case]. Notice that we have
taken $\alpha _{\rm osci/rh}=1$.

\par

If, on the contrary, $T_{\rm n-rel}<T_{\rm d}$, then the $\gamma '$
factor can be expressed as
\begin{equation}
\gamma '\simeq  1.38\times 10^{-15}\,
\left(\frac{N_\sigma}{10^{-3}}\right)\frac{1}{\Omega _{\rm \sigma, osci}}
\left(\frac{g_{\rm \star,dec}}
{10.75}\right)^{-1/4}
\left(\frac{m_\sigma}{100\,{\rm
    TeV}}\right)^{3/2} 
\left(\frac{T_{\rm rh}}{10^9\,{\rm GeV}}\right)^{-1}
\ ,\quad T_{\rm n-rel}<T_{\rm d}\, , 
\end{equation}
where, this time, $\rho_{\sigma\leftarrow\phi}^{<_{\rm d}}$ has been
obtained from Eq.~(\ref{eq:rhosigmaphisimple}). In these equations,
$\Omega_{\sigma,\rm osci}$ should be understood as corresponding to
the oscillations of the modulus. It does not include, in particular,
the moduli produced through inflaton decay.

\par

To study the effect of such modulus production through inflaton decay,
we first assume that the potential is purely quadratic, \ie $c=0$. The
results are presented in Fig.~\ref{fig:iso-H2ip-1}. Compared to
Fig.~\ref{fig:iso-quad-m2-tot}, one finds that the big-bang
nucleosynthesis constraints now exclude all moduli masses below
$100\,$TeV. This is expected insofar as the amount of moduli energy
density produced through inflaton decay is sufficient to disrupt
big-bang nucleosynthesis; since this amount does not depend on
$\sigma_{\rm inf}$, contrary to the amount of energy density stored in
moduli oscillations, the big-bang nucleosynthesis constraints also do
not depend on $\sigma_{\rm inf}$. Note that the inflaton may also
decay into gravitinos, with similar consequences for big-bang
nucleosynthesis, see~\cite{Kawasaki:2006gs}.

\par

The constraints from isocurvature fluctuations are pushed to larger
values of $\sigma_{\rm inf}$, since the factor $\gamma$ becomes small
when the energy density produced through inflaton decay far exceeds
that stored in modulus oscillations. Conversely, a larger value of
$\sigma_{\rm inf}$ yields a larger value of $\Omega_{\sigma, \rm
  osci}$ hence a larger value of $\gamma$.  Assuming for simplicity
$T_{\rm n-rel}>T_{\rm d}$, one can check that the suppression of
isocurvature fluctuations becomes effective for
\begin{equation}
\sigma_{\rm inf}\,\ll\, 3\times 10^{-6} M_{\rm Pl}\,
\left(\frac{m_\sigma}{100\,{\rm TeV}}\right)^{1/2}
\left(\frac{m_\phi}{10^{13}\,{\rm GeV}}\right)^{-1/2}
\left(\frac{N_\sigma}{10^{-3}}\right)^{1/2}\ .
\end{equation}
In order to obtain this expression, we have used
Eq.~(\ref{eq:gammaprime}) and have written the condition $\gamma '\gg
1$ (which is equivalent to $\gamma \ll 1$) in the quadratic case,
namely $\Omega _{\rm osci } \sim \sigma_{\rm inf}^2/M_{\rm
  Pl}^2$. This allows to understand, at least qualitatively, the trend
shown in Fig.~\ref{fig:iso-H2ip-1}.
 
\par

\begin{figure*}
  \centering
  \includegraphics[width=0.49\textwidth,clip=true]{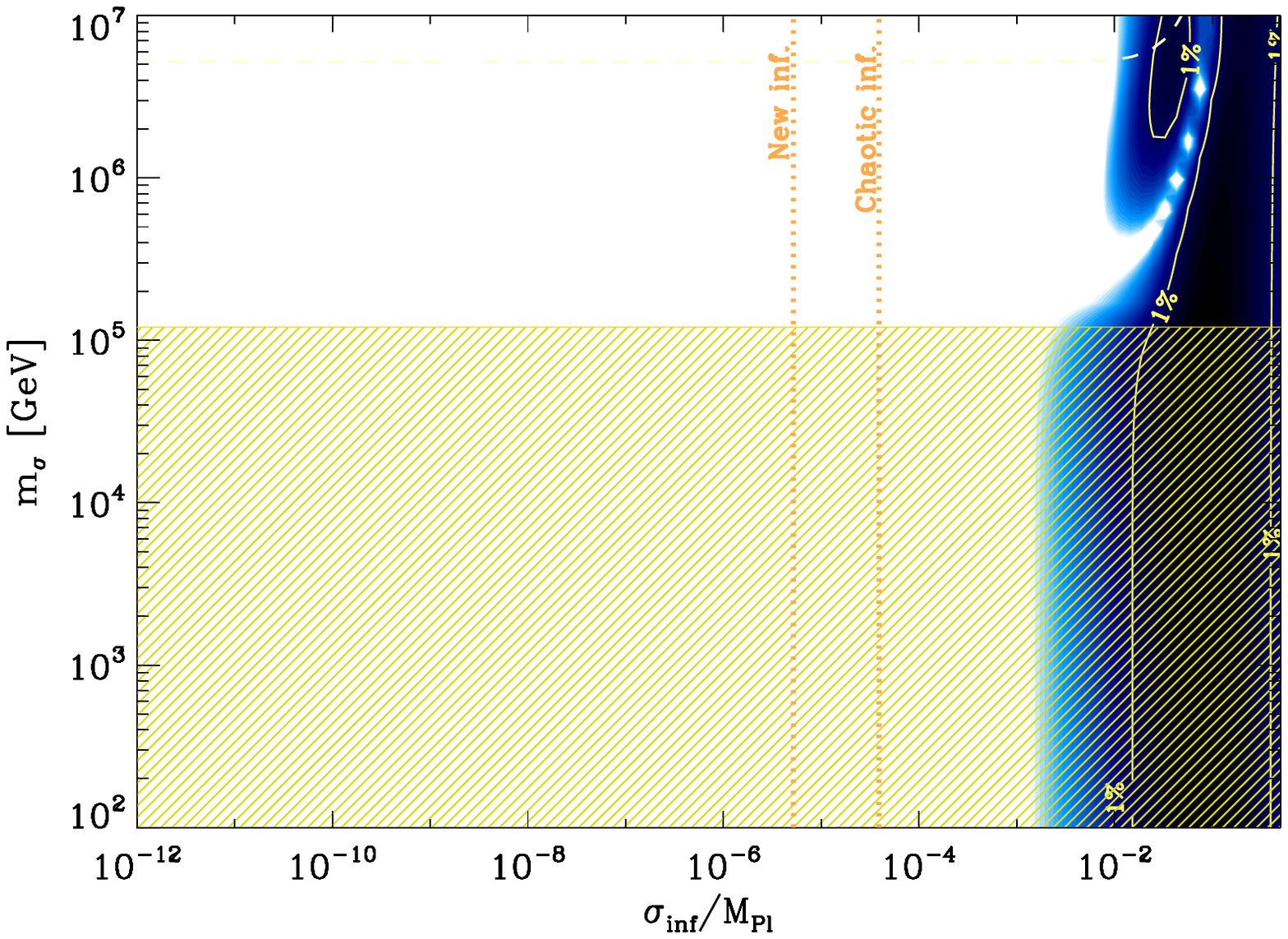}
  \includegraphics[width=0.49\textwidth,clip=true]{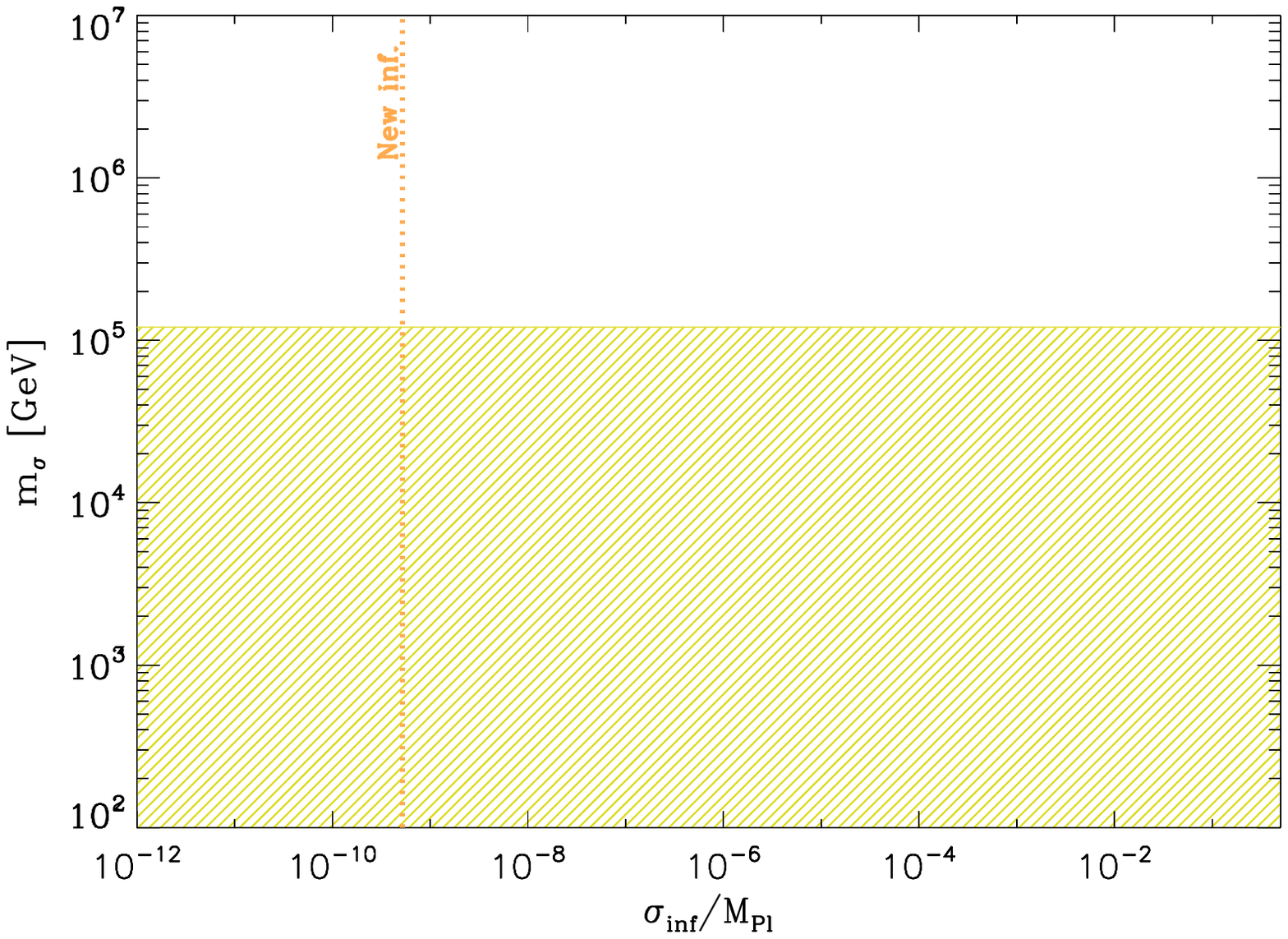}
  \caption[...]{Same as Fig.~\ref{fig:iso-H2-3}, for a receiving a
    positive squared mass Hubble contribution with $c^2=2$, but now
    accounting for inflaton produced moduli with $N_\sigma=10^{-3}$
    and an inflaton mass $m_\phi=10^{13}\,$GeV. Left panel: $H_{\rm
      inf}=10^{13}\,$GeV, $T_{\rm rh}=10^9$\,GeV. Right panel: $H_{\rm
      inf}=10^9\,$GeV, $T_{\rm rh}=10^9$\, GeV. The stochastic effects
    are calculated with $c_{\rm i}^2=2$.} \label{fig:iso-H2ip-2}
\end{figure*}

The effect in the case where the potential receives supergravity
corrections is rather straightforward to guess: big-bang
nucleosynthesis constraints remain unchanged as compared to the above
Fig.~\ref{fig:iso-H2ip-1}, but the contours depicting the amount of
isocurvature fluctuations produced are shifted toward higher values of
$\sigma_{\rm inf}$, as a result of the redshifting of the energy
density stored in modulus oscillations after inflation, yielding a
smaller value for $\gamma$. For instance, considering the case $c^2=2$
as in Fig.~\ref{fig:iso-H2-3}, one obtains the constraints depicted in
Fig.~\ref{fig:iso-H2ip-2}.

Quite interestingly, the production of moduli through inflaton decay,
while reducing the overall amount of isocurvature fluctuation, does
not allow to find a solution to the moduli problem with arbitrarily
high [$\sim {\cal O}(M_{\rm Pl})$ in particular] vev at the end of
inflation if $H_{\rm inf}\,\simeq\,10^{13}\,$GeV. Strictly speaking,
the amount of isocurvature fluctuations produced for this value of the
Hubble constant and $\sigma_{\rm inf}\,\sim\,0.01 - 1\, M_{\rm Pl}$ is
of the order of a few percent, therefore it is not excluded by present
cosmic microwave background data. If present at this level, it could
actually be detected by the upcoming generation of instruments. One
should also emphasize that we have considered a rather conservative
case, in the sense that the modulus is produced at a comparable rate
than other light particles; if $N_\sigma$ is decreased, the amount of
isocurvature fluctuations would increase. At smaller values of $H_{\rm
  inf}$, the isocurvature constraints have disappeared, due to the
combined effect of the partial erasure associated with moduli
production in inflaton decay and a smaller initial isocurvature
fluctuation.  Finally, independently of $H_{\rm inf}$, the production
of moduli through inflaton decay significantly worsens the effect of
moduli on big-bang nucleosynthesis. The success of big-bang
nucleosynthesis now requires both $m_\sigma\,\gtrsim\,100\,{\rm TeV}$
for all $\sigma_{\rm inf}$.

\section{Summary and conclusions}
\label{sec:disc}

Let us first summarize the results obtained. We have shown that the
decay of a generic modulus tends to produce strong isocurvature
fluctuations between dark matter and radiation, or between baryons and
radiation. The amount of isocurvature fluctuations produced,
relatively to the total curvature perturbation, depends on several
parameters: the value of the initial modulus - radiation isocurvature
perturbation (in units of the total initial curvature perturbation),
and the amount of energy density stored in the modulus oscillations at
the time of decay, relatively to that contained in radiation, in
particular.  We have discussed in some detail the evolution of the
modulus energy density and of its perturbations from the end of
inflation onwards, for a variety of possible moduli effective
potentials, assuming the modulus remains light during inflation. We
have then translated the constraints derived from the analysis of
cosmic microwave background data into constraints in the modulus
parameter space $m_\sigma - \sigma_{\rm inf}$. We find that the
constraints associated with the production of isocurvature
fluctuations significantly exceed those from big-bang nucleosynthesis
in this parameter space.  One reason why the constraints obtained
cover most of the $m_\sigma - \sigma_{\rm inf}$ parameter space is
that the modulus will produce a large baryon isocurvature mode if
$\Omega_\sigma^{<_{\rm d}}\,\sim\,1$, while it will produce a dark
matter - radiation isocurvature fluctuation if $\Omega_\sigma^{<_{\rm
    d}}\,\ll\,1$ (but $\Omega_\sigma^{<_{\rm d}}$ large enough to
affect the dark matter).

Evading the constraints from big-bang nucleosynthesis and from the
generation of isocurvature fluctuations requires one of the following
conditions to be satisfied.

Firstly, if the modulus potential is time independent (\ie it does not
receive supergravity contributions at any time), it is mandatory that:
{\it (i)} $H_{\rm inf}\,\ll\,10^9\,$GeV and the modulus initially lies
very close to the minimum of its potential, within
$\sim10^{-10}\,M_{\rm Pl}$ (depending on the modulus mass, here taken
to be $\sim 1\,$TeV, see Fig.~\ref{fig:iso-quad-m2-tot}); or {\it
  (ii)} $m_\sigma\,\gtrsim\,100\,$TeV, $\sigma_{\rm
  inf}\,\gtrsim\,(H_{\rm inf}/10^{13}\,{\rm GeV})\,M_{\rm Pl}$ and $H_{\rm
  inf}\,\ll\,10^{13}\,$GeV. Solution {\it (i)} ensures that the moduli
energy density at the time of its decay is sufficiently small to
affect neither big-bang nucleosynthesis nor the dark matter
perturbations (which turn out to be more sensitive probes than the
former in this region of parameter space). The constraint on the
Hubble parameter corresponds to the requirement that the stochastic
motion of the modulus during inflation remains small enough as
compared to the bound on the final effective displacement of the
modulus vev. Such a solution might be realized in models in which the
modulus is bound to remain close to an enhanced symmetry point in
moduli space due to the friction caused by its coupling to other light
degrees of
freedom~\cite{Dine:1995uk,Dine:2000ds,Kofman:2004yc}. Solution {\it
  (ii)} is typical of particle physics models which achieve a high
modulus mass scale. The constraints on $\sigma_{\rm inf}$ and most
particularly the bound on $H_{\rm inf}$ directly come from the
constraints due to isocurvature fluctuations. If $H_{\rm
  inf}\,\ll\,10^{13}\,$GeV as required by this solution, the tensor
modes should be unobservable by upcoming cosmic microwave background
missions.

Secondly, if one assumes that the modulus potential receives
supergravity corrections after inflation (but remains light during
inflation, as realized in some models of inflation discussed in
Appendix~\ref{sec:appendix:concretemodels}), one has to require
$H_{\rm inf}\,\ll\,10^{13}\,$GeV if the minima of the effective
modulus potential at high energy ($H\,\gg\,m_\sigma$) matches that at
low energy ($H\,\ll\,m_\sigma$). Then the moduli problem can be solved
either for large $\sigma_{\rm inf}\,\sim\,M_{\rm Pl}$, provided
$m_\sigma\,\gtrsim\,100\,$TeV, or for vanishingly small $\sigma_{\rm
  inf}$ (in units of $M_{\rm Pl}$, see the main text for details) and
arbitrary $m_\sigma$. If the minima do not coincide, the situation is
very similar to that discussed for a time independent modulus
potential.

Finally, we have recalled that if the modulus is heavy during
inflation, either because $H_{\rm inf}\,\lesssim\,m_\sigma$, or
because the modulus receives an effective Hubble mass during
inflation, then the isocurvature fluctuation between the inflaton and
the modulus disappears~\cite{Yamaguchi:2005qm}. This provides a
natural solution for the moduli problem at the perturbative level, but
it does not automatically satisfies the constraints on entropy
injection at the time of big-bang nucleosynthesis. First of all,
either $m_\sigma\,\gg\,100\,$TeV, or the minima of the effective
modulus potentials during inflation, after inflation and at low energy
($H\,\ll\,m_\sigma$) match one another. In this latter case, one also
needs to ensure that the quantum jumps of the modulus in its potential
does not result in too large an energy density at the time of big-bang
nucleosynthesis. This can be done using the discussion of
Appendix~\ref{app:stoch}, which discusses the stochastic motion of the
modulus in the presence of a Hubble effective mass $c_{\rm
  i}H$. Strictly speaking the results are valid in the slow-roll
regime, which requires $c_{\rm i}<3/4$, but they should remain valid
as long as as $c_{\rm i}$ is of order unity. For chaotic inflation,
the standard deviation $\langle\delta\sigma^2\rangle^{1/2}$ is given
by the Bunch-Davies expression~(\ref{eq:varxisigma}). This leads to
the bound of Eq.~(\ref{eq:boundchh2}) which gives $\mean{\delta\sigma
  ^2}^{1/2}\simeq 3.0 \times 10^{-6}M_{\rm Pl}$ (here for $c_{\rm
  i}^2=0.5$). For small field inflation, the Bunch-Davies expression
is given by Eq.~(\ref{eq:varsfclarge}). In the case where $\mu\simeq
M_{\rm Pl}$, which is the case if $H_{\rm inf}=10^{13}\, $GeV, then
one can also use Eq.~(\ref{eq:varhcorrmuone}) and this leads to
$\mean{\delta\sigma ^2}^{1/2}\simeq 10^{-6}M_{\rm Pl}$. These values
are in conflict with the big-bang nucleosynthesis constraints if
$m_\sigma\,\sim\,1\,$TeV and the modulus potential becomes time
independent after inflation, since these constraints require
$\sigma_{\rm inf}\,<\,10^{-10}\,M_{\rm Pl}$ (see
Fig.~\ref{fig:iso-quad-m2-tot}). When turned into a bound on the
Hubble constant during inflation, this becomes $H_{\rm
  inf}\,\ll\,10^9\,$GeV, a low value indeed. Even if the modulus
potential receives supergravity corrections after inflation, there is
a potential conflict, since the big-bang nucleosynthesis constraints
impose $\sigma_{\rm inf}\,<\,10^{-6}\,M_{\rm Pl}$ for $c^2=0.5$, but
$\sigma_{\rm inf}\,<\,10^{-5}\,M_{\rm Pl}$ for $c^2=2$, see
Figs.~\ref{fig:iso-H2-1}, \ref{fig:iso-H2-3}.  

We have also considered the consequences of a possible production of
moduli through inflaton decay. Since those moduli inherit the
fluctuations of the inflaton, which correspond to those of radiation,
this tends to reduce the initial modulus - radiation isocurvature
fluctuation, all things being equal. However, it also aggravates the
effect of moduli on big-bang nucleosynthesis, and one generically
finds that all moduli must have masses $m_\sigma \,\gtrsim\,100\,$TeV
independently of $\sigma_{\rm inf}$. Furthermore, one must still
require that the effective minima of the modulus potential during
inflation, after inflation and at low energy match one another, or
$H_{\rm inf}\,\ll\,10^{13}\,$GeV as above.

\par

In short, the moduli problem is worse at the perturbative level. A
clear trend emerges from the above calculation, namely the lower the
inflationary scale, the easier it is to solve the moduli problem. A
significant amount of late time entropy production could alleviate the
moduli problem, provided baryon isocurvature perturbations are not
produced at the decay of the entropy producing component. As mentioned
above, one needs to have a reheating temperature (after entropy
production) higher than the baryogenesis scale; alternatively the
fluctuations carried by the entropy producing fluid could be similar
to those carried by the baryons. Those are certainly non trivial
constraints.

\acknowledgments{This work was supported in part by CNRS-JSPS
  bilateral project of cooperative research and JSPS Grant-in-Aid for
  Scientific Research No. 19340054 (JY).}

\appendix

\section{Expected value for $\sigma_{\rm inf}$}
\label{app:stoch}

We now analyze the stochastic (quantum) behavior of the modulus field
in the case where the total energy density is still dominated by the
vacuum energy of the inflaton field. A similar analysis was performed
in Refs.~\cite{Linde:2005yw} but focusing on the chaotic inflationary
scenario and for negligible initial values of the modulus
field. Moreover, the stochastic nature of the inflaton field was also
ignored. Here, we relax these assumptions and generalize the results
of Refs.~\cite{Linde:2005yw,Lyth:2006gd}.

The problem treated here bears close resemblance with the problem
tackled in Ref.~\cite{Martin:2004ba} where the behavior of the quantum
quintessence field during inflation was studied (the quantum behavior
of the inflaton field being also taken into account). For this reason,
we will follow a similar treatment.

\par

According to the formalism of stochastic inflation, the coarse-grained
inflaton field $\phi $ obeys the following Langevin
equation~\cite{Starobinsky:1986fx,Starobinsky:1994bd}
\begin{equation}
\label{eq:Langevin_inf}
  \frac{{\rm d}\phi}{{\rm d}t}+\frac{V_{\phi}'(\phi)}{3H(\phi)} =
  \frac{H^{3/2}(\phi)}{2\pi}\xi_\phi(t)\, ,
\end{equation}
where $\xi_\phi$ is a white-noise field such that
$\mean{\xi_\phi(t)\xi_\phi(t')}=\delta(t-t')$ and where a prime denotes
a derivative with respect to the field. The stochastic evolution of the
modulus field $\sigma$ is also controlled by a Langevin equation which,
in the slow-roll approximation, reads
\begin{equation}
\label{eq:Langevin_mod}
  \frac{{\rm d}\sigma}{{\rm d}t}+\frac{V_{\sigma}'(\sigma)}{3H(\phi)} =
  \frac{H^{3/2}(\phi)}{2\pi}\xi_\sigma(t)\, ,
\end{equation}
where $\xi_\sigma$ is another white-noise field such that
\begin{equation}
  \mean{\xi_\sigma(t)\xi_\sigma(t')}=\delta(t-t'),\quad
  \mean{\xi_\sigma(t)\xi_\phi(t')}=0\, .
\end{equation}
The solution of the Langevin equation~(\ref{eq:Langevin_mod}) depends
explicitly on $\xi_\sigma$ but also on the inflaton noise $\xi_\phi$
through the coarse-grained field $\phi$. Since the modulus is considered
as a test field, $H$ only depends on $\phi$, hence all primes
superscript mean differentiation with respect to $\phi$; obviously, all
primes superscript on $V_\phi$ (resp. $V_\sigma$) denote differentiation
with respect to $\phi$ (resp. $\sigma$).

\par

In order to find an approximate solution to
Eqs.~\eqref{eq:Langevin_inf} and~\eqref{eq:Langevin_mod}, one may try
to use the same perturbative technique as the one used in
Refs.~\cite{Martin:2005ir, Martin:2005hb}. Therefore, we expand the
inflaton and modulus fields about their classical solution and write
\begin{equation}
\phi(t)=\phi_{\rm cl}(t)+\delta \phi_1+\delta \phi_2 + \cdots \, , \quad   
\sigma(t)=\sigma_{\rm cl}(t)+\delta \sigma_1+\delta \sigma_2 + \cdots \, ,
\end{equation}
where the first terms in the expansions are linear in the noise, the
second ones are quadratic in the noise and so on. As mentioned before,
the approximation made in Ref.~\cite{Linde:2005yw} consists in ignoring
the stochastic nature of the inflaton field, $\phi=\phi_{\rm cl}$.

\par

Let us first examine the solution for the inflaton field. Expanding up
to second order in the equation of motion, we obtain two linear
differential equations for $\dphi_1$ and $\dphi_2$, see
Refs.~\cite{Martin:2005ir, Martin:2005hb}, namely
\begin{equation}
  \frac{{\rm d}\dphi_1}{{\rm d}t}+2M_{\rm Pl}^2H''(\phi _{\rm
  cl}) \dphi_1 = \frac{H^{3/2}(\phi _{\rm cl})}{2\pi}\xi_{\phi}(t)
\end{equation}
and
\begin{eqnarray}
\frac{{\rm d}\dphi_2}{{\rm d}t}+2M_{\rm Pl}^2H''(\phi _{\rm
  cl}) \dphi_2 = -M_{\rm Pl}^2H'''(\phi _{\rm cl})\dphi_1^2 
+
\frac{3}{4\pi}H^{1/2}(\phi _{\rm cl})H'(\phi _{\rm
  cl})\dphi_1\xi_{\phi}(t)\, .
\end{eqnarray}
These equations can be solved by varying the integration constant. Let
us first consider the equation for $\dphi_1$. If the initial conditions
are such that $\dphi_1(t=\tin)=0$, then the solution reads
\begin{equation}
\label{soldphi1}
  \dphi_1(t)=\frac{H'\left[\phi _{\rm
  cl}(t)\right]}{2\pi}\int_\tin^t\de{.5}{\tau}
  \frac{H^{3/2}\left[\phi _{\rm cl}(\tau )\right]}{H'\left[\phi
  _{\rm cl}(\tau )\right]}\xi_{\phi}(\tau)\, .
\end{equation}
We are now in a position where the various correlation functions can be
calculated exactly. Since $\dphi_1$ is linear in the noise $\xi$, the
mean value obviously vanishes $\mean{\dphi_1}=0$. The two-point
correlation function for $\dphi_1$ can be calculated as:
\begin{equation}
\mean{\dphi_1(t_1)\dphi_1(t_2)} = 
\frac{1}{4\pi^2}H'(t_1)H'(t_2)\int_{t_\ini}^{{\rm min}(t_1,t_2)}
{\rm d}\tau\,\frac{H^3(\tau)}{H^{\prime 2}(\tau)}\ .
\end{equation}
Once the inflaton potential has been specified, it is more convenient to
carry out these integrals using the classical value $\phi_{\rm cl}$ as a
variable instead of $\tau$, thanks to the relation derived from the classical trajectory:
\begin{equation}
{\rm d}\phi_{\rm cl}\,=\, -2 H'M_{\rm Pl}^2 \,{\rm d}t\ .
\end{equation}

For instance, the two-point correlation function calculated at the
same time, \ie the variance, reads~\cite{Martin:2005ir, Martin:2005hb}
\begin{equation}
\mean{\dphi_1^2}   = 
  \frac{H^{\prime 2}}{8\pi^2M_{\rm Pl}^2}\,
  \int^{\phi_\ini}_{\phi _{\rm cl}}\de{.7}{\varphi}
  \left(\frac{H}{H'}\right)^3\, .
\end{equation}
Detailed calculations of these integrals will be given below for various
prototypical models of inflation.

\par

We now turn to the equation of motion for the second order
perturbation $\dphi_2$. It can be solved by following exactly the
steps that were described before. Then, the solution can be written
as~\cite{Martin:2005ir, Martin:2005hb}
\begin{eqnarray}
\dphi_2(t) &=& -H'M_{\rm Pl}^2\int _\tin^t \de{.5}\tau
\frac{H'''}{H'}\delta \phi _1^2(\tau ) 
+\frac{3H'}{4\pi}\int_\tin^t\de{.5}{\tau}
H^{1/2}\dphi_1(\tau)\xi_{\phi}(\tau) \, .
\end{eqnarray}
As expected the second order perturbation is quadratic in the
noise. One can easily evaluate the mean value of $\dphi_2(t)$, taking
into account a factor $1/2$ which originates from the fact that the
Dirac $\delta$-function appearing in the noise correlation function is
centered on an integration limit, see Refs.~\cite{Martin:2005ir,
  Martin:2005hb}

\par

Let us now turn to the modulus case when it has a sufficiently flat
potential to acquire an independent quantum noise besides the
inflaton. As for the inflaton case, it is easy to establish that the
equations of motion for the perturbed quantities $\delta \sigma_1$ and
$\delta \sigma_2$ are given by the following
expressions~\cite{Martin:2004ba}
\begin{eqnarray}
\frac{{\rm d}\delta \sigma_1}{{\rm d}t}
+\frac{V_{\sigma}''(\sigma_{\rm cl})}{3H(\phi
_{\rm cl})}\delta \sigma_1 &=& \frac{V_{\sigma}'(\sigma_{\rm cl})
H'(\phi _{\rm cl})}{3H^2(\phi _{\rm cl})}\delta \phi _1
+\frac{H^{3/2}(\phi _{\rm cl})}{2\pi }\xi _\sigma\, , 
\\
\frac{{\rm d}\delta \sigma_2}{{\rm d}t}+\frac{V_{\sigma}''(\sigma_{\rm
 cl})}{3H(\phi
_{\rm cl})}\delta \sigma_2 &=& \frac{V_{\sigma}'(\sigma_{\rm cl})
H'(\phi _{\rm cl})}{3H^2(\phi _{\rm cl})}\delta \phi _2
+\frac{V_{\sigma}'(\sigma_{\rm cl})H''(\phi _{\rm
cl})}{6H^2(\phi _{\rm cl})}\delta \phi _1^2
-\frac{V_{\sigma }'(\sigma_{\rm cl})H'^2(\phi _{\rm
cl})}{3H^3(\phi _{\rm cl})}\delta \phi _1^2
\nonumber \\
& & +\frac{V_{\sigma}''(\sigma_{\rm cl})H'(\phi _{\rm
cl})}{3H^2(\phi _{\rm cl})}\delta \phi _1\delta \sigma_1
-\frac{V_{\sigma}'''(\sigma_{\rm cl})}{6H(\phi _{\rm cl})}\delta \sigma_1^2
+\frac{3H^{1/2}(\phi _{\rm cl})H'(\phi _{\rm
cl})}{4\pi }\delta \phi _1\xi _\sigma\, .\label{eq:seq}
\end{eqnarray}
Although these equations look quite complicated, they can be solved
easily because (by definition) they are linear. Assuming that the
modulus potential does not depend explicitly on time (that is, other by
its dependence on $\sigma$), the solution for $\delta \sigma_1$ reads:
\begin{equation}
\delta \sigma_1(t)=V_{\sigma}'(\sigma_{\rm cl})
\int _\tin ^t \biggl[\frac{H'(\phi _{\rm
cl})}{3H^2(\phi _{\rm cl})}\delta \phi_1 (\tau ) \\
+\frac{H^{3/2}(\phi _{\rm cl})}{2\pi V_{\sigma}'(\sigma_{\rm cl})}
\xi _\sigma(\tau )\biggr] {\rm d}\tau \, ,\label{eq:ds1}
\end{equation}
and, as required, is linear both in the quintessence noise
$\xi_{\sigma}$ and (through $\delta\phi_1$) in the inflaton noise
$\xi$. The above formula is not valid anymore if $V_\sigma$ contains an
explicit dependence on time, for instance if the modulus mass receives
Hubble term corrections.  In this particular case, which will be
discussed further below, one has to extract the explicit time dependence
out of the potential, then proceed as above. Unless otherwise said, we
assume in the following that $V_\sigma$ does not contain any such
explicit dependence on time.

\par

As is obvious, $\delta \sigma_1$ has a vanishing mean value,
$\mean{\delta \sigma_1}=0$, but a non-vanishing variance given by the
sum of two contributions originating from the inflaton and quintessence
noise variances, namely~\cite{Martin:2004ba}
\begin{eqnarray}
\label{varq11}
\mean{\delta \sigma_1^2} &=&\frac{V_{\sigma}'^2(\sigma_{\rm cl})}{9}
\int _\tin^t \int _\tin^t \frac{H'(\tau )}{H^2(\tau )}
\frac{H'(\eta )}{H^2(\eta )}
\mean{\delta \phi _1(\tau )\delta \phi _1(\eta )}
{\rm d}\tau {\rm d}\eta 
+\frac{V_{\sigma}'^2(\sigma_{\rm cl})}{4\pi ^2}
\int _\tin^t \int _\tin^t \frac{H^{3/2}(\tau )}{V_{\sigma}'(\tau )}
\frac{H^{3/2}(\eta )}{V_{\sigma}'(\eta )}
\mean{\xi _\sigma(\tau )\xi _\sigma(\eta )}
{\rm d}\tau {\rm d}\eta \nonumber \\
\\
\label{varq12}
&\equiv & 
\mean{\delta \sigma_1^2}\vert _{\delta\phi_1\delta\phi_1}+
\mean{\delta \sigma_1^2}\vert _{\xi_\sigma\xi_\sigma}\, .
\end{eqnarray}
Let us notice that there is no mixed contribution since the
cross-correlation $\mean{\delta\phi_1\xi_\sigma}=0$. Using the
correlation function of the modulus noise, the term $\mean{\delta
\sigma_1^2}\vert _{\xi_\sigma\xi_\sigma}$ can be further simplified,
namely~\cite{Martin:2004ba}
\begin{equation}
\label{eq:varnoisemod}
\mean{\delta \sigma_1^2}\vert _{\xi_\sigma\xi_\sigma}
=
\frac{V_{\sigma}'^2(\sigma_{\rm cl})}{4\pi ^2}
\int _\tin^t \frac{H^{3}(\tau )}{V_{\sigma}'{}^2(\tau )}
{\rm d}\tau \, .
\end{equation}

\par

Let us now turn to the second order correction. The solution for
$\delta \sigma_2$ can be written as~\cite{Martin:2004ba}
\begin{eqnarray}
\label{sol2}
\delta \sigma_2(t) &=& V_{\sigma}'(\sigma_{\rm cl})\int _\tin ^t 
\biggl\{\frac{H'(\phi _{\rm
cl})}{3H^2(\phi _{\rm cl})}\delta \phi_2 (\tau )
+\left[\frac{H''(\phi _{\rm
cl})}{6H^2(\phi _{\rm cl})}
-\frac{H'^2(\phi _{\rm
cl})}{3H^3(\phi _{\rm cl})}\right]\delta \phi_1^2(\tau )
+\frac{V_{\sigma}''(\sigma_{\rm cl})H'(\phi _{\rm
cl})}{3V_{\sigma}'(\sigma_{\rm cl})H^2(\phi _{\rm cl})}
\delta \phi_1(\tau )\delta \sigma_1(\tau )
\nonumber \\
& & -\frac{V_{\sigma}'''(\sigma_{\rm cl})}{6V_{\sigma}'(\sigma_{\rm cl})
H(\phi _{\rm cl})}
\delta \sigma_1^2(\tau )
+\frac{3}{4\pi }
\frac{H^{1/2}(\phi _{\rm
cl})H'(\phi _{\rm
cl})}{V_{\sigma}'(\sigma_{\rm cl})}\delta \phi _1(\tau )\xi _{\sigma}\biggr\}
{\rm d}\tau \, .
\end{eqnarray}
As expected, one sees that $\delta \sigma_2$ is quadratic in the
noises.

\par

{}From the above expression, one deduces that the mean value of $\delta
\sigma_2$ is non-vanishing and is the sum of various terms~\cite{Martin:2004ba}
\begin{equation}
\label{meanq2}
\mean{\delta \sigma_2}=
\mean{\delta \sigma_2}\vert _{\delta \phi _2} 
+\mean{\delta \sigma_2}\vert _{\delta \phi _1^2} 
+\mean{\delta \sigma_2}\vert _{\delta \phi _1\delta \sigma_1} 
+\mean{\delta \sigma_2}\vert _{\delta \sigma_1^2(\xi _\sigma)}
+\mean{\delta \sigma_2}\vert _{\delta \sigma_1^2(\xi _\phi)}\, ,
\end{equation} 
where the last term in Eq.~(\ref{sol2}) does not contribute because
$\mean{\delta \phi _1\xi _{\sigma}}=0$. Had we not taken into
account the stochastic behavior of the inflaton, only the term
$\mean{\delta \sigma_2}\vert _{\delta \sigma_1^2(\xi _\sigma)}$ would
have contributed.

\par

At this stage it is interesting to compare the previous considerations
to Ref.~\cite{Linde:2005yw}. In particular the above approach is a
perturbative one and a relevant question is its domain of
validity~\cite{Martin:2005ir, Martin:2005hb}. In fact, as shown in
Ref.~\cite{Linde:2005yw}, it turns out that, for the potential
$m_{\sigma }^2\sigma ^2/2$, the Langevin equation can be integrated
exactly. The solution reads
\begin{equation}
\sigma (t)=\sigma _\ini {\rm e}^{-\int _{t_\ini}^t m_{\sigma }^2/(3H){\rm d}\tau}
+\frac{1}{2\pi}\int _{t_\ini}^t H^{3/2}(\tau)\xi_{\sigma}(\tau)
{\rm e}^{\int _{t}^{\tau} m_{\sigma }^2/(3H){\rm d}\eta}{\rm d}\tau \, .
\end{equation}
From this expression, it is easy to compute the variance. One obtains
\begin{equation}
\mean{\sigma ^2}=\frac{1}{4\pi ^2}\int _{t_\ini}^t H^{3}(\tau)
{\rm e}^{2/3\int _{t}^{\tau} m_{\sigma }^2/H{\rm d}\eta}{\rm d}\tau 
=\frac{m_{\sigma }^4\sigma ^2(t)}{4\pi ^2}
\int _{t_\ini}^t \frac{H^3(\tau)}{m_{\sigma }^4\sigma ^2(\tau)}{\rm
d}\tau \, .
\end{equation}
This is exactly the result obtained in
Eq.~(\ref{eq:varnoisemod}). Therefore, although perturbative in nature,
the approach used before has in fact a wider domain of validity and, in
the specific case treated above, can be used even if the corrections are
not small. The perturbative approach is also interesting for two
reasons: firstly, it allows us to take into account the stochastic
behavior of the inflaton field (even if, most of the time, we will show
that these corrections are negligible). Secondly, the method of
Eq.~(\ref{eq:varnoisemod}) rests on one's ability to solve exactly the
Langevin equation which is possible only for a quadratic potential for
$\sigma$. If the potential is different (that is to say, not quadratic),
only the method used here allows us to derive explicit results.

\par

We now turn to the calculation of the various corrections presented
above in the following specific inflaton and modulus potentials.

\subsection{Chaotic $m_\phi^2\phi^2$ inflation}

If we assume that $V_{\phi}=m_{\phi}^2\phi^2/2$, then, in the slow-roll
approximation, the classical evolution of the inflaton field is given by
the following expression
\begin{equation}
\label{solclass}
\frac{\phi_{\rm cl}}{M_{\rm Pl}}=\sqrt{\left(\frac{\phi_\ini}{M_{\rm Pl}}\right)^2 -4N}\ ,
\end{equation}
where $N$ is the number of e-folds defined by $N\equiv \ln
\left(a/a_\ini\right)$, $a_\ini$ being the initial value of the scale
factor at the beginning of inflation, and $\phi_{\rm
  cl}(N=0)=\phi_\ini$. The model remains under control only if the
energy density is below the Planck energy density. This amounts to the
following constraint on the initial conditions $\phi_\ini/M_{\rm Pl}
\lta 8\pi \sqrt{2}M_{\rm Pl}/m_{\phi}$. Inflation stops when the
slow-roll parameter $\varepsilon _1=-\dot{H}/H^2$ is equal to unity
corresponding to $\phi _{\rm end}=\sqrt{2}M_{\rm Pl}$. As a
consequence, one can easily check that the argument of the square root
in Eq.~(\ref{solclass}) remains always positive. The total number of
e-folds during inflation is simply given by $N_{_{\rm
    T}}=(\phi_\ini/M_{\rm Pl})^2/4-1/2$. This number can be huge if
the initial energy density of the inflaton field is close to the
Planck energy density. Finally, the inflaton mass is fixed by the WMAP
normalization
\begin{equation}
\frac{Q^2_{\rm rms-PS}}{T^2}=\frac{1}{480\pi
 ^2\epsilon_{1*}}\frac{H_*^2}{M_{\rm Pl}^2}=\frac{1}{1440\pi
 ^2\epsilon_{1*}}
\frac{V_{\phi*}}{M_{\rm Pl}^4}\, ,
\end{equation}
where the cosmic microwave background quadrupole is given by $Q_{\rm
  rms-PS}/T\simeq 6\times 10^{-6}$ and a star denotes the time at
which the scales of astrophysical interest today crossed the Hubble
radius during inflation. Using the fact that $\epsilon
_{1*}=1/(2N_*+1)$, where $N_*$ is the number of e-folds between the
time at which the physical scales left the Hubble radius during
inflation and the end of inflation, one obtains
\begin{equation}
\frac{m_{\phi}}{M_{\rm Pl}}\simeq \frac{12\pi \sqrt{10}}{2N_*+1}
\frac{Q_{\rm rms-PS}}{T}\simeq 7\times 10^{-6}\, ,
\end{equation}
where we have used $N_*\simeq 50$. This also implies the Hubble
constant at the end of inflation, $H_{\rm
  inf}\,=\,m_\phi/\sqrt{3}\,\simeq\,10^{13}\,$GeV.

\par

Using the perturbative presented before, one can solve
Eq.~(\ref{eq:Langevin_inf}) (let us notice that the Langevin equation
can be solved exactly only in the case of a quartic potential) and
determine the quantum behavior of the inflaton field. Through
straightforward albeit lengthy calculations, one
obtains~\cite{Martin:2005ir, Martin:2005hb}
\begin{equation}
\mean{\delta\phi_1^2} \,=\,
- \frac{m_\phi^2}{192\pi^2 M_{\rm Pl}^4}\,\left(\phi^4_\cl-\phi_\ini^4\right)\ .
\label{eq:correl-phi-m2}
\end{equation}
Note that the classical trajectory obeys $\phi\,<\,\phi_\ini$ during
inflation, so that the above is positive as it should be. Similarly, the
correction to the mean value reads~\cite{Martin:2005ir, Martin:2005hb}
\begin{equation}
\mean{\dphi_2}\,=\,
-\frac{m_\phi^2}{192\pi^2M_{\rm Pl}^4}
\left(\phi^3_\cl-\phi_\ini^3\right)\, .
\label{meanphi}
\end{equation}
As before, since $\phi\,<\,\phi_\ini$ during inflation, the quantity
$\mean{\dphi_2}$ is positive.

\par

Concerning the modulus, we will consider two possible potentials, in
the spirit of previous sections: one with fixed mass
$V_\sigma=m_{\sigma}^2\sigma^2/2$, and one with typical supergravity
corrections.

\subsubsection{Modulus potential: $V_\sigma=m_{\sigma}^2\sigma^2/2$}

Let us first determine the classical trajectory of the modulus during
inflation. Solving Eq.~(\ref{eq:Langevin_mod}) without the noise term
leads to the following solution
\begin{equation}
\sigma _{\rm cl}(N)\,=\,\sigma _\ini\left[\frac{\phi_{\cl}(N)}
{\phi _\ini}\right]^{m_{\sigma }^2/m_{\phi}^2}\, .
\end{equation}
In practice, one has $m_{\phi }\gg m_{\sigma }$ and, therefore, the
modulus evolves slowly during inflation. Let us also notice that
$m_{\sigma }^2/m_{\phi }^2=2/3 (m_{\sigma }/H_{\rm inf, \ini})^2N_{_{\rm
T}}$ and, therefore, the limit $m_\sigma/m_\phi\rightarrow 0$
corresponds to $m_{\sigma}\ll H_{\rm inf,\ini}N_{_{\rm T}}^{-1/2}$. Even
though the modulus is completely frozen in the limit
$m_\sigma/m_\phi\rightarrow 0$, it is necessary to take into account its
evolution when computing the integrals in Eqs.~(\ref{varq12}),
(\ref{sol2}), since its displacement can be non negligible if inflation
lasts long enough. One then finds:
\begin{equation}
\label{eq:varchao1}
\mean{\delta\sigma_1^2}_{\xi_\sigma\xi_\sigma}\,=\,
-\frac{m_\phi^2}{192\pi^2M_{\rm Pl}^4}\frac{1}
{1-m_\sigma^2/\left(2m_\phi^2\right)}\left(\phi^4_\cl - \phi_\ini^{4-2m_\sigma^2/m_\phi^2}
\phi^{2m_\sigma^2/m_\phi^2}_\cl\right)
\ ,
\end{equation}
which, in the limit $m_\sigma/m_\phi\rightarrow 0$ and
$\phi_\cl\,\ll\,\phi_\ini$ reduces to:
\begin{equation}
\label{eq:varchao2}
\mean{\delta\sigma_1^2}_{\xi_\sigma\xi_\sigma}\,\simeq\,
\frac{m_\phi^2\phi_\ini^4}{192\pi^2M_{\rm Pl}^4}=\frac{H_\ini^2}{8\pi
^2}N_{_{\rm T}}\ .
\end{equation}
Note that $H_{\ini}$ refers to the Hubble constant at the onset of
inflation, which differs (by $\sqrt{2N_{_{\rm T}}}$) from the Hubble
constant at the end of inflation, noted $H_{\rm inf}$ and used in the
rest of our analysis.  The two formulas~(\ref{eq:varchao1})
and~(\ref{eq:varchao2}) are identical to Eqs.~(23) and (24) of
Ref.~\cite{Linde:2005yw}. The interpretation of
Eq.~(\ref{eq:varchao2}) is as follows. If the field is light, then the
classical drift in the Langevin equation can be ignored. Then, it is
easy to show that the Langevin equation can be integrated exactly and,
as a consequence, that $\mean{\sigma ^2}=\int H^2{\rm d}N/(4\pi
^2)$. If the Hubble parameter is approximatively constant, then the
previous expression reduces to $\mean{\sigma ^2}=H^2N_{_{\rm T}}/(4\pi
^2)$. However, as well known in the case of chaotic inflation, the
Hubble parameter can change significantly during inflation, Therefore,
the integral has to be evaluated exactly. When this is done, this
produces the additional factor $1/2$ present in
Eq.~(\ref{eq:varchao2}).

\par

Let us now turn to the second contribution originating from the inflaton
noise. It reads:
\begin{equation}
\mean{\delta\sigma_1^2}_{\delta\phi_1\delta\phi_1}\,=\,
-\frac{m_\sigma^2\sigma^2_\cl}{576\pi^2M_{\rm Pl}^4}\frac{m_\sigma^2}{m_\phi^2}\frac{
\left(\phi_\cl-\phi_\ini\right)^3\left(\phi_\cl+3\phi_\ini\right)}{\phi^2_\cl}
\simeq \frac{m_\phi^2\phi_\ini^4}{192\pi^2M_{\rm Pl}^4}\frac12\frac{m_{\sigma
}^4}{m_{\phi}^4}\frac{\sigma ^2_\cl}{M_{\rm Pl}^2}\ ,
\end{equation}
where the last expression is valid at the end of inflation. As a
consequence, we note that
$\mean{\delta\sigma_1^2}_{\delta\phi_1\delta\phi_1}$ is negligible in
comparison with $\mean{\delta\sigma_1^2}_{\xi_\sigma\xi_\sigma}$ since
$m_{\phi }\gg m_{\sigma}$ and $\sigma _\cl/M_{\rm Pl}\ll 1$.

\par

Let us now calculate the correction to the mean value of the modulus
field. The various contributions $\delta\sigma_2$ amount to:
\begin{eqnarray}
\mean{\delta \sigma_2}\vert _{\delta \phi _2} &\,=\,&
\frac{ m_{\sigma }^2\sigma_\cl}{192\pi^2M_{\rm Pl}^4}
\left(\frac{1}{2}\phi^2_\cl-\frac{3}{2}\phi_\ini^2+\frac{\phi_\ini^3}
{\phi_\cl}\right)\, ,\nonumber \\
\mean{\delta \sigma_2}\vert _{\delta \phi _1^2} &\,=\,& 
-\frac{m_\sigma^2\sigma_\cl}{192\pi^2M_{\rm Pl}^4}\,\left(\frac{1}{2}\phi^2_\cl + 
\frac{1}{2}\frac{\phi_\ini^4}{\phi^2_\cl} - \phi_\ini^2\right)\, ,\nonumber\\
\mean{\delta \sigma_2}\vert _{\delta \phi _1\delta \sigma_1} 
&\,=\,& -\frac{m_\sigma^2\sigma_\cl}{192\pi^2M_{\rm Pl}^4}\frac{m_\sigma^2}
{m_\phi^2}\,\left(\frac{1}{6}\phi^2_\cl + \frac{4}{3}\frac{\phi_\ini^3}{\phi_\cl} - 
\frac{1}{2}\frac{\phi_\ini^4}{\phi^2_\cl} 
- \phi_\ini^2\right)\, ,\nonumber\\
\mean{\delta \sigma_2}\vert _{\delta \sigma_1^2(\xi _\sigma)}&=& 0\, ,
\end{eqnarray}
the last result being obtained because $V_{\sigma }'''=0$ in our
case. Using again the fact that, as inflation proceeds, $\phi _\cl\ll
\phi_\ini$, one finally obtains
\begin{equation}
\mean{\delta \sigma_2}\simeq -\frac{m_{\sigma }^2\sigma_\cl\phi_\ini^4}
{384\pi^2M_{\rm Pl}^4\phi^2_\cl}\, .
\end{equation}
Noticing that $\mean{\delta \sigma_2}$ is maximal at the end of
inflation, \ie for $\phi_\cl=\sqrt{2}M_{\rm Pl}$, one can easily
demonstrate that the first order correction is always the dominant
one.

\par

Therefore, using the value of the inflaton mass obtained from the WMAP
normalization, the expression of the initial value of the inflaton
field in terms of the total number of e-folds, and setting $N_{_{\rm
    T}}\,\gtrsim\,60$, one obtains the following lower bound:
\begin{equation}
\mean{\delta \sigma_1^2}^{1/2}\,\gtrsim\, 3.9 \times
 10^{-5}\, M_{\rm Pl}.\label{eq:boundchm2}
\end{equation}
The above considerations are valid provided one does not enter the
regime of eternal inflation where the perturbative treatment of the
quantum behavior of the inflaton field breaks down. Eternal inflation
starts if $\phi_\ini \gta (24)^{1/4}(m_{\phi}/M_{\rm
  Pl})^{-1/2}$. Therefore, the above calculations are applicable if
$N_{_{\rm T}}\lta 2.8\times 10^6$ which implies that
$\sqrt{\mean{\delta \sigma_1^2}}/M_{\rm Pl}\lta 1.8$. As was studied
in this paper, such values of $\sigma $ are anyway excluded since the
amount of entropy perturbations is too large to be compatible with the
cosmic microwave background data.
 
\par

Finally, it is also interesting to investigate what happens if one
considers a more complicated potential for the modulus. As mentioned
before, the perturbative approach used here allows us to determine the
stochastic behavior of $\sigma $ even if the potential is not
quadratic. Therefore, let us consider the case where
\begin{equation}
V_{\sigma }(\sigma)=\frac{\lambda _n}{n!}\frac{\sigma ^n}{M_{\rm
 Pl}^{n-4}}\, ,
\end{equation}
where $\lambda _n $ is a dimensionless constant. The integration of the
classical equation of motion leads to the following solution
\begin{equation}
\frac{\sigma _{\rm cl}}{M_{\rm Pl}}=
\left[\left(\frac{\sigma _{\ini}}{M_{\rm Pl}}\right)^{2-n}
+(2-n)\frac{\lambda _n}{(n-1)!}\frac{M_{\rm Pl}^2}{m_{\phi}^2}
\ln \left(\frac{\phi_\cl}{\phi _\ini}\right)
\right]^{1/(2-n)}\, .
\end{equation}
In the limit $\lambda _n\rightarrow 0$, the modulus field is almost
frozen. Then, one can now compute the variance due to the modulus
noise. According to Eq.~(\ref{eq:varnoisemod}), it reads
\begin{eqnarray}
\mean{\delta \sigma_1^2}\vert _{\xi_\sigma\xi_\sigma}
&=&\frac{H_\ini^2}{8\pi ^2}N_{_{\rm T}}\left(
\frac{\sigma_{\rm cl}}{M_{\rm Pl}}\right)^{2-2n}
4^{(3n-4)/(2-n)}\frac{4(n-1)!m_{\phi}^2}{\lambda _n(2-n)M_{\rm Pl}^2}
\exp\left[-\frac{4(n-1)!m_{\phi}^2}{\lambda _n(2-n)M_{\rm Pl}^2}
\left(\frac{\sigma_\ini}{M_{\rm Pl}}\right)^{2-n}\right]\nonumber \\ 
& & \times
\left\{\gamma \left[\frac{4-3n}{2-n}, 
-4\left(\frac{\sigma_\ini}{M_{\rm Pl}}\right)^{2-n}\right]
-\gamma \left[\frac{4-3n}{2-n}, 
-4\left(\frac{\sigma_{\rm cl}}{M_{\rm Pl}}\right)^{2-n}\right]
\right\}\, ,
\end{eqnarray}
where $\gamma (\alpha, x)\equiv \int _0^x t^{\alpha -1}{\rm e}^{-t}{\rm
d}t$ is the incomplete gamma function.

\subsubsection{Modulus potential: $V_\sigma=c^2_{\rm i}H^2\sigma^2/2$}

In this subsection, one considers the case where the modulus mass
receives supergravity corrections of the form $c^2_{\rm i}H^2$,
assuming $c_{\rm i}<3/4$, during inflation. A comment is in order at
this point. One uses the notation $c_{\rm i}$ in order to emphasize
the fact that the supergravity corrections to the modulus potential
are not necessarily the same during inflation and during the
post-inflationary epoch. As a consequence, one expects $c_{\rm i}\neq
c$. This is of course the same for the minimum of the potential which,
during inflation, is not necessarily equal to $\sigma _0$, the
minimum of the potential in the post-inflationary epoch. In the rest
of this appendix, we work in terms of the field displacement (with
respect to the inflationary minimum) rather than in terms of the field
itself.

\par

Then, if one assumes that the supergravity corrections dominate, then
the modulus classical motion reads:
\begin{equation}
\sigma_{\rm cl}\,=\,\sigma_\ini\,\exp\left[-\frac{c^2_{\rm i}}{3}
\int_{t_\ini}^t{\rm d}\tau\, H(\tau)\right]=
\exp\left[\frac{c^2_{\rm i}}{3M_{\rm Pl}^2}
\int_{\phi_\ini}^{\phi_\cl}{\rm d}\varphi\, \frac{V(\varphi)}{V'(\varphi)}\right]
\ .
\end{equation}
and for $m_\phi^2\phi^2$ inflation, this gives:
\begin{equation}
\sigma_{\rm cl}\,=\,\sigma_\ini\,\exp\left[\frac{c^2_{\rm i}}{12M_{\rm Pl}^2}
\left(\phi^2_\cl-\phi_\ini^2\right)\right]\ .
\end{equation}
Very quickly the argument of the exponential becomes $-c^2_{\rm
  i}\phi_\ini^2/(12 M_{\rm Pl}^2)\simeq -c_{\rm i}^2N_{_{\rm
    T}}/3\simeq -(m_{\sigma }/H_{\rm inf,\ini})^2N_{_{\rm T}}/3$ and,
therefore, as previously, one recovers that the massless condition is
given by $m_{\sigma }\ll H_{\rm inf,\ini}N_{_{\rm T}}^{-1/2}$. Since
$N_{_{\rm T}}>60$, this condition is now always violated (at least
provided that $c_{\rm i}\gta 0.22$).

\par

As mentioned above, Eqs.~(\ref{eq:ds1}), (\ref{sol2}) need to be
corrected to account for the explicit time dependence of the
potential, which enters through the $H$ prefactor. Making this
dependence more explicit in Eq.~(\ref{eq:seq}), one solves these
equations as
\begin{equation}
\delta\sigma_1\,=\,\frac{c^2_{\rm i}}{3}\sigma_{\rm cl}\,\int_{t_\ini}^t\,{\rm d}\tau H'(\tau)
\delta\phi_1(\tau)\,+\,\int_{t_\ini}^t\,{\rm d}\tau\,
\frac{H^{3/2}(\tau)}{2\pi}
\exp\left[\frac{c^2_{\rm i}}{3}\int_{t}^{\tau}\,{\rm d}\eta \,H(\eta)\right]
\xi_\sigma\left(\tau\right)\ ,
\end{equation}
and hence
\begin{equation}
\label{eq:varxisigma}
\mean{\delta\sigma_1^2}_{\xi_\sigma\xi_\sigma}\,=\,\frac{m_\phi^2}
{16\pi^2M_{\rm Pl}^2c^2_{\rm i}}\left[\frac{6M_{\rm Pl}^2}{c^2_{\rm i}} + \phi^2_\cl -
\left(\frac{6M_{\rm Pl}^2}{c^2_{\rm i}} + 
\phi_\ini^2\right){\rm e}^{-\frac{c^2_{\rm i}}{6M_{\rm
Pl}^2}\left(\phi_\ini^2-\phi^2_\cl\right)}\right]
\simeq \frac{3H^4}{8\pi ^2c^2_{\rm i}H^2}\left(1+\frac{6M_{\rm Pl}^2}
{c^2_{\rm i}\phi ^2_\cl}\right)\ .
\end{equation}
Therefore, one obtains the Bunch-Davies result (with the mass
$c^2_{\rm i}H^2$) corrected by a factor the value of which at the end
of inflation is $1+3/c^2_{\rm i}$. This result agrees with that
obtained in Ref.~\cite{Linde:2005yw}. In the limit $c_{\rm
  i}\rightarrow 0$, one of course recovers the result
Eq.~(\ref{eq:varchao2}).

\par

The contribution to the variance due to the inflaton noise is given by
\begin{equation}
\mean{\delta\sigma_1^2}_{\delta\phi_1\delta\phi_1}\,
=\,-\frac{c^4_{\rm i}\sigma^2_\cl m_\phi^2}{3456\pi^2M_{\rm Pl}^8}\left(
\frac{1}{30}\phi^6_\cl - \frac{1}{2}\phi^2_\cl\phi_\ini^4 
+ \frac{4}{5}\phi_\cl\phi_\ini^5 - \frac{1}{3}\phi_\ini^6\right)\ .
\end{equation}
This term is negligible compared to the previous one since the vev of
the modulus is small (in Planck units).

\par

Let us now turn to the corrections $\delta\sigma_2$ to the mean
value. As explained before, one needs to modify the results above to
take into account the fact that the mass is explicitly
time-dependent. Straightforward calculations lead to
\begin{eqnarray}
\mean{\delta\sigma_2}_{\delta\phi_2}&\,=\,& \frac{c^2_{\rm i}}{3}\sigma_\cl\int_{t_\ini}^{t}\,
{\rm d}\tau\,H'(\tau)\mean{\delta\phi_2(\tau)}\ ,\nonumber\\
\mean{\delta\sigma_2}_{\delta\phi_1^2}&\,=\,& \frac{c^2_{\rm i}}{3}\sigma_\cl\int_{t_\ini}^{t}\,
{\rm d}\tau\,\left[\frac{H''(\tau)}{2}-\frac{H^{\prime 2}(\tau)}{H(\tau)}\right]\,
\mean{\delta\phi_1^2}\,\nonumber\\
\mean{\delta\sigma_2}_{\delta\phi_1\delta\sigma_1}&\,=\,&\frac{c^2_{\rm i}}{3}\int_{t_\ini}^{t}\,
H'(\tau)\,{\rm d}\tau\,{\rm e}^{\frac{c^2_{\rm i}}{3}\,\int_{t}^{\tau}\,{\rm d}\eta\,H(\eta)}\,\,
\mean{\delta\phi_1(\tau)\delta\sigma_1(\tau)}\ ,\nonumber\\
\mean{\delta\sigma_2}_{\delta\phi_1\xi_\sigma}&\,=\,&0\ .
\end{eqnarray}
This can be integrated to give:
\begin{eqnarray}
\mean{\delta\sigma_2}_{\delta\phi_2}&\,=\,& \frac{c^2_{\rm i}\sigma _\cl m_\phi^2}{1152\pi^2
M_{\rm Pl}^6}\left(\frac{1}{4}\phi^4_\cl-\phi_\cl \phi_\ini^3+\frac{3}{4}\phi_\ini^4\right)
\ ,\nonumber\\
\mean{\delta\sigma_2}_{\delta\phi_1^2}&\,=\,& -\frac{c^2_{\rm i}\sigma _\cl m_\phi^2}{1152\pi^2
M_{\rm Pl}^6}\left[\frac{1}{4}\phi^4_\cl-\phi_\ini^4\log\left(\frac{\phi_\cl }{\phi_\ini}\right)
-\frac{1}{4}\phi_\ini^4\right]
\,\nonumber\\
\mean{\delta\sigma_2}_{\delta\phi_1\delta\sigma_1}&\,=\,& 
-\frac{c^4_{\rm i}\sigma _\cl m_\phi^2}{6912\pi^2M_{\rm Pl}^8}
\left(\frac{1}{30}\phi^6_\cl-\frac{1}{2}\phi^2_\cl\phi_\ini^4
+\frac{4}{5}\phi_\cl \phi_\ini^5-\frac{1}{3}\phi_\ini^6\right)
\ ,\nonumber\\
\mean{\delta\sigma_2}_{\delta\phi_1\xi_\sigma}&\,=\,&0\ .
\end{eqnarray}
Given that $\sigma _\cl /M_{\rm Pl}\ll 1$, it is easy to see that
these contributions are subdominant. Therefore, the main contribution
is the one given by Eq.~(\ref{eq:varxisigma}). Expressed at the end of
inflation ($\phi_{\rm end}=\sqrt{2}M_{\rm Pl}$) and normalized to the
cosmic microwave background, this expression leads to the following
constraint on the value of the modulus:
\begin{equation}
  \mean{\delta \sigma_1^2}^{1/2}\,\gtrsim\, 0.79 \times
  10^{-6}\frac{\sqrt{3+c^2_{\rm i}}}{c^2_{\rm i}}\, M_{\rm Pl}\, .\label{eq:boundchh2}
\end{equation}

\subsection{Small field and hybrid inflation}

In this section, we turn to another type of inflationary model. We now
consider a potential of the form:
\begin{equation}
V_\phi\,=\,M^4\left[1+ \epsilon\left(\frac{\phi}{\mu}\right)^p\right]\ ,
\end{equation}
with $\epsilon=\pm1$. Such a potential gives rise to small field
inflation if $\epsilon=-1$ or hybrid inflation if $\epsilon=+1$.
Since all integrals cannot be carried out exactly in this case, we
provide the results to leading order in $\phi/\mu$. Out of simplicity,
we use the notation $\Phi\,\equiv\,\phi_\cl/\mu$ and similarly for
$\Phi_\ini$.
 
\par

Let us now discuss small field inflation in more details. For this
model, the slow-roll trajectory is only known implicitly. It can be
expressed as 
\begin{equation}
N=\frac{1}{2p}\frac{\mu ^2}{M_{\rm Pl}^2}\left(\Phi_\ini^2-\Phi ^2
+\frac{2}{p-2}\Phi_\ini^{2-p}
-\frac{2}{p-2}\Phi ^{2-p}
\right)\, .
\end{equation}
If $p=2$ the singular terms must be replaced by a logarithm. From the
above formula, one deduces that, given that $\phi_\ini\ll \phi_{\rm
end}$, the total number of e-folds can be written as
\begin{equation}
N_{_{\rm T}}\simeq \frac{\mu^2}{M_{\rm Pl}^2}\frac{1}{p(p-2)}
\Phi_\ini^{2-p}\, .
\end{equation}
In this class of models, the end of inflation occurs by violation of the
slow-roll conditions. If $\mu/M_{\rm Pl}\ll 1$, it happens at $\Phi_{\rm
end}\simeq [2\mu ^2/(p^2M_{\rm Pl}^2)]^{1/(2p-2)}$. In this regime, the
two first slow-roll parameters are given by
\begin{equation}
\label{eq:srsftwo}
\epsilon _{1*}\simeq \exp\left[-4N_*\left(\frac{M_{\rm Pl}}{\mu}
\right)^2\right]\, ,\quad \epsilon _{2*}=4\left(\frac{M_{\rm
Pl}}{\mu}\right)^2\, ,
\end{equation}
for $p=2$, while, for $p\neq 2$, one has
\begin{equation}
\label{eq:srsfnottwo}
\epsilon _{1*}\simeq \frac{p^2}{2}\left(\frac{M_{\rm
Pl}}{\mu}\right)^2\left[N_*p(p-2)\left(\frac{M_{\rm
Pl}}{\mu}\right)^2\right]^{-2(p-1)/(p-2)}\, ,\quad \epsilon
_{2*}=\frac{2}{N_*}\frac{p-1}{p-2}\, .
\end{equation}
If $\mu/M_{\rm Pl}\gta 1$, then the above formulas are no longer valid
and the slow-roll parameters must be evaluated numerically. In
particular, in the limit $\mu/M_{\rm Pl}\gg 1$, one has $\phi_{\rm
end}\simeq \mu$.

\par

Our next step is to deduce an expression of the mass scale $M$. One
obtains
\begin{equation} 
\left(\frac{M}{M_{\rm Pl}}\right)^4=720\pi ^2p^2\frac{Q^2_{\rm
 rms-PS}}{T^2}
\left[p(p-2)N_*\right]^{-2(p-1)/(p-2)}\left(\frac{\mu}{M_{\rm Pl}}
\right)^{2p/(p-2)}\, ,
\end{equation}
where we have assumed $p\neq 2$. This expression is valid only if
$\mu/M_{\rm Pl}\ll 1$. If $\mu/M_{\rm Pl}\gta 1$, one can show, see
Ref.~\cite{Martin:2006rs}, that the end of inflation occurs when the
vacuum expectation value of the inflaton field is of the order of $\mu
$ and, as a consequence, that $\left(M/M_{\rm Pl}\right)^4\sim {\cal
  O}\left(Q_{\rm rms-PS}^2/T^2\right)$. Let us also notice that,
contrary to the case of chaotic inflation, the Hubble parameter is
approximatively constant during inflation and given by $H_{\rm
  inf}\simeq M^2/(\sqrt{3}M_{\rm Pl})$. Therefore, if
$\mu\,\gtrsim\,M_{\rm Pl}$, $H_{\rm inf}\,\simeq\,10^{13}\,$GeV, but
$H_{\rm inf}$ can be much lower if $\mu\,\ll\,M_{\rm Pl}$.

\par

Let us now discuss the constraints on the free parameters $p$ and
$\mu$. As shown in Ref.~\cite{Martin:2006rs}, there is no prior
independent constraint on the index $p$. However, if one adopts the
theoretical prejudice that the vacuum expectation value of the
inflaton field must be smaller than the Planck mass, then $\mu $ must
be smaller than $M_{\rm Pl}$ and then one can demonstrate that the
case $p=2$ is slightly disfavored by the cosmic microwave background
data. In small field inflation, the energy scale of inflation can be
very low. But, it can not be smaller than, say, the TeV scale,
$V_*/\mpl^4\gta 10^{-64}$. If $p=2$, this implies that $\mu/M_{\rm
  Pl}\gta 1.25$. In this case, the formulas~(\ref{eq:srsftwo}) are no
longer valid but one can show that the spectral index is still
compatible with the data in the regime $\mu /M_{\rm Pl}\gta 1$ or even
$\mu /M_{\rm Pl}\gg 1$. If $p\neq 2$, the parameter $\mu$ is globally
unconstrained. The constraint mentioned before implies
\begin{equation}
\frac{\mu}{M_{\rm Pl}}\gta \left\{\frac{4\times 10^{-64}}{45p^2}
\left(\frac{Q_{\rm rms-PS}}{T}\right)^{-2}
\left[p\left(p-2\right)N_*\right]^{2(p-1)/(p-2)}\right\}^{(p-2)/(2p)}\, .
\end{equation}
This leads to $\mu /M_{\rm Pl}\gta 0.16$ for $p=2.1$, $\mu /M_{\rm
  Pl}\gta 3.4\times 10^{-5}$ for $p=2.5$, $\mu /M_{\rm Pl}\gta
1.5\times 10^{-8}$ for $p=3$, $\mu /M_{\rm Pl}\gta 9.9\times 10^{-13}$
for $p=4$ etc \dots. In this case, one sees that $\mu/M_{\rm Pl}$ can
be small. For $p\neq 2$, Eqs.~(\ref{eq:srsfnottwo}) indicate that
$\epsilon _{2*}$ no longer depends on $\mu$. The spectral index
remains compatible with the data in this regime because $\epsilon
_{2*}\sim 0.04 (p-1)/(p-2)$ still lies in the $2\sigma $ contour
whatever the value of $p$. As already mentioned, if, on the contrary,
$\mu/M_{\rm Pl}\gg 1$, Eqs.~(\ref{eq:srsfnottwo}) are no longer valid
but one can also show that the model is still in agreement with the
data. The conclusion is that, provided the energy scale of inflation
is above the TeV scale, the two regimes $\mu/M_{\rm Pl}\ll 1$ and
$\mu/M_{\rm Pl}\gta 1$ are still compatible with the data and, hence,
the parameter $\mu$ remains basically
unconstrained~\cite{Martin:2006rs}.

\par

Let us now briefly discuss hybrid inflation. A crucial difference with
small inflation is that inflation no longer stops by violation of the
slow-roll conditions but by instability. This means that there is one
more additional parameter namely the value of the inflaton field at
which the instability occurs. This makes the analysis more complicated
since the parameter space is enlarged. For this reason, although we give
all the necessary expressions, we have chosen in this paper to skip a
detailed investigation of this case. 

\par

Straightforward calculations lead to the following results for the
variance and the mean value of the inflaton field
\begin{equation}
\mean{\delta\phi_1^2}\,\simeq\,-\frac{M^4\mu^2}{12\pi^2M_{\rm Pl}^4}
\frac{1}{\epsilon p(4-3p)}\,\left(\Phi^{2-p} - \Phi_\ini^{4-3p}\Phi^{2p-2}\right)\ .
\end{equation}
Similarly,
\begin{equation}
\mean{\delta\phi_2}\,=\,\frac{M^4\mu}{24\pi^2M_{\rm Pl}^4}\frac{1}{\epsilon p(4-3p)}\left[
\frac{p-2}{2}\Phi^{1-p} + \frac{4-3p}{2}\Phi^{p-1}\Phi_\ini^{2-2p}
+(p-1)\Phi^{2p-3}\Phi_\ini^{4-3p}\right]\ .
\end{equation}
These results agree with those of Ref.~\cite{Martin:2005ir} to leading order.

\par

As regards to the modulus, we again consider two possible
potentials. We consider these two possibilities in the following.

\subsubsection{Modulus potential: $V_\sigma=m_{\sigma}^2\sigma^2/2$}

The classical trajectory of the modulus reads:
\begin{equation}
\label{eq:trajecsf}
\sigma_\cl\,=\,\sigma_\ini\exp\left[\frac{\mu^2m_\sigma^2}{\epsilon p (2-p)M^4}
\left(\Phi^{2-p}-\Phi_\ini^{2-p}\right)\right]\ .
\end{equation}
Let us notice that the argument of the exponential is always
negative. We now discuss the case of small field inflation. The
argument of the exponential in Eq.~(\ref{eq:trajecsf}) is always
dominated by the following term
\begin{equation}
\label{eq:argexp}
\frac{\mu^2m_\sigma^2}{p (p-2)M^4}\Phi _\ini^{2-p}=\frac13 
\left(\frac{m_{\sigma}}{H_{\rm inf}}\right)^2N_{_{\rm T}}\, .
\end{equation}
The argument of the exponential can be small or large depending on the
total number of e-folds during inflation and on whether the modulus
field is light or heavy. Using the WMAP normalization, one can also
express the argument of the exponential in terms of the parameters of
the model. In the regime $\mu/M_{\rm Pl}\ll 1$, one obtains
\begin{equation}
\frac{\mu^2m_\sigma^2}{p (p-2)M^4}\Phi _\ini^{2-p}=\frac13 
\left(\frac{m_{\sigma}}{H_{\rm inf}}\right)^2N_{_{\rm T}}
\simeq 6.6\times
 10^{-21}N_{_{\rm T}}\left(\frac{m_{\sigma}}{10^5\mbox{GeV}}\right)^2
\frac{\left[p(p-2)N_*\right]^{2(p-1)/(p-2)}}{p^2}
\left(\frac{\mu}{M_{\rm
       Pl}}\right)^{-2p/(p-2)}\, .
\end{equation}
Let us give a few examples for different values of $p$ with
$m_{\sigma}=10^5\mbox{GeV}$ and $N_*=50$. For $p=2.1$, this gives $43.78
N_{_{\rm T}}(\mu/M_{\rm Pl})^{-42}$, for $p=2.5$, one has $ 6.29 \times
10^{-11}N_{_{\rm T}}(\mu/M_{\rm Pl})^{-10}$, for $p=3$, one obtains $
3.71\times 10^{-13}N_{_{\rm T}}(\mu/M_{\rm Pl})^{-6}$ and for $p=4$,
this leads to $ 2.64 \times 10^{-14}N_{_{\rm T}}(\mu/M_{\rm Pl})^{-4}$.
Let us be more precise and give some numbers. If $p=2.5$ one can easily
check that the argument of the exponential is always large. But this
conclusion can be modified if one considers other values of the
parameters. For instance, if $p=3$ and $1.5\times 10^{-8}\lta \mu/M_{\rm
Pl}\lta 8.4\times 10^{-3}N_{_{\rm T}}^{1/6}$, then the argument of the
exponential is large but, if $8.4\times 10^{-3}N_{_{\rm T}}^{1/6}\lta
\mu/M_{\rm Pl}\ll 1$, it is no longer the case. Let us also notice that
the previous considerations are valid only for values of $N_{_{\rm T}}$
such that the number $8.4\times 10^{-3}N_{_{\rm T}}^{1/6}$ remains small
otherwise the formulas used here would not be valid. This means
$N_{_{\rm T}}\ll 1.84 \times 10^{6}$ which is not so restrictive. The
same analysis is true for $p=4$, the corresponding intervals being
$9.9\times 10^{-13}\lta \mu/M_{\rm Pl}\lta 4\times 10^{-4}N_{_{\rm
T}}^{1/4}$ and $4\times 10^{-4}N_{_{\rm T}}^{1/4}\lta \mu/M_{\rm Pl}\ll
1$. In the present case, the total number of e-folds must satisfy
$N_{_{\rm T}}\ll 3.9 \times 10^{9}$. On the other hand, in the regime
where $\mu/M_{\rm Pl}\gta 1$, the argument of the exponential can now be
written as
\begin{equation}
\frac{\mu^2m_\sigma^2}{p (p-2)M^4}\Phi _\ini^{2-p}
\simeq 1.87 \times
 10^{-18}\left(\frac{m_{\sigma}}{10^5\mbox{GeV}}\right)^2N_{_{\rm T}}\, .
\end{equation}
and is small, unless we choose very large values of $N_{_{\rm T}}$.

\par

We are now in a position where one can compute the variance of the
stochastic motion using Eq.~(\ref{varq12}). This leads to the following
result
\begin{equation}
\mean{\delta\sigma_1^2}_{\xi_\sigma\xi_\sigma}\,=\,\frac{M^8}{24\pi^2M_{\rm Pl}^4m_\sigma^2}
\left\{1-\exp\left[\frac{2m_\sigma^2\mu^2}{\epsilon p (2-p)M^4}\left(\Phi^{2-p}-
\Phi_\ini^{2-p}\right)\right]\right\}\ ,
\end{equation}
Let us discuss this result. If the argument of the exponential in
Eq.~(\ref{eq:trajecsf}) is large (in absolute value), then the
exponential becomes negligible and
\begin{equation}
\label{eq:varianceBD}
\mean{\delta\sigma_1^2}_{\xi_\sigma\xi_\sigma}\,\simeq \,\frac{M^8}{24\pi^2M_{\rm Pl}^4m_\sigma^2}
=\frac{3H_{\rm inf}^4}{8\pi ^2m_{\sigma}^2}\, ,
\end{equation}
where we used $H_{\rm inf}^2\simeq M^4/(3M_{\rm Pl}^2)$. It was suggested in
Ref.~\cite{Linde:2005yw} that this formula describes the small field case. But, as
was noticed before, the argument of the exponential in
Eq.~(\ref{eq:trajecsf}) can also be small and, in this case, one has
\begin{equation}
\label{eq:variancenonBD}
\mean{\delta\sigma_1^2}_{\xi_\sigma\xi_\sigma}\,\simeq
 \,\frac{M^4\mu^2}{12\pi ^2p(2-p)M_{\rm
 Pl}^4}\left(\Phi^{2-p}_{\rm end}-\Phi_\ini^{2-p}\right)\simeq
 \frac{H_{\rm inf}^2}{4\pi ^2}N_{_{\rm T}}\, .
\end{equation}
We notice that this last expression is similar to
Eq.~(\ref{eq:varchao2}). The only difference is that we do not have
the presence of an additional factor $1/2$ with respect to the
standard case. This is because, in the case of small field inflation,
the Hubble parameter is almost constant and, hence, one obtains the de
Sitter result. We conclude that if $m_{\sigma }\lta H_{\rm
  inf}/\sqrt{N_{_{\rm T}}}$, then the variance is given by
Eq.~(\ref{eq:variancenonBD}) while if $H_{\rm inf}/\sqrt{N_{_{\rm
      T}}}\lta m_{\sigma}\lta H_{\rm inf}$, it is given by the
Bunch-Davis expression~(\ref{eq:varianceBD}). Finally, if
$m_{\sigma}>H_{\rm inf}$, then the variance becomes negligible.

\par

As was done for the case of chaotic inflation, one can also express the
variance in terms of the parameters of the model. Let us start with the
case $\mu\ll M_{\rm Pl}$. In the regime where the Bunch-Davis term
dominates, the variance can be written as
\begin{equation}
\frac{\mean{\delta\sigma_1^2}_{\xi_\sigma\xi_\sigma}}{M_{\rm Pl}^2}\,\simeq \,1.63 \times 10^{11}p^4
\left[p(p-2)N_*\right]^{-4(p-1)/(p-2)}\left(\frac{m_{\sigma}}{10^5\mbox{GeV}}\right)^{-2}
\left(\frac{\mu}{M_{\rm Pl}}\right)^{4p/(p-2)}\, .
\end{equation}
For $p=2.5$ and $m_{\sigma}=10^5\mbox{GeV}$, this gives
$\mean{\delta\sigma_1^2}_{\xi_\sigma\xi_\sigma}/M_{\rm Pl}^2\simeq
1.79\times 10^{-9}(\mu/M_{\rm Pl})^{20}$ which implies that $7.6\times
10^{-99}\lta \mean{\delta\sigma_1^2}_{\xi_\sigma\xi_\sigma}/M_{\rm
Pl}^2\lta 1.79\times 10^{-29}$. On the other hand, if $p=3$ and
$m_{\sigma}=10^5\mbox{GeV}$, this leads to
$\mean{\delta\sigma_1^2}_{\xi_\sigma\xi_\sigma}/M_{\rm Pl}^2\simeq
5.15\times 10^{-5}(\mu/M_{\rm Pl})^{12}$ which means that $6.7\times
10^{-99}\lta \mean{\delta\sigma_1^2}_{\xi_\sigma\xi_\sigma}/M_{\rm
Pl}^2\lta 6.35\times 10^{-30}N_{_{\rm T}}^2$. Finally, if $p=4$ and
$m_{\sigma}=10^5\mbox{GeV}$, this means that
$\mean{\delta\sigma_1^2}_{\xi_\sigma\xi_\sigma}/M_{\rm Pl}^2\simeq
0.01(\mu/M_{\rm Pl})^{8}$ which implies that $9.2\times 10^{-99}\lta
\mean{\delta\sigma_1^2}_{\xi_\sigma\xi_\sigma}/M_{\rm Pl}^2\lta
6.5\times 10^{-30}N_{_{\rm T}}^2$. We see that we always find a very
small contribution. It is easy to see that it originates from the fact
that the energy scale of inflation is very small and that we always have
$M^4/M_{\rm Pl}^4\ll m_{\sigma }/M_{\rm Pl}$. Let us now consider the
other regime, given by Eq.~(\ref{eq:variancenonBD}) and where the
Bunch-Davis term is subdominant. In this case, the variance can be
expressed as
\begin{equation}
\frac{\mean{\delta\sigma_1^2}_{\xi_\sigma\xi_\sigma}}{M_{\rm
 Pl}^2}\,\simeq \,2.16 \times 10^{-9}
N_{_{\rm T}}p^2
\left[p(p-2)N_*\right]^{-2(p-1)/(p-2)}
\left(\frac{\mu}{M_{\rm Pl}}\right)^{2p/(p-2)}\, .
\end{equation}
Therefore, if $m_{\sigma}=10^5\mbox{GeV}$ and $p=3$, one obtains
$\mean{\delta\sigma_1^2}_{\xi_\sigma\xi_\sigma}/M_{\rm Pl}^2\simeq
3.84\times 10^{-17}N_{_{\rm T}}(\mu/M_{\rm Pl})^{6}$. This gives the
following range of values: $1.35\times 10^{-29}N_{_{\rm T}}^2\lta
\mean{\delta\sigma_1^2}_{\xi_\sigma\xi_\sigma}/M_{\rm Pl}^2\lta
3.8\times 10^{-23}N_{_{\rm T}}$. This leads to a significant constraint
only if the number of e-folds is large (remembering that, for $p=3$, the
formulas applies only if $N_{_{\rm T}}\ll 1.84 \times 10^{6}$, see
above). Taking this upper limit, one arrives at
$\sqrt{\mean{\delta\sigma_1^2}_{\xi_\sigma\xi_\sigma}}/M_{\rm Pl}\simeq
8.3 \times 10^{-9}$ which, besides being an extreme case, remains small
in comparison to what was found in the case of large field models.  If
$p=4$, one has $\mean{\delta\sigma_1^2}_{\xi_\sigma\xi_\sigma}/M_{\rm
Pl}^2\simeq 5.4\times 10^{-16}N_{_{\rm T}}(\mu/M_{\rm Pl})^{4}$ and the
same conclusion can be reached. However, the extreme case evoked before
becomes more significant as one obtains
$\sqrt{\mean{\delta\sigma_1^2}_{\xi_\sigma\xi_\sigma}}/M_{\rm Pl}\simeq
1.4 \times 10^{-5}$.

\par

Let us also consider the regime where $\mu/M_{\rm Pl}\gta 1$ or
$\mu/M_{\rm Pl}\gg 1$. As mentioned above, in this situation, the
argument of the exponential in Eq.~(\ref{eq:trajecsf}) is always small
and $H_{\rm inf}\,\sim\,10^{13}\,$GeV. This means that one is always
in the case where the Bunch-Davies term is subdominant. Then, given
that $N_{_{\rm T}}>60$, one obtains the robust lower limit:
\begin{equation}
\mean{\delta\sigma_1^2}^{1/2}_{\xi_\sigma\xi_\sigma}\,\gtrsim\,
4.2 \times 10^{-6}\, M_{\rm Pl}\ .
\end{equation}

\par

Let us now turn to the other contribution to the variance originating
from the fact that the inflaton noise. It can be expressed as
\begin{eqnarray}
\mean{\delta\sigma_1^2}_{\delta\phi_1\delta\phi_1}\,&=&\,
-\frac{\mu^4m_\sigma^4\sigma^2_\cl}{24\pi^2M_{\rm Pl}^4M^4}\frac{1}{\epsilon p (4-3p)}\Biggl[
\frac{1}{(4-2p)(4-p)}\Phi^{4-p} + \frac{4-3p}{p^2(4-2p)}\Phi^{p}\Phi_\ini^{4-2p}
-\frac{1}{2p^2}\Phi^{2p}\Phi_\ini^{4-3p} \nonumber \\ & & 
-\frac{4-3p}{2p^2(4-p)}\Phi_\ini^{4-p}\Biggr]\ .
\end{eqnarray}
The analysis is complicated by the fact that the dominant term depends
on whether $p<4$ or $p>4$. Since, previously, we have mainly considered
situations where $p<4$, we will restrict ourselves to this case. In
addition, very large values of $p$ appears rather unnatural form a high
energy physics point of view. Therefore, in the case of small field
inflation, the dominant term can be written as
\begin{equation}
\mean{\delta\sigma_1^2}_{\delta\phi_1\delta\phi_1}\,\simeq \,
\frac{\mu^4m_\sigma^4\sigma^2_\cl}{24\pi^2M_{\rm
Pl}^4M^4}\frac{1}{p(4-3p)(4-2p)(4-p)}\Phi^{4-p}_{\rm end} \ .
\end{equation}
In the regime where $\mu/M_{\rm Pl}\ll 1$, the term can can be
re-expressed as 
\begin{eqnarray}
\frac{\mean{\delta\sigma_1^2}_{\delta\phi_1\delta\phi_1}}{M_{\rm Pl}^2}\,&\simeq &\,
4.7\times
10^{-50}\frac{\left[p(p-2)N_*\right]^{2(p-1)/(p-2)}}{p^3(4-3p)(4-2p)(4-p)}
\left(\frac{2}{p^2}\right)^{(4-p)/(2p-2)}\left(\frac{m_{\sigma}}{10^5\mbox{GeV}}\right)^4
\left(\frac{\sigma _\cl}{M_{\rm Pl}}\right)^2\nonumber \\ & & \times \left(\frac{\mu}{M_{\rm
 Pl}}\right)^{p(p-4)/[(p-2)(p-1)]}\, .
\end{eqnarray}
If $p=2.5$ and $m_{\sigma }=10^5\mbox{GeV}$, then one obtains
$\mean{\delta\sigma_1^2}_{\delta\phi_1\delta\phi_1}/M_{\rm Pl}^2\simeq
1.9\times 10^{-41}(\sigma_\cl/M_{\rm Pl})^2(\mu/M_{\rm Pl})^{-5}$. But we
have seen before that, if $p=2.5$, the argument of the exponential in
Eq.~(\ref{eq:trajecsf}) is always large (and negative). This means that
the modulus is exponentially killed during inflation and, hence,
negligible at the end of inflation. In this case, the contribution
$\mean{\delta\sigma_1^2}_{\delta\phi_1\delta\phi_1}/M_{\rm Pl}^2$ is
also negligible. If $p=3$, the story is slightly different. Indeed, for
$m_{\sigma }=10^5\mbox{GeV}$, one obtains
$\mean{\delta\sigma_1^2}_{\delta\phi_1\delta\phi_1}/M_{\rm Pl}^2\simeq
6\times 10^{-44}(\sigma_\cl/M_{\rm Pl})^2(\mu/M_{\rm Pl})^{-3/2}$. If
$1.5\times 10^{-8}<\mu/M_{\rm Pl}< 8.4 \times 10^{-8}N_{_{\rm
T}}^{1/6}$, the argument of the exponential in Eq.~(\ref{eq:trajecsf})
is large and the same conclusion as before applies. But when $8.4 \times
10^{-8}N_{_{\rm T}}^{1/6}<\mu/M_{\rm Pl}\ll 1$, the argument is small
and the modulus is almost frozen, $\sigma _\cl\simeq \sigma _\ini$. In this
case, this implies that $1.9\times 10^{-42}(\sigma_\ini/M_{\rm
Pl})^2\lta \mean{\delta\sigma_1^2}_{\delta\phi_1\delta\phi_1}/M_{\rm
Pl}^2 \lta 7.8\times 10^{-41}N_{_{\rm T}}^{1/4}(\sigma_\ini/M_{\rm
Pl})^2$. We see that this contribution remains very small.

\par

Let us now consider the case where $\mu /M_{\rm Pl}>1$. As already
mentioned, this means that $\phi_{\rm end}\simeq \mu$. In addition, we
have shown before that, in this regime, the modulus is almost frozen
during inflation. As a consequence, the variance due to the inflaton
noise can be expressed as
\begin{equation}
\frac{\mean{\delta\sigma_1^2}_{\delta\phi_1\delta\phi_1}}{M_{\rm Pl}^2}\,\simeq \,
3.3\times
10^{-66}\frac{1}{p(4-3p)(4-2p)(4-p)}
\left(\frac{m_{\sigma}}{10^5\mbox{GeV}}\right)^4
\left(\frac{\sigma _\ini}{M_{\rm Pl}}\right)^2\left(\frac{\mu}{M_{\rm
 Pl}}\right)^{4}\, .
\end{equation}
We conclude that this contribution is negligible unless one takes very
large and unrealistic values of $\mu$.

\par

After having estimated the variance, one can now calculate the
correction to the mean value. The non-zero contributions to
$\mean{\delta\sigma_2}$ read:
\begin{eqnarray}
\mean{\delta \sigma_2}\vert _{\delta \phi _2} &\,=\,&
\frac{\mu^2m_\sigma^2\sigma_\cl}{96\pi^2M_{\rm Pl}^4}\frac{1}{\epsilon p (4-3p)}\left[
\Phi^{2-p} -\frac{4-3p}{p}\Phi^p\Phi_\ini^{2-2p} 
-\Phi^{2p-2}\Phi_\ini^{4-3p} +\frac{4-3p}{p}\Phi_\ini^{2-p}\right]
\, ,\\
\mean{\delta \sigma_2}\vert _{\delta \phi _1^2} &\,=\,& 
\frac{\mu^2m_\sigma^2\sigma_\cl}{48\pi^2M_{\rm Pl}^4}\frac{1}{\epsilon p (4-3p)}\left[
\frac{p-1}{2-p}\Phi^{2-p} - \frac{1}{2}\Phi^{2p-2}\Phi_\ini^{4-3p} + 
\frac{4-3p}{2(2-p)}\Phi_\ini^{2-p}\right]
\, ,\\
\mean{\delta \sigma_2}\vert _{\delta \phi _1\delta \sigma_1}&\,=\,&
-\frac{\mu^4m_\sigma^4\sigma_\cl}{48\pi^2M_{\rm Pl}^4M^4}\frac{1}{\epsilon p (4-3p)}\Biggl[
\frac{1}{(4-2p)(4-p)}\Phi^{4-p} + \frac{4-3p}{p^2(4-2p)}\Phi^p\Phi_\ini^{4-2p} - 
\frac{1}{2p^2}\Phi^{2p}\Phi_\ini^{4-3p} \nonumber \\ & & 
- \frac{4-3p}{2p^2(4-p)}\Phi_\ini^{4-p}\Biggr]\, .
 \end{eqnarray}
 We notice that the corrections to the mean value are of the same
 order of magnitude as
 $\mean{\delta\sigma_1^2}^{1/2}_{\delta\phi_1\delta\phi_1}$. Therefore,
 following the above analysis, one can safely conclude that these
 contributions are negligible.

\subsubsection{Modulus potential: $V_\sigma=c^2_{\rm i}H^2\sigma^2/2$}

If the modulus mass is dominated by a Hubble term contribution from
supergravity effects, one obtains the following results for the
classical trajectory and the quantum corrections:
\begin{equation}
\label{eq:trajecsfc}
\sigma_\cl\,=\,\sigma_\ini\exp\left[\frac{c^2_{\rm i}\mu^2}{3M_{\rm Pl}^2}\frac{1}{\epsilon p (2-p)}
\left(\Phi^{2-p}-\Phi_\ini^{2-p}\right)\right]\ .
\end{equation}
At this stage, one can reproduce the discussion of the previous section
and study the argument of the exponential. It is easy to show that it
can be expressed as 
\begin{equation}
-\frac{c^2_{\rm i}\mu ^2}{3M_{\rm Pl}^2p(p-2)}\Phi_\ini^{2-p}\,\simeq\,
-\frac{c^2_{\rm i}}{3}N_{_{\rm T}}\, .
\end{equation}
We see that the above formula is nothing but Eq.~(\ref{eq:trajecsf})
with the time dependent mass $c_{\rm i}H$. The argument of the
exponential can be large or small depending on the parameter $c_{\rm
  i}$ and the total number of e-folds. But, as already mentioned in
the section devoted to chaotic inflation, if $c_{\rm i}\gta 0.22 $,
then it is always greater than one given the fact that $N_{_{\rm
    T}}>60$.

\par

The variance of the first order correction is made of the combination of
the following two terms, as above. The term due to the modulus noise
reads
\begin{equation}
\mean{\delta\sigma_1^2}_{\xi_\sigma\xi_\sigma}\,=\,
\frac{M^4}{8\pi^2M_{\rm Pl}^2c^2_{\rm i}}\,\left\{1-\exp\left[
\frac{2c^2_{\rm i}\mu^2}{3M_{\rm Pl}^2}\frac{1}{\epsilon p (2-p)}\left(\Phi^{2-p}-
\Phi_\ini^{2-p}\right)\right]\right\}\ .
\end{equation}
If the argument of the exponential in Eq.~(\ref{eq:trajecsfc}) is large,
then one has
\begin{equation}
\label{eq:varsfclarge}
\mean{\delta\sigma_1^2}_{\xi_\sigma\xi_\sigma}\,\simeq \,
\frac{M^4}{8\pi^2M_{\rm Pl}^2c^2_{\rm i}}=\frac{3H_{\rm inf}^4}{8\pi ^2
\left(c_{\rm i}H_{\rm inf}\right)^2}\, ,
\end{equation}
and one recovers the Bunch-Davis term with a  mass
$cH_{\rm inf}$. On the contrary, if the argument of the exponential in
Eq.~(\ref{eq:trajecsfc}) is small, then the expression of the variance
reads
\begin{equation}
\mean{\delta\sigma_1^2}_{\xi_\sigma\xi_\sigma}\,\simeq \,
\frac{M^4N_{_{\rm T}}}{12\pi^2M_{\rm Pl}^2}=\frac{H_{\rm inf}^2}{4\pi ^2}N_{_{\rm T}}\, ,
\end{equation}
and one recovers the de Sitter result. In particular, the term
$c^2_{\rm i}$ has cancelled out, as, in the corresponding situation,
the modulus mass $m_{\sigma}$ cancelled out in the previous section.

\par

As before, one can also express the above results directly in terms of
the relevant parameters. Let us start with the situation where the
argument of the exponential is large. In the regime $\mu/M_{\rm Pl}\ll
1$, this gives
\begin{equation}
\label{eq:varhcorrmusmall}
\frac{\mean{\delta\sigma_1^2}_{\xi_\sigma\xi_\sigma}}{M_{\rm
 Pl}^2}\,\simeq \,
3.24\times
10^{-9}\frac{p^2}{c^2_{\rm i}}\left[p(p-2)N_*\right]^{-2(p-1)/(p-2)}
\left(\frac{\mu}{M_{\rm Pl}}\right)^{2p/(p-2)}\, .
\end{equation}
If $p=2.5$, this gives
 $\mean{\delta\sigma_1^2}_{\xi_\sigma\xi_\sigma}/M_{\rm Pl}^2\simeq 3.39
 \times 10^{-19}c^{-2}_{\rm i}\left(\mu/M_{\rm Pl}\right)^{10}$ and if $p=3$,
 one has $\mean{\delta\sigma_1^2}_{\xi_\sigma\xi_\sigma}/M_{\rm
 Pl}^2\simeq 5.76 \times 10^{-17}c^{-2}_{\rm i}\left(\mu/M_{\rm
 Pl}\right)^{6}$. Given that the above estimates are valid in the regime
 where $\mu/M_{\rm Pl}$ is small, we conclude that the variance is
 always very small unless $c_{\rm i}$ takes tiny and unrealistic values. 

\par

In the regime $\mu/M_{\rm Pl}\gtrsim 1$, one obtains:
\begin{equation}
\label{eq:varhcorrmuone}
\mean{\delta\sigma_1^2}^{1/2}_{\xi_\sigma\xi_\sigma}\,\simeq \,
6.8\times 10^{-7}\frac{M_{\rm Pl}}{c_{\rm i}^2}\, .
\end{equation}
Contrary to the previous case, the above formula indicates that the
quantum effects can now be significant, especially when the parameter
$c_{\rm i}$ is small ($c_{\rm i}<0.22$).

\par

If the argument of the exponential in Eq.~(\ref{eq:trajecsfc}) is
small, then it is easy to see that the variance is now equal to
$2N_{_{\rm T}}c^2_{\rm i}/3$ times the variance given in
Eq.~(\ref{eq:varsfclarge}). In the regime $\mu/M_{\rm Pl}\ll 1$, the
factor $N_{_{\rm T}}c^2_{\rm i}$ is unlikely to compensate the smallness of
the variance obtained in Eq.~(\ref{eq:varsfclarge}) (unless the total
number of e-folds is huge), and we conclude that the quantum effects
can become arbitrarily small in this regime.

\par

On the other hand, if $\mu/M_{\rm Pl}\gta 1$, then one obtains
$\mean{\delta\sigma_1^2}_{\xi_\sigma\xi_\sigma}/M_{\rm Pl}^2\simeq
3\times 10^{-13}N_{_{\rm T}}$ and, given the fact that the total
number of e-folds must be larger than $60$, one obtains the following
lower bound
\begin{equation}
  \mean{\delta\sigma_1^2}^{1/2}_{\xi_\sigma\xi_\sigma}\,\gtrsim\, 4.2
\times 10^{-6}\, M_{\rm Pl}\, .
\end{equation}

\par

Let us now consider the second contribution to the variance due to the
inflaton noise. Its expression can be written as
\begin{eqnarray}
\mean{\delta\sigma_1^2}_{\delta\phi_1\delta\phi_1}\,&=&\,
-\frac{c^4_{\rm i}\sigma^2_\cl M^4\mu^4}{216\pi^2M_{\rm Pl}^8}\frac{1}{\epsilon p (4-3p)}
\Biggl[\frac{1}{(4-2p)(4-p)}\Phi^{4-p} + \frac{4-3p}{2p^2(2-p)}\Phi^p\Phi_\ini^{4-2p}
-\frac{1}{2p^2}\Phi^{2p}\Phi_\ini^{4-3p} \nonumber \\
& &
- \frac{4-3p}{2p^2(4-p)}\Phi_\ini^{4-p}\Biggr]\ .
\end{eqnarray}
As discussed in the previous section in a similar context, the amplitude
of this term depends on whether $p<4$ or $p>4$. Here we restrict
ourselves to the case $p<4$. Moreover, if the modulus significantly
evolves during inflation, then the above contribution becomes
negligible. So we assume that $c^2_{\rm i}N_{_{\rm T}}/3\ll 1$ which implies
that $c _{\rm i}$ is small. In this situation, if $\mu/M_{\rm Pl}\ll 1$, one has
\begin{eqnarray}
\mean{\delta\sigma_1^2}_{\delta\phi_1\delta\phi_1}\,&=&\,
\frac{c^4_{\rm i}\sigma^2_\cl M^4\mu^4}{216\pi^2M_{\rm Pl}^8}\frac{1}{p (4-3p)(4-2p)(4-p)}
\frac{1}{(4-2p)(4-p)}\Phi^{4-p}_{\rm end} \nonumber \\
&\simeq &
1.2 \times
10^{-10}\frac{pc^4_{\rm i}}{(4-3p)(4-2p)(4-p)}\left(\frac{2}{p^2}\right)^{(4-p)/(2p-2)}
\left[p(p-2)N_*\right]^{-2(p-1)/(p-2)}\left(\frac{\sigma_\ini}{M_{\rm
 Pl}}\right)^2
\nonumber \\ & & \times
\left(\frac{\mu}{M_{\rm Pl}}\right)^{(5p^2-8)/[(p-2)(p-1)]}M_{\rm Pl}^2\, .
\end{eqnarray}
This term is obviously tiny, in particular because the power of the term
$\mu/M_{\rm Pl}$ is positive. Hence, the contribution due to the
inflaton noise can be neglected.

\par

The case where $\mu/M_{\rm Pl}\gta 1$ remains to be
studied. Straightforward considerations lead to
\begin{eqnarray}
\frac{\mean{\delta\sigma_1^2}_{\delta\phi_1\delta\phi_1}}{M_{\rm
 Pl}^2}\,&\simeq &\,
1.68 \times 10^{-14}
\frac{c^4_{\rm i}}{p (4-3p)(4-2p)(4-p)}
\left(\frac{\sigma_\ini}{M_{\rm
 Pl}}\right)^2\left(\frac{\mu}{M_{\rm Pl}}\right)^{4}\, .
\end{eqnarray}
This contribution is small unless $\mu/M_{\rm Pl}\gg 1$. 

\par

Finally, the (non-zero) second order corrections to the mean value of
the modulus vacuum expectation value read:
\begin{eqnarray}
\mean{\delta \sigma_2}\vert _{\delta \phi _2} &\,=\,&
\frac{c^2_{\rm i}\sigma_\cl \mu^2 M^4}{288\pi^2M_{\rm Pl}^6}\frac{1}{\epsilon p(4-3p)}\left[
\Phi^{2-p} - \frac{4-3p}{p}\Phi^p\Phi_\ini^{2-2p} - \Phi^{2p-2}\Phi_\ini^{4-3p}
+\frac{4-3p}{p}\Phi_\ini^{2-p}\right]
\, ,\nonumber \\
\mean{\delta \sigma_2}\vert _{\delta \phi _1^2} &\,=\,& 
\frac{c^2_{\rm i}\sigma_\cl \mu^2 M^4}{288\pi^2M_{\rm Pl}^6}\frac{1}{\epsilon p(4-3p)}\left[
\frac{2(p-1)}{2-p}\Phi^{2-p}  - \Phi^{2p-2}\Phi_\ini^{4-3p} + 
\frac{4-3p}{2-p}\Phi_\ini^{2-p}\right]
\, ,\nonumber\\
\mean{\delta \sigma_2}\vert _{\delta \phi _1\delta \sigma_1}&\,=\,&
-\frac{c^4_{\rm i}\sigma_\cl \mu^4 M^4}{432\pi^2M_{\rm Pl}^8}\frac{1}{\epsilon p(4-3p)}\Biggl[
\frac{1}{(4-2p)(4-p)}\Phi^{4-p} + \frac{4-3p}{2p^2(2-p)}\Phi^p\Phi_\ini^{4-2p} - 
\frac{1}{2p^2}\Phi^{2p}\Phi_\ini^{4-3p} \nonumber \\
& & - \frac{4-3p}{2p^2(4-p)}\Phi_\ini^{4-p}\Biggr]
\, .\nonumber\\
& & 
\end{eqnarray}
It is clear that the order of magnitude of those terms is similar to
$\mean{\delta\sigma_1^2}^{1/2}_{\delta\phi_1\delta\phi_1}$. Hence
one concludes that the corrections to the mean value can be safely
neglected, at least when $\mu/M_{\rm Pl}$ is not too large.

\section{Details of the solutions to the equations of motion in
  presence of Hubble scale mass corrections}
\label{app:sol}

In this appendix, we study in details the solutions of the modulus
equation of motion in the post-inflationary epoch in the case where
the potential receives supergravity corrections.

\subsection{Case $+c^2H^2/2$}
\label{subsec:pluscase}

The solution to the equation of motion has been obtained by Lyth \&
Stewart~\cite{Lyth:1995ka}. The field equation can indeed be put in the
following form:
\begin{equation}
(m_\sigma t)^2\tilde{\sigma}'' + (m_\sigma t)\tilde{\sigma}' + \left[
    (m_\sigma t)^2 + p^2c^2 - \frac{(3p-1)^2}{4}\right]\tilde{\sigma}\,=\,
p^2c^2\sigma_0 (m_\sigma t)^{(3p-1)/2}\ ,
\end{equation} 
with the following definition:
\begin{equation}
\tilde{\sigma}\,\equiv\, (m_\sigma t)^{(3p-1)/2}\sigma\ .
\end{equation}
A prime denotes derivative with respect to $m_{\sigma}t$, while $p$
characterizes the global equation of state of the Universe: $H=p/t$,
with $p=2/3$ for matter domination or $p=1/2$ for radiation
domination. The above field equations takes the form of a Lommel
equation. Out of convenience, we write the general solution as the sum
of a homogeneous solution with constants $\alpha_1$ and $\alpha_2$,
and a particular solution given in terms of a hypergeometric function:
\begin{eqnarray}
\sigma(t)&\,=\,& \alpha_1 (m_\sigma t)^{-(\mu+1)} \left[J_\nu(m_\sigma
  t) + J_{-\nu}(m_\sigma t)\right] +
 \alpha_2 (m_\sigma t)^{-(\mu+1)}\left[J_\nu (m_\sigma t) 
- J_{-\nu}(m_\sigma t)\right]
+ \nonumber\\
&&\,+\sigma_0\,\, {}_1F_2\left[1\,;\,\,\frac{\mu-\nu+3}{2}
\,;\,\,\frac{\mu+\nu+3}{2}\,;\,\,
-\frac{(m_\sigma t)^2}{4}\right]\ ,\label{eq:solLommelapp}
\end{eqnarray}
where we have defined
\begin{equation}
\mu\,=\,\frac{3p-3}{2}\, ,\quad\nu^2\,=\,(\mu+1)^2 - p^2c^2\ .
\end{equation}
The initial conditions are defined at $t=p/H_{\rm inf}$ as follows:
$\sigma\,=\,\sigma_{\rm inf}$ and $\sigma'\,\simeq\,0$. As a
consequence, the coefficients $\alpha_1$ and $\alpha_2$ can then be
expressed as:
\begin{eqnarray}
\label{eq:al1}
\alpha_1&\,=\,&\left(\sigma_{\rm
  inf}-\sigma_0\right)\left(\frac{p\,m_\sigma}{H_{\rm
    inf}}\right)^{\mu+1}\left[
  \frac{\Gamma(1+\nu)}{2^{2-\nu}}\left(1+\frac{\mu+1}{\nu}\right)
\left(\frac{p\,m_\sigma}{H_{\rm
      inf}}\right)^{-\nu} +
  \frac{\Gamma(1-\nu)}{2^{2+\nu}}\left(1-\frac{\mu+1}{\nu}\right)
  \left(\frac{p\,m_\sigma}{H_{\rm
      inf}}\right)^{\nu}\right]\,\\ 
\label{eq:al2}
\alpha_2&\,
=\,&\left(\sigma_{\rm
  inf}-\sigma_0\right)\left(\frac{p\,m_\sigma}{H_{\rm
    inf}}\right)^{\mu+1}\left[\frac{\Gamma(1+\nu)}{2^{2-\nu}}
  \left(1+\frac{\mu+1}{\nu}\right)\left(\frac{p\,m_\sigma}{H_{\rm
      inf}}\right)^{-\nu} -
  \frac{\Gamma(1-\nu)}{2^{2+\nu}}\left(1-\frac{\mu+1}{\nu}\right)
  \left(\frac{p\,m_\sigma}{H_{\rm inf}}\right)^{\nu}\right]\ .
\end{eqnarray}
Obviously, if $\sigma_{\rm inf}=\sigma_0$, meaning that the vev of
$\sigma $ at the end of inflation corresponds with the high energy
minimum, then $\alpha_1=\alpha_2=0$. The solution then reduces to the
particular solution.

\par

We now study the behavior of the solutions according to the values of
the parameters. Let us start with the situation where $\nu $ is real,
namely $c < (\mu+1)/p$. At
early times, $m_\sigma t\,\ll \,1$ (hence $H\,\gg \,m_\sigma$) and at
late times, $m_\sigma t\,\gg\,1$ (hence $H\,\ll \,m_\sigma$), the
particular and homogeneous solutions evolve differently. At early
times, the particular solution remains constant, 
\begin{equation}
  \sigma _{\rm part}(t)\simeq \sigma _0\, , \quad m_{\rm \sigma}t\ll 1\, .
\end{equation}
Therefore, before the onset of oscillations, the particular solution
does not undergo any redshift. On the contrary, at late times, one
obtains
\begin{equation}
\label{eq:partlate} 
\sigma_{\rm part}(t)\,\simeq\,\sigma_0 \left(m_{\sigma }t\right)^{-\mu
 -3/2}\frac{2^{\mu+3/2}}{\sqrt{\pi}}
\Gamma\left({\nu\over2}+{\mu+3\over2}\right)\Gamma
\left(-{\nu\over2}+{\mu+3\over2}\right)\sin\left(m_{\sigma }t
+\frac{\pi}{4}-\frac{\mu +1}{2}\pi\right)\, , \quad m_{\sigma}t\gg 1\, ,
\end{equation}
and, in this regime, one has that $\sigma_{\rm part}(t)\propto
a^{-3/2}$ regardless of $p$. Since, in the limit $m_\sigma t\,\gg
\,1$, the supergravity corrections are by definition negligible, one
can write $\rho _{\sigma }\simeq m_{\sigma }^2\sigma _{\rm part,
  osci}^2(a/a_{\rm osci})^{-3}/2$. From the above formula, one deduces
that the quantity $\sigma_{\rm part,osci}$ can be expressed as
\begin{equation}
\sigma_{\rm part,osci}\simeq \frac{2^{\mu+3/2}}{\sqrt{\pi}}
\Gamma\left({\nu\over2}+{\mu+3\over2}\right)\Gamma
\left(-{\nu\over2}+{\mu+3\over2}\right)\sigma _0\, .
\end{equation}
Let us notice that the modulus energy density at $m_{\sigma }t\sim 1$
(or $H\simeq m_{\sigma}$) is not given by $m_{\sigma }^2\sigma _{\rm
  part, osci}^2/2$ only. Indeed, at $m_{\sigma }t\sim 1$, the term
$c^2H^2$ can not be neglected and participate to the energy density at
the onset of oscillations. Therefore, strictly speaking, $m_{\sigma
}^2\sigma _{\rm part, osci}^2/2$ only represents the contribution of
the term $m_{\sigma }^2\sigma ^2/2$ at $H=m_{\sigma}$. However what
really matters is the modulus energy density at the time of dark
matter freeze out. In this regime, the supergravity corrections are
negligible. As a consequence, one only needs to take into account the
contribution due to the term $m_{\sigma}^2\sigma ^2/2$ and to express
it in terms of its value at the onset of oscillations. Then the
contribution of the particular solution to the energy density
contained in the modulus is given by
\begin{equation}
\Omega_{\sigma,\rm part, osci}\,\simeq \, \frac{2^{2\mu+2}}{3\pi}
\Gamma^2\left({\nu\over2}+{\mu+3\over2}\right)\Gamma^2
\left(-{\nu\over2}+{\mu+3\over2}\right)
\left({\sigma_0\over M_{\rm Pl}}\right)^2\ .\label{eq:ooscpart1}
\end{equation}
Note that this result differs from that obtained for the purely
quadratic potential: in the latter case, $\Omega_{\sigma,\rm osci}$ is
controlled by the value of $\sigma$ at the end of inflation, while
here, it is controlled by the value of the local minimum $\sigma_0$
generated by supergravity corrections. 

\par

Another issue concerns the values of the parameters $\mu $ and $\nu $
that should be inserted in the above equation. If $T_{\rm rh}>T_{\rm
  osci}$, then one should use $\mu_{\rm RD}$ and $\nu_{\rm RD}$, where
quantities indexed with ``$_{\rm RD}$'' (resp. ``$_{\rm MD}$'') are to
be evaluated for a radiation dominated (resp. matter dominated) era,
that is to say with $p=1/2$ (resp. $p=2/3$). The justification is the
following one. Just after inflation, the universe is matter dominated
and the solution~(\ref{eq:solLommelapp}) with $p=2/3$ should be used
in order to describe the evolution of $\sigma $. Since, in the present
case, we study a situation where the onset of oscillations happens
after reheating, one is in fact in the regime $m_{\sigma }t\ll 1$
during the whole matter dominated era and, as a result, the particular
solution remains constant and equal to $\sigma _0$ during this
phase. Then, after the reheating is completed, the universe becomes
radiation dominated and, in principle, one should solve again the
equation of motion but, this time, with $p=1/2$ rather than $p=2/3$
and with ``new initial conditions'' inherited from the reheating
era. However, at the beginning of the radiation era, we are still
before the onset of oscillations (\ie still in the regime $m_{\sigma
}t\ll 1$) and the solution with $p=1/2$ is still constant since this
asymptotic behavior does not depend on $p$. Consequently, the
matching of the two solutions is in fact trivial. This is means that,
in the radiation dominated era, the relevant solution is nothing but
the solution~(\ref{eq:solLommelapp}) with $p=1/2$. Therefore, at late
times, it is sufficient to use Eq.~(\ref{eq:partlate}) with $p=1/2$,
hence the above claim.

\par

If $T_{\rm osci}>T_{\rm rh}$, the situation is slightly more
complicated. This times, the onset of oscillations occurs during the
matter dominated phase and the matching between the solution with
$p=2/3$ and $p=1/2$ should be done in the $m_{\sigma }t\gg 1$
regime. However, as before, the two solutions are the same because
$\sigma $ scales as $a^{-3/2}$ independently of $p$, see above. As a
result, one can ignore the matching and still work with the solution
with $p=2/3$ at late times in the radiation dominated era. Therefore,
if $T_{\rm osci}>T_{\rm rh}$, one should use $\mu_{\rm MD}$ and
$\nu_{\rm MD}$.

\par

Let us now consider the homogeneous solution. At early times, it can
be approximated as:
\begin{equation}
\sigma_{\rm hom}(t)\,\simeq\,\frac{1}{2}
\left(\sigma_{\rm inf}-\sigma_0\right)\left(1 + 
\frac{1+\mu}{\nu}\right)
\left(\frac{tH_{\rm inf}}{p}\right)^{-\mu-1+\nu}\quad m_\sigma t \,\ll\,1\ ,
\label{eq:earlyhomnureal}
\end{equation}
where one has used the expressions~(\ref{eq:al1})
and~(\ref{eq:al2}). In particular, the last terms in
Eqs.~(\ref{eq:al1}) and~(\ref{eq:al2}) become negligible which implies
$\alpha _1\simeq \alpha _2$ and, therefore, the terms proportional to
$J_{-\nu}$ cancel out.  One notices that the homogeneous solution does
not remain constant prior to the onset of oscillations as it was the
case for the particular solution. One finds that $\rho_{\rm \sigma,
  kin}$ and $\rho _{\rm \sigma, pot}$ scale as $\propto
t^{2(-\mu-2+\nu)}$. For a matter dominated era, one therefore finds
$\rho_{\sigma}\propto a^{-9/2+3\nu}$, while for a radiation dominated
era one has $\rho_{\sigma }\propto a^{-5+4\nu}$.
On the other hand, the late time evolution of $\sigma$ reads as follows:
\begin{eqnarray}
\sigma_{\rm hom}(t)&\,\simeq\,& (m_\sigma t)^{-\mu-3/2}\Biggl[
  \sqrt{\frac{8}{\pi}}\alpha_1\cos\left(m_\sigma t -
  \frac{\pi}{4}\right)\cos\left(\nu\frac{\pi}{2}\right) +
  \sqrt{\frac{8}{\pi}}\alpha_2\sin\left(m_\sigma t -
  \frac{\pi}{4}\right)\sin\left(\nu\frac{\pi}{2}\right)\Biggr]\, ,
\quad m_\sigma t \gg 1\ .\nonumber \\
\label{eq:latenureal}
\end{eqnarray}
As expected, one finds the standard evolution at late times, with
$\rho_\sigma\,\propto\, a^{-3}$ since $\mu+3/2=3p/2$. In one uses the
expressions of $\alpha _1$ and $\alpha _2$ given in
Eqs.~(\ref{eq:al1}) and~(\ref{eq:al2}), this can be re-written as 
\begin{eqnarray}
\label{eq:homlate}
\sigma_{\rm hom}(t)&\, \simeq\, & 
\left(m_{\sigma }t\right)^{-\mu-3/2}\left(\sigma _{\rm inf}
-\sigma _0\right)\left(\frac{pm_{\rm \sigma}}{H_{\rm inf}}\right)^{\mu+1-\nu}
\sqrt{\frac{2}{\pi}}\frac{\Gamma (1+\nu)}{2^{1-\nu}}
\left(1+\frac{\mu+1}{\nu}\right)\,\nonumber\\
& &\quad\times
\cos\left(m_{\sigma}t-\frac{\pi}{4}-\frac{\pi \nu}{2}\right)\, ,
\quad m_\sigma t\,\gg\,1\ .
\end{eqnarray}
As before, one should distinguish two different cases. If $T_{\rm
  osci}>T_{\rm rh}$, then the onset of oscillations occurs in the
matter dominated era. As a consequence, at the end of the reheating
stage (in the regime $m_\sigma t\,\gg \,1$), the homogeneous solution
is given by
\begin{eqnarray}
\label{eq:homlate2}
\sigma_{\rm hom}(t)\, &\simeq & 
\left(m_{\sigma }t\right)^{-\mu_{\rm MD}-3/2}\sqrt{\frac{2}{\pi}}
\left(\sigma _{\rm inf}-\sigma _0\right)
\left(\frac{p_{\rm MD}m_{\rm \sigma}}{H_{\rm inf}}\right)^{\mu_{\rm MD}+1-\nu_{\rm MD}}
\frac{\Gamma (1+\nu_{\rm MD})}{2^{1-\nu_{\rm MD}}}
\left(1+\frac{\mu_{\rm MD}+1}{\nu_{\rm MD}}\right)
\nonumber \\ & & \times
\cos\left(m_{\sigma}t-\frac{\pi}{4}-\frac{\pi \nu_{\rm MD}}{2}\right)\, .
\end{eqnarray}
Then, one should match this solution to the solution valid in the
radiation dominated era. However, as in the case of the particular
solution, since one is in the regime $m_{\sigma }t\gg 1$, the two
solutions scale as $a^{-3/2}$ and the matching becomes trivial. As a
consequence, one finds
\begin{equation}
  \Omega_{\sigma,\rm hom,osci}\,\simeq \, {1\over 6}\left({\sigma_{\rm
        inf}-\sigma_0 \over M_{\rm Pl}}\right)^2
  \left({p_{\rm MD}m_\sigma\over H_{\rm
        inf}}\right)^{2(\mu_{\rm MD}+1-\nu_{\rm MD})}\frac{2^{2\nu_{\rm MD}-1}}{\pi}
\Gamma ^2(1+\nu_{\rm MD})
  \left(1+\frac{\mu _{\rm MD}+1}{\nu_{\rm MD}}\right)^2
  \ .\label{eq:sc1}
\end{equation}
Finally, the case $T_{\rm rh}>T_{\rm osci}$ remains to be treated. In
this situation, one needs to take into account the change in the
values of $\mu$ and $\nu$ and there is no other choice than performing
the matching explicitly at the time of reheating. This time, the
matching turns out to be non trivial. The whole matter dominated era
occurs in the regime $m_{\sigma}t\ll 1$ and, to leading order [using
again the expressions~(\ref{eq:al1}) and~(\ref{eq:al2})] one obtains
that
\begin{eqnarray}
\label{eq:sigreh}
\sigma \vert_{\rm rh} &\simeq & \frac{\sigma _{\rm inf}-\sigma _0}{2}
\left(1+\frac{\mu_{\rm MD}+1}{\nu_{\rm MD}}\right)
\left(\frac{H_{\rm inf}}{H_{\rm rh}}\right)^{\nu_{\rm MD}-\mu_{\rm MD}-1}\, ,\\
\label{eq:dersigreh}
t\frac{{\rm d}\sigma }{{\rm d}t}\biggl \vert _{\rm rh} &\simeq &
 \frac{\sigma _{\rm inf}-\sigma _0}{2}
\left(1+\frac{\mu_{\rm MD}+1}{\nu_{\rm MD}}\right)\left(-\mu_{\rm MD}-1+\nu_{\rm MD}\right)
\left(\frac{H_{\rm inf}}{H_{\rm rh}}\right)^{\nu_{\rm MD}-\mu_{\rm MD}-2}\, .
\end{eqnarray}
In the radiation dominated era, the solution is given by $\sigma
=\beta _1(m_{\sigma }t)^{-\mu _{\rm RD}-1}J_{\nu_{\rm
    RD}}(m_{\sigma}t)+\beta _2(m_{\sigma }t)^{-\mu _{\rm
    RD}-1}J_{-\nu_{\rm RD}}(m_{\sigma}t)$ and one should determine the
coefficients $\beta_1$ and $\beta_2$ by matching, at reheating, the
above solution to the values~(\ref{eq:sigreh})
and~(\ref{eq:dersigreh}). Straightforward calculations show that
$\beta _1\gg \beta _2$. Then, using the asymptotic behavior of the
Bessel function $J_{\nu _{\rm RD}}(m_{\sigma }t)$ that appears in the
dominant branch of the solution in the regime $m_{\sigma }t\gg 1$, \ie
after the onset of oscillations, one can check that this solution
scales as $a^{-3/2}$, the proportionality coefficient being, as
before, directly related to $\Omega_{\sigma,\rm hom,osci}$. One finds 
\begin{eqnarray}
\Omega_{\sigma,\rm hom,osci}\,& \simeq &\, 
{1\over 6}\left({\sigma_{\rm
    inf}-\sigma_0 \over M_{\rm Pl}}\right)^2
\left({p_{\rm MD}m_\sigma\over H_{\rm rh}}
\right)^{2\left(\mu_{_{\rm RD}}+1-\nu_{_{\rm RD}}\right)}
\left({H_{\rm rh}\over H_{\rm
    inf}}\right)^{2\left(\mu_{_{\rm MD}}+1-\nu_{_{\rm MD}}\right)}
\frac{2^{2\nu_{\rm RD}-3}}{\pi}\frac{\Gamma ^2(1+\nu_{\rm RD})}{\nu_{\rm RD}^2}
\nonumber \\ &\times &
\left(1+\frac{\mu_{\rm RD}+1}{\nu_{\rm RD}}\right)^2
\left(\mu_{\rm MD}-\mu_{\rm RD}+\nu_{\rm MD}+\nu_{\rm RD}\right)^2\ .
\label{eq:sc1b}
\end{eqnarray}
As expected, one can check that, in the limit where all the quantities
labelled ``${\rm RD}$'' becomes equal to their counterparts labelled
``${\rm MD}$'', the above expression exactly reduces to
Eq.~(\ref{eq:sc1}).

\par

Summarizing our result, one has
\begin{eqnarray}
\label{eq:omnup1}
\Omega_{\rm \sigma,osci} &\simeq & \frac{2^{2\mu_{\rm MD}+2}}{3\pi}
\Gamma^2\left({\mu_{\rm MD}+3+\nu_{\rm MD}\over2}\right)\Gamma^2
\left({\mu_{\rm MD}+3-\nu_{\rm MD}\over2}\right)\left({\sigma_0 \over M_{\rm Pl}}\right)^2
\nonumber \\ & & 
+{1\over 6}\left({\sigma_{\rm
        inf}-\sigma_0 \over M_{\rm Pl}}\right)^2
  \left({p_{\rm MD}m_\sigma\over H_{\rm
        inf}}\right)^{2(\mu_{\rm MD}+1-\nu_{\rm MD})}\frac{2^{2\nu_{\rm MD}-1}}{\pi}
\Gamma ^2(1+\nu_{\rm MD})
  \left(1+\frac{\mu _{\rm MD}+1}{\nu_{\rm MD}}\right)^2 \, ,\quad T_{\rm osci}>T_{\rm rh}\, ,\\
\label{eq:omnup2}
\Omega_{\rm \sigma,osci} &\simeq & 
\frac{2^{2\mu_{\rm RD}+2}}{3\pi}
\Gamma^2\left({\mu_{\rm RD}+3+\nu_{\rm RD}\over2}\right)\Gamma^2
\left({\mu_{\rm RD}+3-\nu_{\rm RD}\over2}\right)\left({\sigma_0 \over M_{\rm Pl}}\right)^2
\nonumber \\ & & 
+{1\over 6}\left({\sigma_{\rm
    inf}-\sigma_0 \over M_{\rm Pl}}\right)^2
\left({p_{\rm MD}m_\sigma\over H_{\rm rh}}
\right)^{2\left(\mu_{_{\rm RD}}+1-\nu_{_{\rm RD}}\right)}
\left({H_{\rm rh}\over H_{\rm
    inf}}\right)^{2\left(\mu_{_{\rm MD}}+1-\nu_{_{\rm MD}}\right)}
\frac{2^{2\nu_{\rm RD}-3}}{\pi}\frac{\Gamma ^2(1+\nu_{\rm RD})}{\nu_{\rm RD}^2}
\nonumber \\ & & \times
\left(1+\frac{\mu_{\rm RD}+1}{\nu_{\rm RD}}\right)^2
\left(\mu_{\rm MD}-\mu_{\rm RD}+\nu_{\rm MD}+\nu_{\rm RD}\right)^2\ ,
\quad T_{\rm osci}<T_{\rm rh}\, . 
\end{eqnarray}
Let us emphasize that, in order to obtain the above expressions, we
have assumed that the total energy density at the onset of
oscillations is only made of two pieces, one originating from the
particular solution and the other from the homogeneous solution. This
means that we have neglected the cross terms that, in principle,
should contribute. This is justified by the fact that, in practice,
one solution always dominates the other. Which one is dominating
depends on the region explored in the parameter space. It is also
convenient to define
\begin{equation}
{\cal A}_1\,\equiv\,\frac{2^{\mu_{\rm MD}+3/2}}{\sqrt{\pi}}
\Gamma\left(\frac{\mu_{\rm MD}+\nu_{\rm MD}+3}{2}\right)
    \Gamma\left(\frac{\mu_{\rm MD}-\nu_{\rm MD}+3}{2}\right)\ ,
\label{eq:prefacA1}
\end{equation}
and ${\cal A}_2$ expressed in the same way as ${\cal A}_1$ above, but
with $\mu_{\rm MD}$ and $\mu_{\rm MD}$ replaced by $\mu_{\rm RD}$ and
$\nu_{\rm RD}$; The coefficients ${\cal B}_1$ and ${\cal B}_2$ are
given by
\begin{eqnarray}
{\cal B}_1\,&\equiv&\, \sqrt{\frac{2}{\pi}}\frac{\Gamma
  (1+\nu_{\rm MD})}{2^{1-\nu_{\rm MD}}}\left(1+\frac{\mu_{\rm
      MD}+1}{\nu_{\rm MD}}\right)\ , \\
\label{eq:prefacB1}
{\cal B}_2\,&\equiv& \,\frac{2^{\nu_{\rm RD}-3/2}}
{\sqrt{\pi}}\frac{\Gamma(1+\nu_{\rm RD})}{\nu_{\rm RD}}
\left(1+\frac{\mu_{\rm RD}+1}{\nu_{\rm RD}}\right)
\left(\mu_{\rm MD}-\mu_{\rm RD}+\nu_{\rm MD}+\nu_{\rm RD}\right)\ .
\label{eq:prefacB2}
\end{eqnarray}
With these definitions, the above expressions~(\ref{eq:omnup1})
and~(\ref{eq:omnup2}) exactly match the ones used in the main text,
see Eqs.~(\ref{eq:omeganureal1}) and~(\ref{eq:omeganureal2}).

\par

The other possibility is when $\nu $ is imaginary namely $c >
(\mu+1)/p$. For $p=2/3$, this case corresponds to $c>3/4$. Writing
$\nu=i\hat\nu$, one has $\hat\nu>0$ growing with $c$. This brings in
non-trivial modifications for the scaling of the modulus energy
density. Following the same reasoning as before, one finds that the
contribution due to the particular solution can be expressed as
\begin{equation}
\Omega_{\sigma,\rm part,osci}\,\simeq \, {2^{2\mu +2}\over
  3\pi}\Gamma^2\left({\mu+3\over2}+i{\hat{\nu}\over2}\right)
\Gamma^2\left({\mu+3\over2}-i{\hat{\nu}\over2}
\right)\left({\sigma_0\over m_{\rm
    Pl}}\right)^2\ .\label{eq:ooscpart2}
\end{equation}
As before, in the above equation, one should use quantities labelled
``{\rm MD}'' or ``{\rm RD}'' according to $T_{\rm osci}>T_{\rm rh}$ or
$T_{\rm osci}<T_{\rm rh}$. The difference with
Eq.~(\ref{eq:ooscpart1}) comes from the fact that the numerical
prefactors, which are of order unity if $\nu$ is real, may become
quite small if $\nu$ is pure imaginary and its modulus is large. This
point is elaborated further in more detail in the main text.

\par

Turning to the homogeneous solution, we finds that it scales as
follows. At early times, one no longer finds $\alpha _1\simeq \alpha
_2$ and this his implies that
\begin{equation}
\sigma (t)\simeq t^{-\mu -1}\, , \quad m_\sigma t \,\ll\,1 \, .
\end{equation}
Keeping in mind that $\nu$ is imaginary, this gives the following
scaling of the kinetic energy density: $\rho_{\rm kin, hom}\,\propto\,
t^{2(-\mu-2)}$, and it is easy to see that at early times, the
potential energy density scales similarly. At late times, the
homogeneous solution evolves as
\begin{eqnarray}
\sigma(t)&\,\simeq\,&\left(m_\sigma t\right)^{-\mu-3/2}\Biggl[
  \alpha_1\sqrt{\frac{8}{\pi}}\cos\left(m_\sigma t -
    \frac{\pi}{4}\right){\rm cosh}\left(\frac{\hat\nu\pi}{2}\right)
    +i\alpha_2\sqrt{\frac{8}{\pi}}\sin\left(m_\sigma t -
      \frac{\pi}{4}\right){\rm sinh}\left(\frac{\hat\nu\pi}{2}\right)
\Biggr]\, , \quad m_{\sigma }t\gg 1\, .\nonumber \\
\label{eq:ltimes2}
\end{eqnarray}
As expected and as it was the case before for the case of $\nu $ real,
one can check that $\sigma $ scales as $a^{-3/2}$ regardless of
$p$. At this point, as was done before in the case where $\nu $ was
real, one has to distinguish the case where the onset of oscillations
occurs before or after the reheating. Let us start with the case
$T_{\rm osci}>T_{\rm rh}$. Since the oscillations start in the matter
dominated era, one is interested in the regime $m_{\sigma }t\gg 1$
with $p=2/3$. The corresponding solution is given by
Eq.~(\ref{eq:ltimes2}). Expressing the coefficients $\alpha _1$ and
$\alpha _2$ using Eqs.~(\ref{eq:al1}) and~(\ref{eq:al2}), one obtains
the following result
\begin{eqnarray}
\label{eq:solnucomplex}
\sigma &\simeq & (m_{\sigma }t)^{-\mu_{\rm MD}-3/2}\frac{\sigma _{\rm inf}-\sigma _0}{2}
\sqrt{\frac{8}{\pi}}\left(\frac{p_{\rm MD}m_{\sigma }}{H_{\rm inf}}\right)^{\mu_{\rm MD}+1}
\sqrt{\frac{\pi \hat{\nu}_{\rm MD}}{\sinh(\pi \hat{\nu}_{\rm MD})}}
\sqrt{1+\frac{(\mu_{\rm MD}+1)^2}{\hat{\nu}_{\rm MD}^2}}
\nonumber \\ & & \times
\left[\cos \Upsilon _{\rm MD}\cos\left(m_{\sigma }t-\frac{\pi}{4}\right)
\cosh\left(\frac{\pi \hat{\nu}_{\rm MD}}{2}\right)
+\sin \Upsilon _{\rm MD}\sin \left(m_{\sigma }t-\frac{\pi}{4}\right)
\sinh\left(\frac{\pi \hat{\nu}_{\rm MD}}{2}\right)\right]\, ,
\end{eqnarray}
where $\Upsilon_{\rm MD}\equiv \hat{\nu}_{\rm MD}\ln
[pm_{\sigma}/(2H_{\rm inf})]-\Theta_{\rm MD} -\Psi_{\rm MD}$, $\Theta
_{\rm MD}$ being the phase of the complex number $\Gamma
(1+i\hat{\nu}_{\rm MD})$ and $\Psi _{\rm MD}$ the phase of $1-i(\mu
_{\rm MD}+1)/\hat{\nu}_{\rm MD}$. As usual, this function scales as
$a^{-3/2}$. Then one should match this solution with the solution in
the radiation dominated era (still in the regime $m_{\sigma }t\gg
1$). However, the solution in the radiation dominated era also scales
as $a^{-3/2}$. As a consequence, the matching is trivial. Therefore,
using the above equation in the limit where $\vert \nu \vert \gg 1$
(which simply amounts to replace, in the above expression, $\sinh x$
and $\cosh x$ by ${\rm e}^x/2$), one obtains
\begin{equation}
\Omega_{\sigma,\rm hom, osci}\,\simeq \, 
{1\over 6}\left({\sigma_{\rm
    inf}-\sigma_0 \over M_{\rm Pl}}\right)^2
\left({p _{\rm MD}m_\sigma\over H_{\rm
    inf}}\right)^{2(\mu_{\rm MD}+1)}\hat{\nu}_{\rm MD}
\left[1+\left(\frac{\mu _{\rm MD}+1}{\hat{\nu }_{\rm MD}}\right)^2
\right]\ , \quad T_{\rm osci}>T_{\rm rh}\, ,
\label{eq:sc2}
\end{equation}
where we have ignored the remaining trigonometric functions since they
give contributions of order one. It is remarkable that all the
exponential dependence in $\hat{\nu}$ has cancelled out in this
expression. 

\par

When $T_{\rm rh}> T_{\rm osci}$, the onset of oscillations occurs in
the radiation dominated era and one additional matching is required as
explained previously. At reheating, we have 
\begin{eqnarray}
\label{eq:sigrehcomplexnu}
\sigma \vert_{\rm rh} &\simeq & \left(\sigma _{\rm inf}-\sigma _0\right)
\left(\frac{H_{\rm inf}}{H_{\rm rh}}\right)^{-\mu_{\rm MD}-1}
\sqrt{1+\left(\frac{\mu_{\rm MD}+1}{\hat{\nu}_{\rm MD}}\right)^2}
\cos\left[\hat{\nu}_{\rm MD}
\ln\left(\frac{H_{\rm inf}}{H_{\rm rh}}\right)+\Psi_{\rm MD}\right]\, ,\\
\label{eq:dersigrehcomplexnu}
t\frac{{\rm d}\sigma }{{\rm d}t}\biggl \vert _{\rm rh} &\simeq &
-\left(\sigma _{\rm inf}-\sigma _0\right)
\left(\frac{H_{\rm inf}}{H_{\rm rh}}\right)^{-\mu_{\rm MD}-2}
\hat{\nu }_{\rm MD}
\left[1+\left(\frac{\mu_{\rm MD}+1}{\hat{\nu}_{\rm MD}}\right)^2\right]
\sin\left[\hat{\nu}_{\rm MD}
\ln\left(\frac{H_{\rm inf}}{H_{\rm rh}}\right)\right]\, .
\end{eqnarray}
These equations should be compare with Eqs.~(\ref{eq:sigreh})
and~(\ref{eq:dersigreh}). As before, in the radiation dominated era,
the solution is given by $\sigma =\beta _1(m_{\sigma }t)^{-\mu _{\rm
    RD}-1}J_{\nu_{\rm RD}}(m_{\sigma}t)+\beta _2(m_{\sigma }t)^{-\mu
  _{\rm RD}-1}J_{-\nu_{\rm RD}}(m_{\sigma}t)$, where $\nu _{\rm RD}$
is now a complex number. The coefficients $\beta_1$ and $\beta_2$ are
determined by matching the previous solution to the
values~(\ref{eq:sigrehcomplexnu})
and~(\ref{eq:dersigrehcomplexnu}). Then, the solution is completely
specified. In the regime $m_{\sigma }t\gg 1$, one obtains
\begin{eqnarray}
\label{eq:solnucomplex2}
\sigma (t)&\simeq & (m_{\sigma}t)^{-\mu_{\rm RD}-3/2}\left(\sigma _{\rm inf}-\sigma _0\right)
\sqrt{\frac{2}{\pi}}\left(\frac{H_{\rm inf}}{H_{\rm rh}}\right)^{-\mu _{\rm MD}-1}
\left(\frac{m_{\sigma }p}{H_{\rm rh}}\right)^{\mu _{\rm RD}+1}
\frac{{\rm e}^{\pi \hat{\nu}_{\rm RD}/2}}{2\hat{\nu}_{\rm RD}} 
\sqrt{1+\left(\frac{\mu _{\rm MD}+1}{\hat{\nu }_{\rm MD}}\right)^2}
\sqrt{\frac{\pi \hat{\nu}_{\rm RD}}{\sinh\left(\pi \hat{\nu}_{\rm RD}\right)}}
\nonumber \\ & & \times
\Biggl\{
-\hat{\nu}_{\rm MD}
\sqrt{1+\left(\frac{\mu _{\rm MD}+1}{\hat{\nu }_{\rm MD}}\right)^2}
\sin\left[\hat{\nu}_{\rm MD}\ln \left(\frac{H_{\rm inf}}{H_{\rm rh}}\right)\right]
\sin\left[\hat{\nu}_{\rm RD}\ln \left(\frac{2H_{\rm rh}}{m_{\sigma}p}\right)
+\Theta_{\rm RD}+m_{\sigma }t-\frac{\pi }{4}\right]
\nonumber \\ & & 
+\hat{\nu}_{\rm RD}
\sqrt{1+\left(\frac{\mu _{\rm RD}+1}{\hat{\nu }_{\rm RD}}\right)^2}
\cos\left[\hat{\nu}_{\rm MD}\ln \left(\frac{H_{\rm inf}}{H_{\rm rh}}\right)+\Psi_{\rm MD}\right]
\cos\left[\hat{\nu}_{\rm RD}\ln \left(\frac{2H_{\rm rh}}{m_{\sigma}p}\right)
+\Theta_{\rm RD}+\Psi_{\rm RD}+m_{\sigma }t-\frac{\pi }{4}\right]
\Biggr\}\, .\nonumber \\
\end{eqnarray}
As expected, this solution scales as $a^{-3/2}$. Again, it is
interesting to observe how any exponential dependence in
$\hat{\nu}_{\rm RD}$ is in fact exactly cancels out. It is also
interesting to study the case where all the quantities labelled
``${\rm RD}$'' becomes equal to their counterparts labelled ``${\rm
  MD}$''. In this case, lengthy calculations using trigonometric
identities show that the term in the curled bracket reduces to
\begin{eqnarray}
\biggl\{-\biggr\}=
\hat{\nu}_{\rm MD}
\sqrt{1+\left(\frac{\mu _{\rm MD}+1}{\hat{\nu }_{\rm MD}}\right)^2}
\cos \Psi _{\rm MD}
\left[\cos \Upsilon _{\rm MD}\cos\left(m_{\sigma }t-\frac{\pi}{4}\right)
+\sin \Upsilon _{\rm MD}\sin \left(m_{\sigma }t-\frac{\pi}{4}\right)
\right]\, ,
\end{eqnarray}
and recalling the definition of $\Psi _{\rm MD}$ (which implies that
$\cos \Psi _{\rm MD}$ exactly cancels the square root in the above
formula), it is easy to demonstrate that Eq.~(\ref{eq:solnucomplex2})
exactly reduces to Eq.~(\ref{eq:solnucomplex}). Finally, one obtains
\begin{eqnarray}
\Omega_{\rm \sigma,osci} &\simeq & \frac{1}{6}\frac{(\sigma _{\rm inf}-\sigma _0)^2}{M_{\rm Pl}^2}
\left(\frac{H_{\rm inf}}{H_{\rm rh}}\right)^{-2\mu _{\rm MD}-2}
\left(\frac{m_{\sigma }p_{\rm MD}}{H_{\rm rh}}\right)^{2\mu _{\rm RD}+2}
\frac{1}{\hat{\nu}_{\rm RD}}\left[1+\left(\frac{\mu _{\rm MD}+1}{\hat{\nu}_{\rm MD}}\right)^2
\right]\biggl\{-\biggr\}^2\, , \quad T_{\rm rh}> T_{\rm osci}\, .
\end{eqnarray}
This completes our analysis of the case where $\nu $ is complex.

\par

As before, the total $\Omega_{\sigma,\rm osci}$ is roughly given by
the sum of $\Omega_{\sigma,\rm hom, osci}$ and $\Omega_{\sigma,\rm
  part, osci}$. Summarizing our results 
one obtains
\begin{eqnarray}
\label{eq:omnum1}
\Omega_{\rm \sigma,osci} &\simeq & \frac{2^{2\mu_{\rm MD}+2}}{3\pi}
\Gamma^2\left({\mu_{\rm MD}+3\over2}+i\frac{\hat{\nu}_{\rm MD}}{2}\right)\Gamma^2
\left({\mu_{\rm MD}+3\over2}-i\frac{\hat{\nu}_{\rm MD}}{2}\right)
\left({\sigma_0 \over M_{\rm Pl}}\right)^2
\nonumber \\ & & 
+{1\over 6}\left({\sigma_{\rm
    inf}-\sigma_0 \over M_{\rm Pl}}\right)^2
\left({p _{\rm MD}m_\sigma\over H_{\rm
    inf}}\right)^{2(\mu_{\rm MD}+1)}\hat{\nu}_{\rm MD}
\left[1+\left(\frac{\mu _{\rm MD}+1}{\nu _{\rm MD}}\right)^2
\right] \, , \quad T_{\rm osci}>T_{\rm rh}\, ,
\\
\label{eq:omnum2}
\Omega_{\rm \sigma,osci} &\simeq & 
\frac{2^{2\mu_{\rm RD}+2}}{3\pi}
\Gamma^2\left({\mu_{\rm RD}+3\over2}+i\frac{\hat{\nu}_{\rm RD}}{2}\right)\Gamma^2
\left({\mu_{\rm RD}+3\over2}-i\frac{\hat{\nu}_{\rm RD}}{2}\right)
\left({\sigma_0 \over M_{\rm Pl}}\right)^2
\nonumber \\ & & 
+\frac{1}{6}\frac{(\sigma _{\rm inf}-\sigma _0)^2}{M_{\rm Pl}^2}
\left(\frac{H_{\rm inf}}{H_{\rm rh}}\right)^{-2\mu _{\rm MD}-2}
\left(\frac{m_{\sigma }p_{\rm MD}}{H_{\rm rh}}\right)^{2\mu _{\rm RD}+2}
\frac{1}{\hat{\nu}_{\rm RD}}\left[1+\left(\frac{\mu _{\rm MD}+1}{\hat{\nu}_{\rm MD}}\right)^2
\right]\biggl\{-\biggr\}^2\, , \quad T_{\rm rh}> T_{\rm osci}\, .
\end{eqnarray}
These equations are the counterparts of Eqs.~(\ref{eq:omeganureal1})
and (\ref{eq:omeganureal2}) in the case where the quantity $\nu$ is
complex. As was done before, it is convenient to define a prefactor
${\cal A}_3$ (the equivalent of ${\cal A}_1$, see before) by
\begin{equation}
{\cal A}_3\,\equiv\, \frac{2^{\mu_{\rm MD} +1}}{\sqrt{\pi}}
\Gamma\left({\mu_{\rm MD}+3\over2}+i{\hat{\nu}_{\rm MD}\over2}\right)
\Gamma\left({\mu_{\rm MD}+3\over2}-i{\hat{\nu}_{\rm MD}\over2}\right)\ .
\label{eq:prefacA3}
\end{equation}
As before, if $T_{\rm osci}<T_{\rm rh}$, one should rather use a
prefactor ${\cal A}_4$ defined as ${\cal A}_3$ above, but with all
$\mu_{\rm MD}$ and $\nu_{\rm MD}$ replaced by their values for the
radiation dominated era, $\mu_{\rm RD}$ and $\nu_{\rm RD}$. Finally, for the 
homogeneous solution, one introduces the coefficients
\begin{eqnarray}
{\cal B}_3\,&\equiv\,& \hat{\nu}_{\rm MD}^{1/2}
\left[1+\left(\frac{\mu _{\rm MD}+1}{\hat{\nu }_{\rm MD}}\right)^2
\right]^{1/2}\ ,\label{eq:prefacB3}\\
{\cal B}_4\,&\equiv&\,\frac{1}{\sqrt{\hat{\nu}_{\rm  RD}}}
\left[1+\left(\frac{\mu _{\rm MD}+1}{\hat{\nu}_{\rm  MD}}\right)^2
\right]^{1/2}\biggl\{-\biggr\}\ .\label{eq:prefacB4}
\end{eqnarray}
With definition The above expressions~(\ref{eq:omnum1})
and~(\ref{eq:omnum2}) reduces exactly to the expressions used in the
main text, namely Eqs.~(\ref{eq:omeganucomplex1})
and~(\ref{eq:omeganucomplex2}).

\subsection{Case $-c^2H^2/2$}
\label{subsec:minuscase}

As explained in the main text, the potential can be written under the
approximate form:
\begin{equation}
V(\sigma)\,\simeq\,\frac{1}{2}\tilde c^2 H^2\left(\sigma-\sigma_n\right)^2\,+\,
{\lambda_n\over (n+4)!}\,{(\sigma-\sigma_n)^{4+n}\over M_{\rm Pl}^n}\ ,\label{eq:potmc2}
\end{equation}
with:
\begin{equation}
\sigma_n\,=\,\left[\frac{(n+3)!}{\lambda_n}c^2\,H^2M_{\rm Pl}^n\right]^{\frac{1}{n+2}}\ ,
\quad
\tilde c^2\,=\,(n+2)c^2\ .
\end{equation}
The vev $\sigma_n$ indicates the (time-dependent) local minimum of the
effective potential. 

As discussed in the main text, the potential can be further
approximated by
$V(\sigma)\,\simeq\,\lambda_n\sigma^{n+4}/[(n+4)!M_{\rm Pl}^n]$ in the
limit $\sigma\,\gg\,\sigma_n$.  Then one can show, as follows, that
the field will evolve to values of order $\sigma_n$ in a fraction of
an e-fold of order $(\sigma/\sigma_n)^{-(n+2)/2}$. In the following,
we write $\sigma_{n,\rm inf}$ the value $\sigma_n(H=H_{\rm
  inf})$. Consider the field to be initially at rest, $\dot\sigma_{\rm
  inf}=0$. Then, neglecting the damping term $H\dot\sigma$ in the
equation of motion, one can rewrite this latter equation as
\begin{equation}
  \left. H^{-2}\frac{\ddot\sigma}{\sigma}\right\vert_{H=H_{\rm inf}}\,\simeq\,
-c^2\left(\frac{\sigma_{\rm inf}}{\sigma_{n,\rm inf}}\right)^{n+2}\ ,
\end{equation}
which shows that the field initially evolves on the e-folding scale
announced above. As soon as the field starts to evolve, it converts
its potential energy into kinetic energy; since this occurs in a
fraction of an e-fold, expansion damping can be ignored and the
conversion is nearly fully efficient: $\rho_{\sigma,\rm
  kin}\,\simeq\,V(\sigma_{\rm inf})-V(\sigma)$. Once the kinetic energy
dominates over the potential energy, $\rho_{\sigma,\rm
  kin}\,\propto\,a^{-6}$, and one finds the approximate solution:
\begin{equation}
\sigma\,\simeq\,\sigma_{\rm inf} - \frac{2\sqrt{2V(\sigma_{\rm inf})}}{3H_{\rm inf}}
\left(1 - \frac{H}{H_{\rm inf}}\right)\ ,
\end{equation}
assuming matter domination after inflation. The factor
$\sqrt{2V(\sigma_{\rm inf})}$ is introduced to approximate the value
$\dot\sigma$ at the time at which the kinetic energy of the modulus
dominates over the potential energy. The ratio $\sqrt{V(\sigma_{\rm
    inf})}/H_{\rm inf}\,\approx\,\sigma_{\rm inf}(\sigma_{\rm inf}/\sigma_{n,\rm
  inf})^{(n+2)/2}$, therefore the above solution shows that $\sigma$
evolves down to $\sigma_n$ on the aforementioned e-folding scale.

When $\sigma\,\ll\,\sigma_n$ (and $H\,\gg\,m_\sigma/c$), one can
approximate the potential (\ref{eq:potmc2}) with the low order term
$+\tilde c^2H^2(\sigma-\sigma_n)^2$. In this case, the
solution of the equation of motion for $\sigma$ is the sum of a
particular and an homogeneous solution, namely
\begin{equation}
\label{eq:solminushcorr}
\sigma (t)=\alpha_n\sigma _n(t)+\sigma_{\rm hom}(t)\ ,
\end{equation}
with the definition
\begin{equation}
\alpha_n\,\equiv\,\left[\frac{4}{(n+2)^2} +
\frac{2}{n+2}-\frac{6p}{n+2}+p^2\tilde c^2\right]^{-1}p^2\tilde c^2\ .
\end{equation}
The homogeneous solutions can be expressed as follows. Reintroducing
the notation: $ \nu^2\,=\,(\mu+1)^2-p^2\tilde c^2\ , $ the solutions
read differently according to whether $\nu$ is real or imaginary. Let
us start with the case $\nu$ real. In this case, the solution
normalized to negligible initial kinetic energy and initial value of
the modulus $\sigma_1$ reads~\cite{Dimopoulos:2003ss}:
\begin{equation}
\sigma_{\rm hom}(t)\,=\,\sigma_1
\left(\frac{H}{H_{\rm inf}}\right)^{\mu+1}\left[
{1\over 2}\left(1+\frac{\mu+1}{\nu}\right)
\left(\frac{H}{H_{\rm inf}}\right)^{-\nu}
+
{1\over 2}\left(1-\frac{\mu+1}{\nu}\right)
\left(\frac{H}{H_{\rm inf}}\right)^{\nu}\right]\ .
\end{equation}
The time dependence is contained in the dependence of $H(t)$. Of
course $\sigma_1$ must be chosen so that the sum of $\sigma_{\rm
  hom}+\sigma_{\rm part}$ takes the right value at the end of
inflation. The question of the initial conditions in this case is
discussed in more detail in the main text, see
Sec.~\ref{subsubsec:potwithminus}. Depending on the value of $n$,
either the particular solution or the homogeneous solutions may
dominate at late times. In general, however, one should expect the
homogeneous solution to decay less rapidly, since this corresponds to:
$n < 2(-\mu+\nu)/(\mu+1 - \nu)$, and during matter domination,
$\mu=-1/2$.

\par

Let us now study the case where $\nu$ is imaginary. As usual, we write
$\nu =i\hat{\nu}$ and, then, the solution
reads~\cite{Dimopoulos:2003ss}:
\begin{equation}
\sigma_{\rm hom}(t)\,=\,\sigma_1\left(\frac{H}{H_{\rm inf}}
\right)^{\mu+1}\left\{
\cos\left[\hat{\nu}\log\left(\frac{H}{H_{\rm inf}}\right)\right] 
- \frac{\mu+1}{\hat{\nu}}
\sin\left[\hat{\nu}\log\left(\frac{H}{H_{\rm inf}}\right)\right]\right\}\ .
\end{equation}
The homogeneous solution scales less rapidly with time than the
particular solution if $n<-2\mu/(\mu+1)$.

\section{Some Concrete models of moduli evolution}
\label{sec:appendix:concretemodels}

In this appendix, we present specific supergravity examples of
inflationary scenarios which give different predictions of modulus
evolution. Since the moduli fields are scalar fields with a flat
potential which eventually get a supersymmetry-breaking mass of order
of the gravitino mass $m_{3/2}$, the simplest example is the
supersymmetry-breaking field or the Polonyi field itself (denoted $S$
in what follows). Therefore, we consider a system consisting of an
inflaton sector and a separate Polonyi sector. In the rest of this
appendix, we assume that the Polonyi sector is characterized by a
minimal K\"ahler potential, $K_S=|S|^2$, and a superpotential given by
$W_S\equiv M_S^2\left(\beta+S\right)$. Here $\beta$ is a constant
chosen as $\beta=\left(2-\sqrt{3}\right)M_{\rm Pl}$ so that the
potential energy vanishes at its minimum
$S_{\min}=\left(\sqrt{3}-1\right)M_{\rm Pl}$ when the inflaton sector
is absent. At the potential minimum the gravitino mass reads
$m_{3/2}M_{\rm Pl}=\left\langle{\rm e}^{K_{S_{\rm
        min}}/2}W_{S_{\min}}\right\rangle=M_S^2{\rm
  e}^{2-\sqrt{3}}$. Below we mostly take $M_{\rm Pl}=1$ but recover it
when appropriate.

\subsection{Small-field inflation models induced by the F-term}
\label{subsec:Fsmallfield}

First we consider the case inflation is driven by a F-term according
to a small-field scenario. We denote the fields in the inflaton sector
by $I$ collectively although this sector usually contains two or more
fields. We assume a minimal K\"ahler potential $K_I$ and a
superpotential $W_I$ that we do not specify in details. In the
following, we define the F-term scalar potential of each sector by
\begin{equation}
  V^{(F)}_j={\rm e}^{K_j}\lkk |D_jW_j|^2-3|W_j|^2 \rkk, 
\end{equation}
where $D_jW_l\equiv \partial_jW_l+\partial_jKW_l$. Our fundamental
assumption is that the inflaton and modulus sectors are
separated. Technically, this means $K=K_I+K_S$ and $W=W_I+W_s$. Then,
the total potential reads 
\begin{eqnarray}
\label{eq:pottotal} 
V_{\rm tot}^{(F)} &=& {\rm e}^{K}\lkk
|D_IW|^2+|D_SW|^2-3|W|^2\rkk =
{\rm e}^{|S|^2}V_I^{(F)}(I)+{\rm e}^{K_I}V_S^{(F)}(S)+V_{\rm int}^{(F)}(I,S)\, , 
\end{eqnarray}
where the interaction potential is given by 
\begin{eqnarray} 
V_{\rm int}^{(F)}&=&
{\rm e}^{|S|^2+K_I} \lkk
D_IW_I(\partial_IK_IW_S)^{\dagger}+(D_IW_I)^{\dagger}\partial_IK_IW_S
+|\partial_IK_IW_S|^2
+D_SW_SSW_I^{\dagger}+(D_SW_S)^{\dagger}S^{\dagger}W_I\right. \nonumber\\
&&\left. +|S|^2|W_I|^2-3W_I^{\dagger}W_S -3W_IW_S^{\dagger}\rkk.  
\end{eqnarray}
Note that the energy scales of $V_I$ and $V_S$ are given by
$V_I(I)\approx 3H_I^2M_{\rm Pl}^2$ and $V_S(S)=M_S^4\approx
m_{3/2}^2M_{\rm Pl}^2$ where $H_I$ is the Hubble parameter during
inflation. Let us also remark that the structure of
Eq.~(\ref{eq:pottotal}) is in fact typical of
supergravity~\cite{Nilles:1983ge}: although the two sectors are
separated, there is an interaction between them because they
``communicate'' through gravity. In the Minimal Supersymmetric
Standard Model (MSSM), this is a possible way to transmit the breaking
of supersymmetry to the observable sector~\cite{Nilles:1983ge}. The
same mechanism has also been used to argue that the quintessence and
inflaton fields must interact~\cite{Brax:2005uf} or to compute the
interaction of dark energy with the observable sector, see
Ref.~\cite{Brax:2006dc}.

\par

Since we are assuming a small-field inflation here one can easily see
that $|D_IW_I| > |W_I|$, and $V_I\approx |D_IW_I|^2\approx 3H_I^2
M_{\rm Pl}^2$.  In this case the modulus acquires a mass $\sim
V_I/M_{\rm Pl}^2 \approx 3H_I^2$ during inflation. After inflation,
$|I|$ takes a value $\sim M_{\rm Pl}$ and oscillates around the
potential minimum $I \equiv I_{\min}$ with $V_I(I_{\min})=0$, so that
$W_I$ and $D_IW_I$ are of the same order of magnitude on average.  If
the field oscillation is dominated by a quadratic potential, we find
the time average of the potential energy $\overline{V_I(I)}=\rho_{\rm
  tot}(t)/2=3H^2(t)M_{\rm Pl}^2/2$, so that we can parameterize
$\overline{D_IW_I}\equiv c_1HM_{\rm Pl}$, $\overline{W_I}\equiv
c_2HM_{\rm Pl}^2$ with $c_1$ and $c_2$ parameters of order of unity.
If we take $K_I=|I-I_{\min}|^2$, the effective potential for $S$
averaged over an inflaton oscillation period reads
\begin{eqnarray}
 \overline{V}_{\rm tot}(S,I_{\min})&=&
\lkk\lmk\frac{3}{2}+c_2^2\rmk H^2-f_1m_{3/2} H\rkk
\left| S- \frac{f_2m_{3/2}M_{\rm Pl} H}{\lmk 3/2+c_2^2\rmk H^2
-f_1m_{3/2} H}\right|^2 \nonumber\\
&&-\frac{|f_2|^2m_{3/2}^2M_{\rm Pl}^2H^2}{\lmk 3/2+c_2^2\rmk H^2
-f_1m_{3/2} H}+V_S(S)+\cdots.
\end{eqnarray}
Here $f_1$ and $f_2$ are numerical coefficients defined by
$f_1=2\beta(c_2+c_2^\ast)e^{\sqrt{3}-2}$
and $f_2=2c_2e^{\sqrt{3}-2}$, respectively.
Thus in the early field oscillation stage when $H$ is larger
than $m_{3/2}$,
the modulus has a mass larger than $H$ and its minimum is
located at
\begin{equation}
 S_{\min}\approx \frac{f_2m_{3/2}M_{\rm Pl} H}{\lmk 3/2+c_2^2\rmk H^2
-f_1m_{3/2} H}.
\end{equation}
As $H$ decreases, it eventually settles down to the absolute minimum
$S_{\min}=(\sqrt{3}-1)M_{\rm Pl}$ with a mass $\sqrt{2}m_{3/2}$ as
determined by $V_S(S)$.  Hence this model can be regarded as an
example of models where the modulus acquires a mass larger than the
Hubble parameter during both inflation and the subsequent field
oscillation regimes with a shift of the minimum of order of $M_{\rm
  Pl}$ typically.

\subsection{D-term hybrid inflation}
\label{subsec:dtermhybrid}

Next we consider the case in which inflation is induced by the D-term
potential. As the simplest realization we consider a hybrid inflation
model originally proposed in
Refs.~\cite{Halyo:1996pp,Binetruy:1996xj}.  Although this model is not
observationally viable any more due to the largeness of the cosmic
string tension produced after inflation, a number of remedies have
been proposed \eg in Ref.~\cite{Urrestilla:2004eh}.  In order to avoid
inessential complexity, we stick to the original model here to adopt
the superpotential with three chiral superfields $W_I=\lambda
Q\phi_+\phi_-$ together with $W_S=M_S^2(\beta +S)$ as in the previous
subsection.  $\phi_{\pm}$ has a U(1) gauge charge $\pm 1$, while $Q$
is neutral. Assuming the minimal K\"ahler potential,
$K=|Q|^2+|\phi_+|^2+|\phi_-|^2 +|S|^2$, the scalar potential is given
by $V_{\rm tot}=V_{\rm tot}^{(F)}+V^{(D)}$, with
\begin{eqnarray}
V_{\rm tot}^{(F)}&=& {\rm e}^K\biggl[
\lambda^2\left(|Q\phi_+|^2+|Q\phi_-|^2+|\phi_+\phi_-|^2
+3|Q\phi_+\phi_-|^2\right)
+\left(|\phi_+|^2+|\phi_-|^2+|S|^2+|Q|^2\right)
\left\vert\lambda Q\phi_+\phi_- +M_S^2(\beta+S)\right \vert^2
\nonumber \\ & &
-3M_S^4\beta^2+M_S^4-2M_S^4\beta (S+S^{\dagger})-M_S^4|S|^2
+\lambda M_S^2SQ^{\dagger}\phi_+^{\dagger}\phi_-^{\dagger}
+\lambda M_S^2S^{\dagger}Q\phi_+\phi_-\biggr] \, , \nonumber \\
V^{(D)}&=& \frac{g^2}{2}\left(|\phi_-|^2-|\phi_+|^2-\xi\right)^2, 
\label{VD} 
\end{eqnarray}
where the last term is the D-term contribution with $\xi>0$ being the
Fayet-Illiopoulos (FI) term. For $|Q|>g\sqrt{\xi}/\lambda\equiv Q_{\rm
  cri}$, the potential minimum with respect to $\phi_{\pm}$ is found
at $\phi_{\pm}=0$ and inflation is induced by the D-term energy
density $g^2\xi^2/2$. Without $W_S$ the F-term potential vanishes in
this regime and the motion of the inflaton is governed by a potential
generated by quantum corrections.  Then terms of the potential
relevant to $S$ are given by
\begin{eqnarray} 
V_{\rm  tot}^{(F)}(S,Q,\phi_{\pm}=0)=M_S^4{\rm e}^{|S|^2+|Q|^2}
\left[\left(|S|^2+|Q|^2\right)|\beta+S|^2-3\beta^2-2\beta 
\left(S+S^{\dagger}\right)-|S|^2+1\right]\, , 
\end{eqnarray}
so that the modulus acquires an extra
mass-squared, 
\begin{eqnarray}
\delta m_{S\rm eff}^2 \sim \frac{M_S^4}{M_{\rm Pl}^4}|Q|^2
\sim\frac{|Q|^2}{M_{\rm Pl}^2}m_{3/2}^2,
\end{eqnarray}
which is smaller than the original one as long as 
$|Q|$ is smaller than $M_{\rm Pl}$.  The shift of the potential
minimum of $S$ is of order of $|Q|^2/M_{\rm Pl}$.  Thus in this model
the modulus mass remains much smaller than the Hubble parameter
during inflation, and the shift of the minimum may also be 
much smaller than $M_{\rm Pl}$
depending on the value of $Q$.  Note, however, that the field
configuration of $S$ is determined by long-wave quantum fluctuations
in this case.

\par

As $Q$ becomes smaller than $Q_{\rm cri}$ the instability occurs with
respect to $\phi_-$ and inflation is terminated.  In this regime the
potential for $S$ acquires an additional term, $\delta V_S(S) ={\rm
  e}^{|S|^2+\cdots}\lambda^2|Q\phi_-|^2$.  This term induces a
correction to the modulus mass term 
\begin{equation} 
\delta m_{S \rm eff}^2=
\frac{\lambda^2|Q|^2|\phi_-|^2}{M_{\rm Pl}^2}
=\frac{\lambda^2|Q|^2}{M_{\rm Pl}^2}\lmk \xi
-\frac{\lambda^2}{g^2}|Q|^2\rmk \, ,
 \label{delta}
\end{equation} 
where we have inserted the minimum of $|\phim|^2$ for $Q<Q_{\rm cri}$.
The last expression takes its maximum at $|Q|^2=|Q_{\rm cri}|^2/2$ and
is given by $\delta m_{S \rm eff}^2=g^2\xi^2/4$. This means that after
inflation the effective mass of the Polonyi field rises significantly
to the level comparable to the Hubble parameter at that time.  When
this induced mass term is operative, the potential minimum for $|S|$
takes a value ${\cal O}\left(m_{3/2}^2/H^2\right)M_{\rm Pl}$ which
departs from the eventual minimum by $\sim M_{\rm Pl}$.

\subsection{Hybrid inflation with both F- and D-term contributions}
\label{subsec:hybridFD}

Next we consider a variant of D-term hybrid inflation model with a
non-vanishing F-term potential as an example of models where the
modulus mass remains much smaller than the Hubble parameter during
inflation but it acquires a large correction of order $H$ just after
inflation. Specifically we consider sneutrino hybrid inflation model
proposed in Ref.~\cite{Kadota:2005mt} with the minimal K\"ahler
potential.  The effects of non-minimal K\"ahler potential are
discussed, for example, in Ref.~\cite{Murayama:1993xu} for the
sneutrino inflation and in Ref.~\cite{Seto:2005qg} for D-term
inflation.

\par

The inflaton sector of the model consists of three species of chiral
superfields, $N_i^c$ containing (s)neutrinos ($i=1,2,3$), and
$\phi_{\pm}$ with U(1) gauge charge $\pm 1$ as in the above model.
The relevant part of the superpotential reads 
\begin{equation}
W_I =\frac{\lambda}{M_{\rm Pl}} N_i^cN_i^c \phi_- \phi_+ + \frac12
M_iN_i^cN_i^c\, , 
\end{equation}
besides the interaction between $N^c_i$ and lepton and Higgs fields.
One can always choose a basis for $N_i^c$ so that their mass matrix is
diagonal, and we take $N_i^c$ to be Majorana mass eigenstate fields
with real mass $M_i$. We assume, without loss of generality, the
inflaton sneutrino is the lightest heavy sneutrino $M_1\ll M_2,M_3$
and we are interested in the lower range of the preferred values of
heavy neutrino masses $M_i=10^{10}\sim 10^{15}\GeV$. D-term
contribution to the scalar potential with non-vanishing FI term
$\xi>0$ is given by $V_D(\phi_+,\phi_-)$ in Eq.~(\ref{VD}) as before.

The large amplitude of inflaton sneutrino gives a large effective mass
to the slepton and the Higgs field $H_u$ and they stay at the origin
and do not affect the inflation dynamics.  $\phi_+$ has a positive
mass during and after inflation and it stays at the origin all the
time.  For the discussion of inflationary dynamics, therefore, we
discuss the evolution of $\sn$ and $\phim$.  The tree-level potential
for the inflaton sector is then given by
\begin{eqnarray}
V_I^{(F)}+V_I^{(D)}&=& {\rm e}^{K_I}\lkk M^2|\sn|^2+\frac{1}{4}M^2|\sn|^4
+\frac{1}{4}M^2|\sn|^6 +\lmk\frac{\lambda}{M_{\rm Pl}}\rmk^2|\sn|^4|\phim|^2
+\frac{1}{4}M^2|\sn|^4|\phim|^2\rkk \nonumber \\ & & 
+ \frac{g^2}{2}\left(\xi-|\phim|^2\right)^2\, .
\end{eqnarray}
We expect $\lambda \gg M/M_{\rm Pl}$, and therefore find $\phim=0$ for
$\left\vert\sn \right\vert^4>g^2\xi M_{\rm Pl}^4/\lambda^2\equiv \snc^4$ during inflation.
$\phim$ destabilizes for $\sn< \snc $ to reach its minimum at
$\phim=\sqrt{\xi}$ after inflation.  We find $\snc\lesssim 2M_{\rm Pl}$
for cosmologically relevant parameter values \cite{Kadota:2005mt}.

The correction to the Polonyi mass during inflation can be found from
the F-term potential with $\phi_{\pm}=0$. The interaction term reads
\begin{eqnarray}
V_{\rm int}\left(I,S\right)&=& M_S^2{\rm e}^{K_I+|S|^2}\biggl\{\lmk
M\sn+\frac{1}{2}M\left\vert\sn\right\vert^2\sn\rmk \sn \left(\beta+S^{\dagger}\right)+
\lmk
M\sn^{\dagger}+\frac{1}{2}M\left\vert\sn\right\vert^2\sn^{\dagger}\rmk \sn^{\dagger} 
\left(\beta+S\right)
\nonumber \\ & & 
+M_S^2\left\vert\sn^{\dagger}\left(\beta+S\right)\right\vert^2
+\lkk 1+S^{\dagger}\left(\beta+S\right)\rkk S
\frac{1}{2}M\left(\sn^{\dagger}\right)^2  
+\lkk 1+S\left(\beta+S^{\dagger}\right)\rkk S^{\dagger}
\frac{1}{2}M\sn^2 
\nonumber \\ & &
-\frac{3}{2}M\left(\sn^{\dagger}\right)^2\left(\beta+S\right)
-\frac{3}{2}M\sn^2\left(\beta+S^{\dagger}\right)\biggr\}
+{\rm e}^{K_I+|S|^2}|S|^2\left\vert\frac{1}{2}M\sn^2\right\vert^2.
\end{eqnarray}
For $\sn\sim M_{\rm Pl}$ we find the last term has the largest
contribution to the effective modulus mass, ${\cal O}(M)$.  This
correction, however, is still much smaller than the Hubble parameter
during inflation thanks to the assumption that inflation is driven by
the D-term potential energy which ensures that
\begin{equation}
V_I^{(D)}=\frac{g^2}{2}\xi^2 \gg V_I^{(F)}\supset M^2|\sn|^2\, .  
\end{equation}
As $\sn$ gets smaller than $\snc$, a phase transition occurs to make
$\phim$ nonvanishing and inflation is terminated. Its minimum is
located at 
\begin{equation} 
\left\vert\phim \right\vert^2\simeq  \xi -\lmk\frac{\lambda}
{gM_{\rm Pl}}\rmk^2|\sn|^4\, .  
\end{equation}
We therefore find 
\begin{equation}
{\rm e}^{|S|^2}V_I^{(F)}\supset {\rm e}^{|S|^2+K_I} \lmk\frac{\lambda}{M_{\rm
    Pl}}\rmk^2\left\vert\sn\right \vert^4\left\vert \phim \right\vert^2 
={\rm e}^{|S|^2+K_I}
\lmk\frac{\lambda}{M_{\rm Pl}}\rmk^2|\sn|^4\lkk \xi
-\lmk\frac{\lambda}{gM_{\rm Pl}}\rmk^2|\sn|^4\rkk \leq
{\rm e}^{|S|^2+K_I}\frac{1}{4}g^2\xi^2\, , 
\end{equation} 
where the equality holds at
$\left\vert\sn\right\vert^4=\left\vert\snc\right \vert^4/2$ in the
last inequality.  Thus one can see that in the early field oscillation
regime after inflation, the modulus acquires an effective mass of
order of the Hubble parameter during inflation.

\bibliography{references}

\end{document}